\documentclass[useAMS,usenatbib]{mn2e}

\usepackage{natbib,aas_macros,epsfig,graphicx,rotating,lscape,amssymb}

\title[Searching for streams in CORAVEL and RAVE]{Is the sky falling? Searching for stellar 
  streams in the local Milky Way disc in the CORAVEL and RAVE surveys}
\author[Seabroke et al.]{G. M. Seabroke$^{1}$\thanks{E-mail: gs310@ast.cam.ac.uk}, 
G. Gilmore$^{1}$, 
A. Siebert$^{2}$,
O. Bienaym\'{e}$^{3}$, 
J. Binney$^{4}$,
\newauthor
J. Bland-Hawthorn$^{5}$,
R. Campbell$^{6}$,
K. C. Freeman$^{7}$, 
B. Gibson$^{8}$,
\newauthor   
E. K. Grebel$^{9,10}$,
A. Helmi$^{11}$, 
U. Munari$^{12}$, 
J. F. Navarro$^{13}$,
Q. A. Parker$^{6,5}$,
\newauthor
A. Siviero$^{12}$,
M. Steinmetz$^{2}$,
F. G. Watson$^{5}$,  
R. F. G. Wyse$^{14}$, 
T. Zwitter$^{15}$,
\newauthor
J. Pe\~narrubia$^{13}$,
M.~C. Smith$^{1}$,
M. Williams$^{7}$\\
$^{1}$Institute of Astronomy, University of Cambridge, Madingley Road,  
Cambridge, CB3 0HA, UK\\
$^{2}$Astrophysikalisches Institut Potsdam, An der Sterwarte 16, D-14482
Potsdam, Germany\\
$^{3}$Observatoire de Strasbourg, 11 Rue de L'Universit\'{e}, 67000  
Strasbourg,
France\\
$^{4}$Rudolf Peierls Centre for Theoretical Physics, University of  
Oxford, 1
Keble Road, Oxford OX1 3NP, UK\\
$^{5}$Anglo-Australian Observatory, P.O. Box 296, Epping, NSW 1710,  
Australia\\
$^{6}$Macquarie University, Sydney, NSW 2109, Australia\\
$^{7}$RSAA, Mount Stromlo  
Observatory, Cotter Road, Weston Creek, Canberra, ACT 72611, Australia.\\
$^{8}$University of Central Lancashire, Preston PR1 2HE, UK\\
$^{9}$Astronomisches Institut, Universit\"{a}t Basel, Venusstrasse 7,
Binningen CH-4102, Switzerland\\
$^{10}$Astronomisches Rechen-Institut, Zentrum f\"ur Astronomie der
Universit\"at Heidelberg, M\"onchhofstr.\ 12-14, D-69120 Heidelberg,\\
Germany\\
$^{11}$Kapteyn Astronomical Institute, University of Groningen, P.O. Box  
800, 9700 AV Groningen, The Netherlands\\
$^{12}$INAF Osservatorio Astronomico di Padova, Via dell'Osservatorio  
8, Asiago I-36012, Italy\\
$^{13}$University of Victoria, P.O. Box 3055, Station CSC, Victoria, BC  
V8W 3P6, Canada\\
$^{14}$Johns Hopkins University, 366 Bloomberg Center, 3400 North Charles
Street, Baltimore, MD 21218, USA\\
$^{15}$Department of Physics, University of Ljubljana, Jadranska 19,  
Ljubljana, Slovenia}

\begin{document}

\date{Accepted . Received ; in original form }

\pagerange{\pageref{firstpage}--\pageref{lastpage}} \pubyear{2007}

\maketitle

\label{firstpage}

\begin{abstract}
We have searched for in-falling stellar streams on to the local Milky Way disc
 in the CORAVEL and RAVE surveys.  The CORAVEL survey consists of local dwarf
 stars (N\"ordstrom et al. Geneva-Copenhagen survey) and local Famaey et
 al. giant stars.  We select RAVE stars with radial velocities that are
 sensitive to the Galactic vertical space velocity (Galactic latitude $b < -45^{\circ}$).  Kuiper statistics have been employed to
 test the symmetry of the Galactic vertical velocity
 distribution functions in these samples for evidence of a net vertical flow that could be
 associated with a (tidal?) stream of stars with vertically coherent
 kinematics.  
  In contrast to the `Field of Streams' found in
 the outer halo, we find that the local volumes of the solar neighbourhood sampled by
 the CORAVEL dwarfs (complete within $\sim$3 $\times 10^{-4}$ kpc$^{3}$),
 CORAVEL giants (complete within $\sim$5 $\times 10^{-2}$ kpc$^{3}$) and RAVE
 (5-15 per cent complete within $\sim$8 kpc$^{3}$) are devoid of any
 vertically coherent streams
 containing hundreds of stars.  This is sufficiently sensitive to allow our RAVE
 sample to rule out the passing of the tidal stream of the disrupting
 Sagittarius (Sgr) dwarf galaxy through the solar neighbourhood.  This agrees with the most recent 
 determination of its orbit and dissociates it from the Helmi et al.
 halo stream.  Our constraints on the absence
 of the Sgr stream near the Sun could prove a useful tool for discriminating
 between Galactic potential models.   The lack of a net
 vertical flow through the solar neighbourhood in the CORAVEL giants and RAVE
 samples argues against the Virgo overdensity crossing the
 disc near the Sun.  There are no vertical streams in the CORAVEL giants and
 RAVE samples with
 stellar densities $\gtrsim$1.6 $\times 10^{4}$ and 1.5 $\times 10^{3}$ stars
 kpc$^{-3}$ respectively and therefore no evidence for locally enhanced dark matter. 
 \end{abstract}

\begin{keywords}
solar neighbourhood; Galaxy: disc; Galaxy: formation; Galaxy: kinematics and
dynamics; Galaxy: structure; dark matter
\end{keywords}

\section{Introduction}

Tidal streams are filaments of debris, containing stars, gas and possibly dark
matter (DM), that have become stripped from a disrupting satellite dwarf
galaxy (or star cluster) along its orbit with respect to the centre-of-mass of a
more massive galaxy.  The instantaneous position and velocity of a star
and the Galactic gravitational potential determines its past orbit.  For stars
in tidal streams, one can most reliably know their positions and velocities in the past.  This is because stars in tidal streams all originate, at one time in the past, from a single progenitor on a specific orbit.  It is this property of tidal streams that makes them excellent tracers of the
Galactic potential \citep{johnston1999,murali1999} while the degree of tidal-stream coherence provides a powerful constraint on the lumpiness of the
Galactic potential \citep{johnston2002}.  

The disruption timescales of
merging galaxies (\citealt{helmi1999a}, hereafter H99a) means the discovery
of new tidal streams, e.g. the Orphan stream
\citep{belokurov2006apaper,belokurov2007bpaper,grillmair2006}, can also lead
to the discovery of new satellites, e.g. Ursa Major II (UMa II,
\citealt{zucker2006paper,fellhauer2007}), which are of relevance to the `missing satellite problem' \citep{moore1999,klypin1999,benson2002}.  These discoveries offer more evidence of the hierarchical structure formation of the Milky Way Galaxy and provide accretion history constraints for simulations of this process (e.g. \citealt{abadi2003}).  If a tidal stream containing DM is passing through the solar neighbourhood, it would provide a `cold' flow of DM particles through the numerous direct detection experiments on Earth, increasing the possibility of a positive direct detection \citep{freese2004}.  

The first evidence that there may be tidal streams in the solar neighbourhood
was the discovery of two streams (\citealt{helmi1999b}, hereafter H99b) in
kinematic surveys of the local stellar halo \citep{beers1995,chiba1998}.  This
discovery was subsequently confirmed by \citet{chiba2000} and
\citet{kepley2007}.  

The most spectacular example of tidal streams is from the disruption of the Sagittarius (Sgr) dwarf spheroidal (dSph) galaxy, originally discovered by
\citet{ibata1994}.  Estimates of the central mass-to-light ratios of dwarf galaxies are high,
suggesting they and plausibly their tidal streams are all DM-dominated.  The most extensive sky
panorama of the Sgr stream is traced by M
giants \citep{majewski2003} from the Two-Micron All-Sky Survey (2MASS).  The stream is seen very clearly in the Southern
Galactic hemisphere, as well as heading towards the North Galactic Pole
(NGP). \citet{majewski2003} found the stream lies along a well-defined orbital
plane about the Galactic centre and speculated that the foreshortening of the
stream towards the NGP means it heads back towards the Galactic plane. The Sun
lies within $\sim$1 kpc of that plane, which is within the width of the
stream, leading \citet{majewski2003} to propose the possibility that the
stream crosses the Galactic plane near or in the solar neighbourhood.
\citet{freese2004} postulated that one of the H99b streams is part of the Sgr stream passing through the solar neighbourhood and that its DM density should be measurable in direct detection experiments.

\citet{law2005} modelled the Sgr stream and found that its two arms pass
through a similar position in the plane of the sky, near the Virgo overdensity
(VOD).  \citet{martinez-delgado2007} suggest that the VOD is part
of the Sgr stream, which crosses the Milky Way plane in the solar
neighbourhood, and show that the \citet{law2005} model passes through the
observed location of the VOD and the VOD stellar density is similar to the
model predictions.  However, there is a growing weight of evidence against the
Sgr stream existing in the solar neighbourhood.  Dynamical analyses of the
multiple wraps of the Sgr stream and the newly discovered Orphan stream found
in the `Field of Streams'  \citep{belokurov2006apaper} suggest that neither
the Sgr \citep{fellhauer2006a} nor Orphan streams \citep{fellhauer2006b} pass
near the Sun.  More recently, \citep{newberg2007} used Sloan Digital Sky
Survey (SDSS) imaging and Sloan Extension for Galactic Understanding and
Exploration (SEGUE) spectroscopic observations of F turnoff and blue
horizontal branch (BHB) stars to show that the Sgr stream is neither
coincident with the VOD nor passes through the solar neighbourhood.  Instead,
they find that it misses the Sun by 15 kpc, passing through the plane well outside the solar circle.

\citet{juric2005paper} used SDSS photometry to identify the VOD -- the largest
 clump of tidal debris so far detected in the outer halo.  It covers $\sim$1000
 deg$^{2}$ above the Galactic plane ($5 < Z < 15$ kpc), over the solar position in the Galaxy
 ($R \sim$ 7 kpc).  They interpreted it as a tidal stream or an invading dwarf
 galaxy.  \citet{juric2005paper} did not detect any downturn in the star
 counts towards lower Galactic latitudes ($b$), indicating the VOD could
 extend closer to the Galactic plane than the observations probe ($b >
 60^{\circ}$).  

With the advent of SDSS Data Release 5 and the discoveries of the VOD and Orphan stream, \citet{fuchs2006} re-analysed star counts from their Calar Alto Deep Imaging Survey (CADIS) data.  In retrospect,
 they find that overdensities in the vertical-density distribution of stars
 can be associated with the VOD in the 13-h CADIS field and possibly the Monoceros
 \citep{newberg2002} and Orphan streams in the 9-h CADIS field.
Simulations by \citet{fellhauer2007} show the 9-h field falls on the second wrap of the backward
 orbit of UMa II before it passes along the Orphan stream.  In this direction,
 \citet{fellhauer2007} demonstrated that the Orphan stream is $>$30 kpc from the
 Sun and its orbit does not bring it much closer to the solar neighbourhood (see their fig. 2).

The Monoceros stream is seen in the data of \citet{juric2005paper} at $3 \lesssim Z
\lesssim 5$ kpc and $R \sim$ 16 kpc.  Two
 other overdensities are also visible in the top right plot of \citet{juric2005paper} fig. 20 at ($R$, $Z$) $\sim$ (1.5, 6.5)
 and (9, 0.8) kpc (the latter is seen most clearly in the second row, right
 plot).  None of these four overdensities are seen to exist in the solar neighbourhood.  \citet{penarrubia2005} have modelled the Monoceros stream and their
simulations show that part of it may be seen in the direction of the 9-h field 
at heliocentric distances of the overdensity in the 9-h field (3-21 kpc),
again far from the solar neighbourhood.



Unlike in the Galactic disc, tidal streams in the outer halo are not
phase-mixed quickly due to its dynamical time-scale of $>$1 Gyr and so remain
coherent in configuration space for longer than in the disc.  The low stellar
density in the outer halo enables large tidal streams to be identified as
stellar overdensities in photometric surveys like the SDSS.  The shorter
dynamical timescales and much higher stellar densities cause tidal streams in
the inner Galaxy to quickly lose their coherence in configuration space but
they remain coherent in velocity space (H99a).  

Until recently, the full six-dimensional phase space of the solar
neighbourhood had not been systematically surveyed.  This was because, although the {\it
  Hipparcos} satellite mission \citep{esa1997} provided accurate parallaxes
and proper motions for $\sim$118 000 stars, these stars generally
lacked radial velocities (RVs).  Therefore, a large European consortium obtained kinematically unbiased RVs of {\it Hipparcos} stars of spectral type later than about F5 \citep{udry1997}.  Multi-epoch RVs of $\sim$45 000 stars were measured with the two CORrelation RAdial VELocities
(CORAVEL) photoelectric cross-correlation spectrometers
\citep{baranne1979,mayor1985}.  
There have been two public data releases and three published
analyses of stars in the CORAVEL database: the Geneva-Copenhagen
survey of 16 682 nearby CORAVEL F-G dwarfs, available via VizieR, is
described and analysed in \citet{nordstrom2004}, hereafter N04, and recently re-analysed in \citet{holmberg2007}; and a catalogue of the local
kinematics of 6691 CORAVEL K-M giants, also available via VizieR, is described
and analysed by \citet{famaey2005}, hereafter F05.

The RAdial Velocity Experiment (RAVE, \citealt{steinmetz2006paper}) is a spectroscopic survey measuring the RVs and stellar atmosphere parameters (temperature,
  metallicity and surface gravity) of up to one million stars in the range of magnitudes probing Galactic scales between the very local CORAVEL RV surveys and the more distant SEGUE RV survey.  RAVE started in 2003, using the Six-Degree Field (6dF) multi-object spectrograph on the 1.2-m UK Schmidt Telescope of the
  Anglo-Australian Observatory in Australia.  
  
The recent, timely availability of the CORAVEL and RAVE data sets permit
direct searches for the presence of tidal streams in the solar neighbourhood.
\citet{helmi2006} have already searched the N04 CORAVEL dwarf data set for
signatures of past accretion on to the local disc.  They find that stars with
a common progenitor should show distinct correlations between orbital
apocentre (A), pericentre (P) and $Z$ angular momentum ($J_{Z}$).  In APJ-space, their analysis reveals a statistically significant excess of stars on orbits of common eccentricity, analogous to the pattern expected for merger debris.  They identify three coherent groups with distinct metallicity and age distributions that they assert correspond to the remains of disrupted satellites.

In this paper, we (quite literally) take an orthogonal approach by simply
posing the question: is there any net vertical flow through the solar
neighbourhood?  This is a different question to that explored by
\citet{gould2003} and \citet{bell2007}: we are not attempting to constrain the
  amount of phase-space substructure.  The well-established  Galactic stellar
components in the solar neighbourhood (thin and thick discs and inner halo)
are kinematically symmetric about the Galactic plane.  Therefore, any net
vertical flow could reasonably be associated with a tidal stream of stars with
vertically coherent kinematics.  Hence, within the sample volumes of the
CORAVEL and RAVE surveys, we can directly test whether the Sgr stream and VOD pass near the Sun and consequently whether direct DM detection experients should expect enhanced modulated signals from these or any other streams.

We organize the paper by analysing the data sets in order of increasing
sampled volume.  Ergo, we begin with the CORAVEL dwarfs in Section
\ref{s:dwarfs}.  In Section \ref{s:binary},  we use this sample to show that
the effect of binarity is negligible in the analysis of velocity distribution functions, which is important for our purpose here as well as many future science applications of RAVE and SEGUE data.  We introduce our statistical test to identify streams in Section \ref{s:dwarf_lsr}.  This test is applied to the CORAVEL giants in Section \ref{s:giants} and all the RAVE stars in Section \ref{s:rave}.  

This is not the first science paper to make use of the RAVE data
set. \citet{smith2007paper} used high-velocity RAVE stars to constrain the
local Galactic escape speed.  Veltz et al. (2007) used 2MASS to select RAVE
stars with $0.5 \le J - K_{S} \le 0.7$, which corresponds to K3-K7 dwarfs and
G3-K1 (red clump) giants, to identify kinematic and density discontinuities
between the thin disc, thick disc and inner halo towards the South Galactic
Pole (SGP).  We introduce the RAVE stellar populations in Section
\ref{s:ravepop}.  We discuss our results in Section \ref{s:discussion} before
ending on an historical aside in Appendix A.

\section{CORAVEL dwarfs}
\label{s:dwarfs}

\subsection{Derivation of the space velocities of the CORAVEL dwarfs}
\label{s:dwarfs_dev}

The Geneva-Copenhagen survey is complete, all-sky, magnitude-limited and
 kinematically unbiased.  Its observational input catalogue was selected from a compilation
 of catalogues available in the literature with Str\"omgren $uvby\beta$
 photometry of nearby F and G stars, mainly from the surveys by
 \citet{olsen1983,olsen1993,olsen1994a,olsen1994b}.  The input catalogue
 was observed using both CORAVELs.  Their fixed, late-type cross-correlation template
 spectra match the spectra of the majority of the input catalogue
 stars.  The multi-epoch RVs (generally two or more) have a modal mean error of the mean RVs of 0.25 km s$^{-1}$ (see N04 fig. 3).

 The vast majority of these stars have proper motions in the {\it
 Tycho-2} catalogue \citep{hog2000}.  This catalogue was constructed by
 combining the {\it Tycho} star-mapper measurements of the {\it
 Hipparcos} satellite with the Astrographic Catalogue based on
 measurements in the Carte du Ciel and other ground-based catalogues. 
 The typical mean error in the total proper motion vector is 1.8 mas yr$^{-1}$.

 The primary source of distance for these stars is {\it Hipparcos}
 trigonometric parallax. This is adopted if its relative
 error ($\sigma_{\pi}/\pi$) is accurate to 13 per cent or better,
 otherwise the photometric distance calibrations for F and G dwarfs by
 \citet{crawford1975} and \citet{olsen1984} are used, with an
 uncertainty of only 13 per cent.  Distances are not provided for stars
 with unreliable {\it Hipparcos} parallax ($\sigma_{\pi}/\pi >$ 13 per cent) and when
 photometric distances cannot be calculated.  This occurs when the star
 is missing the necessary photometry and/or it falls outside the
 photometric calibrations.  The absence of a distance estimate or
 RV measurement reduces the size of the sample with
 all full six dimensional phase-space information to 13 240 stars.   The space velocity
 components in the Galactic cardinal directions, $U$ (towards the Galactic centre), $V$ (in the direction of
 Galactic rotation) and $W$ (towards the NGP), are computed for all the
 stars with (mean) RVs, proper motions and distances. 

\subsection{Orbital angular momenta of the CORAVEL dwarfs}

\begin{figure}
        \psfig{figure=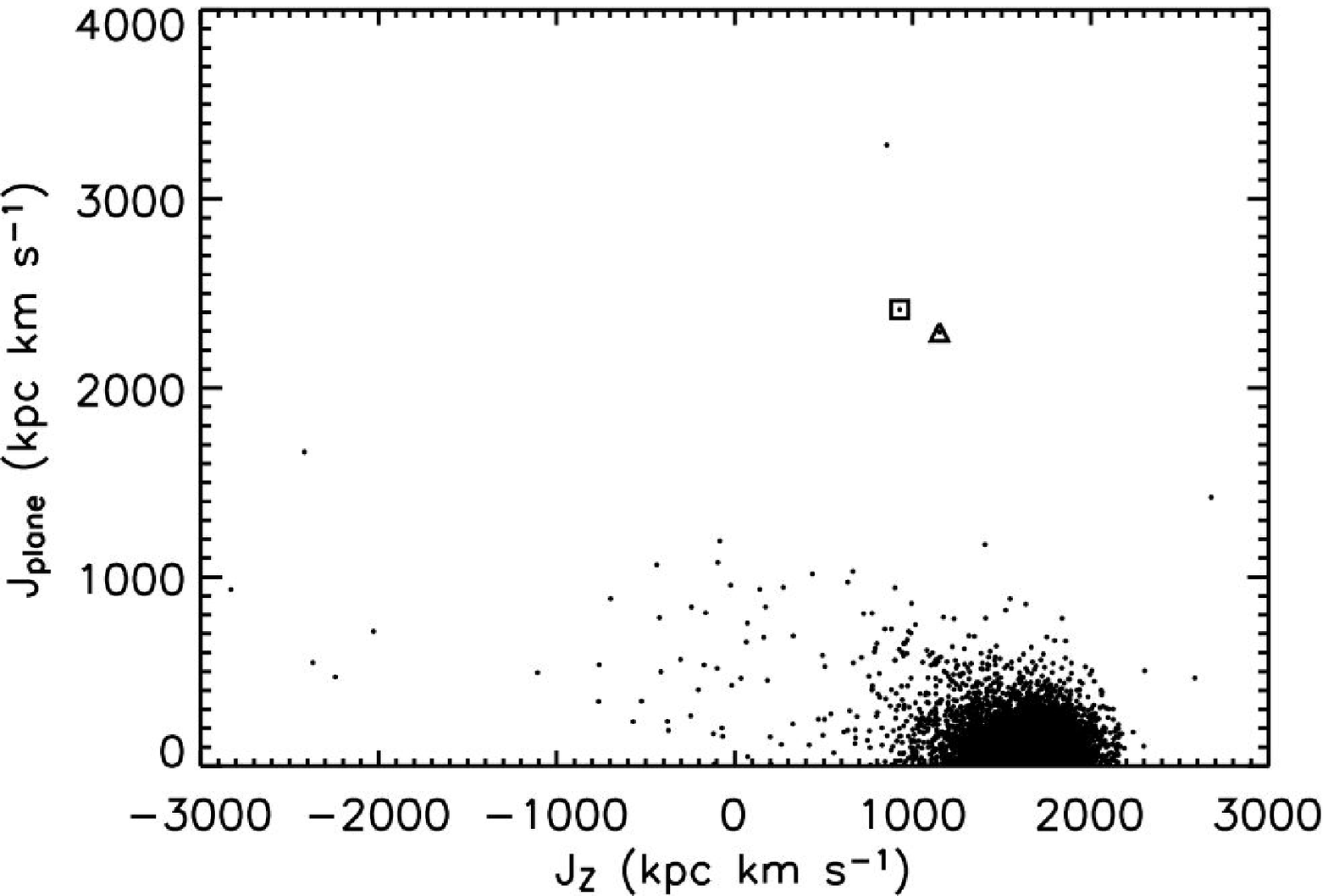,width=1.\columnwidth} 
        \psfig{figure=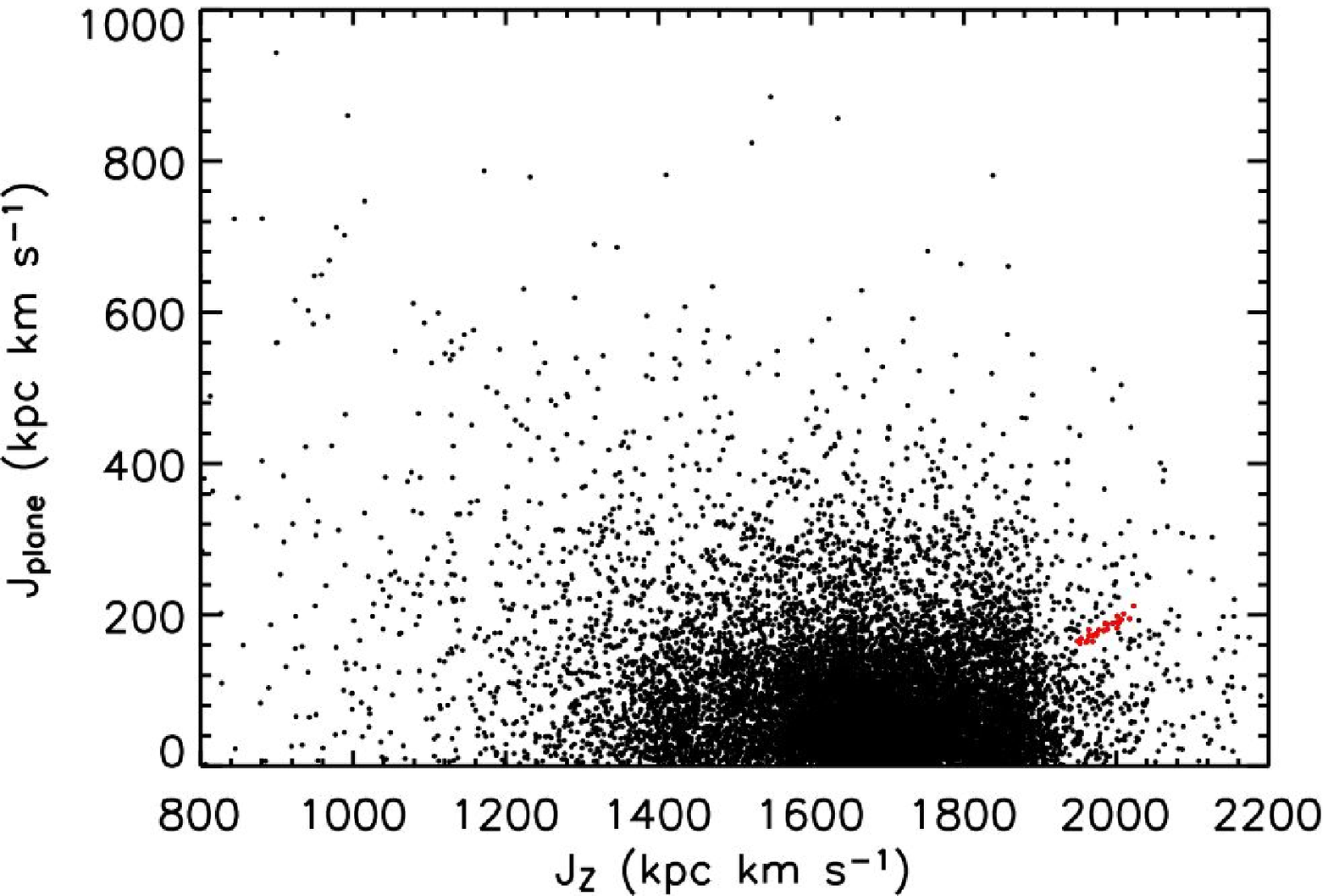,width=1.\columnwidth}
         \caption{Distribution of CORAVEL dwarfs in the plane of orbital angular momentum components, where $J_{plane} = \sqrt{J_{X}^{2} + J_{Y}^{2}}$.  {\it Top}: The triangle is star HD 175305 and the square is  CD $-$80 328.  {\it Bottom}: The red dots are stars in an apparent linear structure.}  
         \label{fig:gencopj}
          \end{figure}
          
The availability of distances to each CORAVEL dwarf allows their
Galactic-centred cartesian co-ordinates ($X$, $Y$, $Z$) to be calculated,
where ($X_{\odot}$, $Y_{\odot}$, $Z_{\odot}$) = (8, 0, 0) kpc, which in
combination with their ($U$, $V$, $W$) relative to the Galactic Standard of
Rest (GSR) permits their components of orbital angular momentum ($J$) to be
resolved into

\begin{eqnarray}
J_{X} &=& YW_{GSR} - ZV_{GSR},\\
J_{Y} &=& ZU_{GSR} - XW_{GSR},\\
J_{Z} &=& XV_{GSR} - YU_{GSR},
\label{equ:j}
\end{eqnarray}
\noindent where 

\begin{equation}
U_{GSR} = U_{LSR} =  U + U^{\odot}_{LSR},
\label{equ:ugsr}
\end{equation}

\begin{equation}
V_{GSR} = V_{LSR} + V_{rot} = V + V^{\odot}_{LSR} + V_{rot},
\label{equ:vgsr}
\end{equation}

\begin{equation}
W_{GSR} = W_{LSR} = W + W^{\odot}_{LSR},
\label{equ:wgsr}
\end{equation}

\noindent where ($U^{\odot}_{LSR}$, $V^{\odot}_{LSR}$, $W^{\odot}_{LSR}$) =
(10.00, 5.25, 7.17) km s$^{-1}$ \citep{dehnen1998a} is the solar motion, decomposed into its cardinal directions, relative to the Local Standard of Rest (LSR), and $V_{rot} = 220$ km s$^{-1}$ is the amplitude of Galactic rotation towards $l=90^{\circ}$ and $b=0^{\circ}$ (IAU 1985 convention; see \citealt{kerr1986}).  Fig. \ref{fig:gencopj} shows the CORAVEL dwarf sample is dominated by
high-angular momentum disc stars with only $\sim$100 low-angular momentum
halo stars.  

Two stars on prograde high-inclination orbits (the triangle and square in the top plot of Fig. \ref{fig:gencopj}) share similar angular momenta as the H99b streams (cf. their fig. 2). The triangle in the top plot of Fig. \ref{fig:gencopj} is HD 175305 and has a photometric metallicity of $-1.39$ dex.  It is also in the \citet{beers1995} and \citet{chiba1998} catalogues with spectroscopic metallicities of  $-1.42$ and $-1.54$ dex respectively.  Its kinematics, $(U, V, W)_{GSR} = (-58, 142, -286)$ km
s$^{-1}$, means it is one of the stars in the H99b stream
moving towards the SGP.  \citet{beers2000} and \citet{kepley2007} explicitly cite this star as a member of one of the H99b streams.   

The square in the top plot of Fig. \ref{fig:gencopj} is CD $-$80 328 and has a photometric metallicity of $-1.98$ dex.  It is not in the \citet{beers1995} nor \citet{chiba1998} catalogues.  Its very low metallicity is spectroscopically confirmed by \citet{beers1999} to be $-2.09$ dex.  Its kinematics, $(U, V, W)_{GSR} = (-193, 117, 303)$ km s$^{-1}$, and metallicity suggest it could be a member of the H99b stream
moving towards the NGP.   However, its kinetic energy is too large to be consistent with the energies of the other members of the stream (listed in table 7 of \citealt{chiba2000} and the energies were calculated using table 4 of \citealt{beers2000}).  Therefore, we consider this star to be a possible outlier, like the different outlier found by H99b.   The red dots in the bottom plot of Fig. \ref{fig:gencopj} apparently align in a linear structure.   The mode of their photometric metallicities ($-0.6$ dex) agrees with the second group of tidal debris found by \citet{helmi2006}.  However, the younger and wider age distribution of the linear structure argues against a tidal origin, suggesting a chance alignment is more probable.

The lack of strong clumping and apparent smoothness of the distribution in
Fig. \ref{fig:gencopj} suggests there are not any significant, coherent tidal
streams in the sample (the tidal debris identified by \citealt{helmi2006} does
not manifest itself in coherent tidal streams).  The \citet{helmi2006} APJ-space
analysis of the CORAVEL dwarfs is similar to the
H99a entropy technique (also used
in H99b), since it partitions the parameter space and counts the number of
stars in each cell.  Therefore, we do not re-apply these techniques to the
CORAVEL dwarfs in Fig. \ref{fig:gencopj}.  In addition, these techniques cannot be applied to
RAVE data as so far we have not derived accurate distances to RAVE stars (see
Section \ref{s:rave}).  Instead, we apply the RAVE data analysis technique
employed in Section \ref{s:rave} to the CORAVEL samples, in order to characterize the solar neighbourhood $W$ distribution and to allow comparison between the CORAVEL and RAVE results.  First, the next section demonstrates that RAVE data can be used on an equal footing with the CORAVEL data.

\subsection{Effect of Binarity on $W$ velocity distribution}
\label{s:binary}

      \begin{figure}
\centering
        \psfig{figure=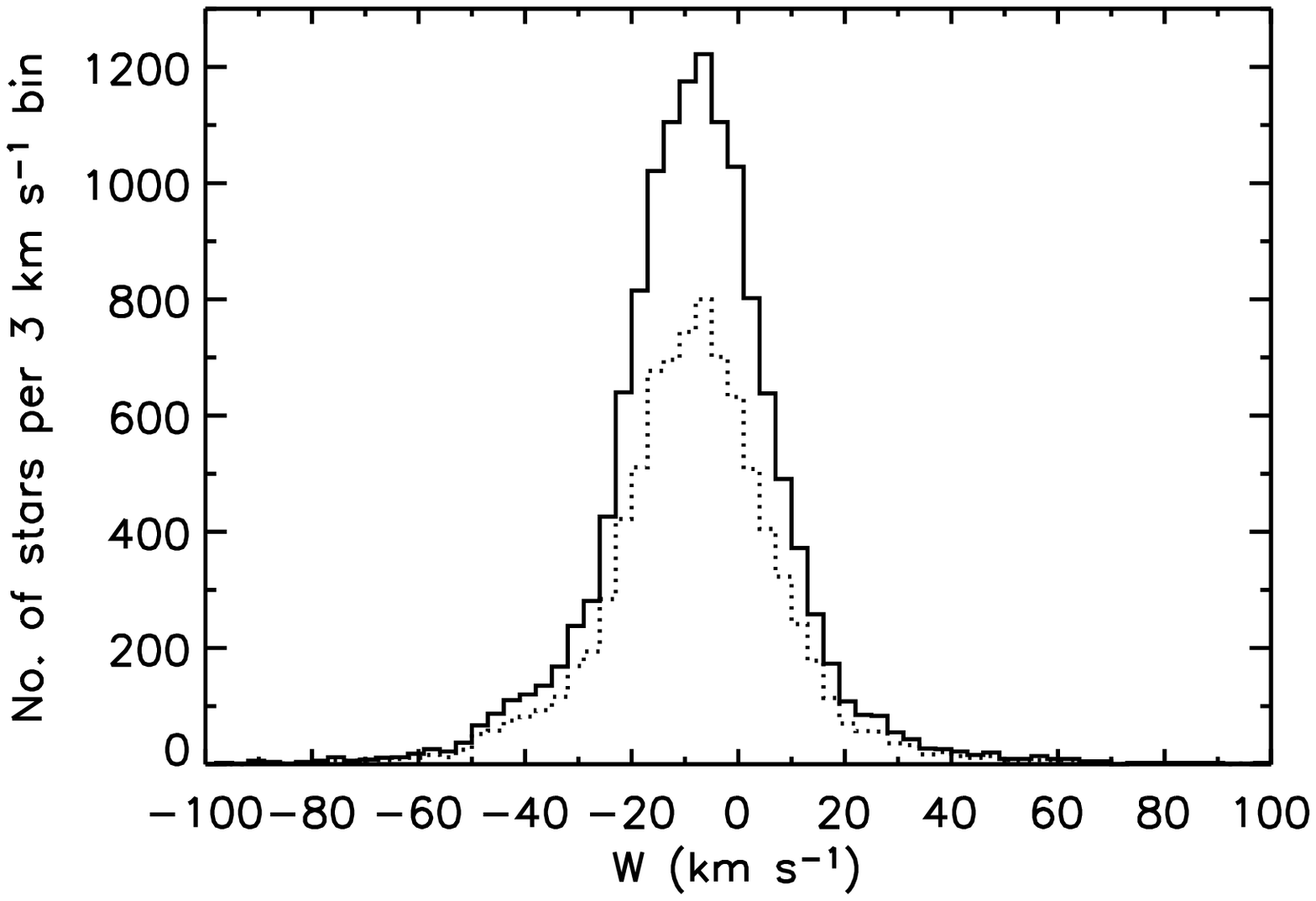,width=1.\columnwidth}
        \psfig{figure=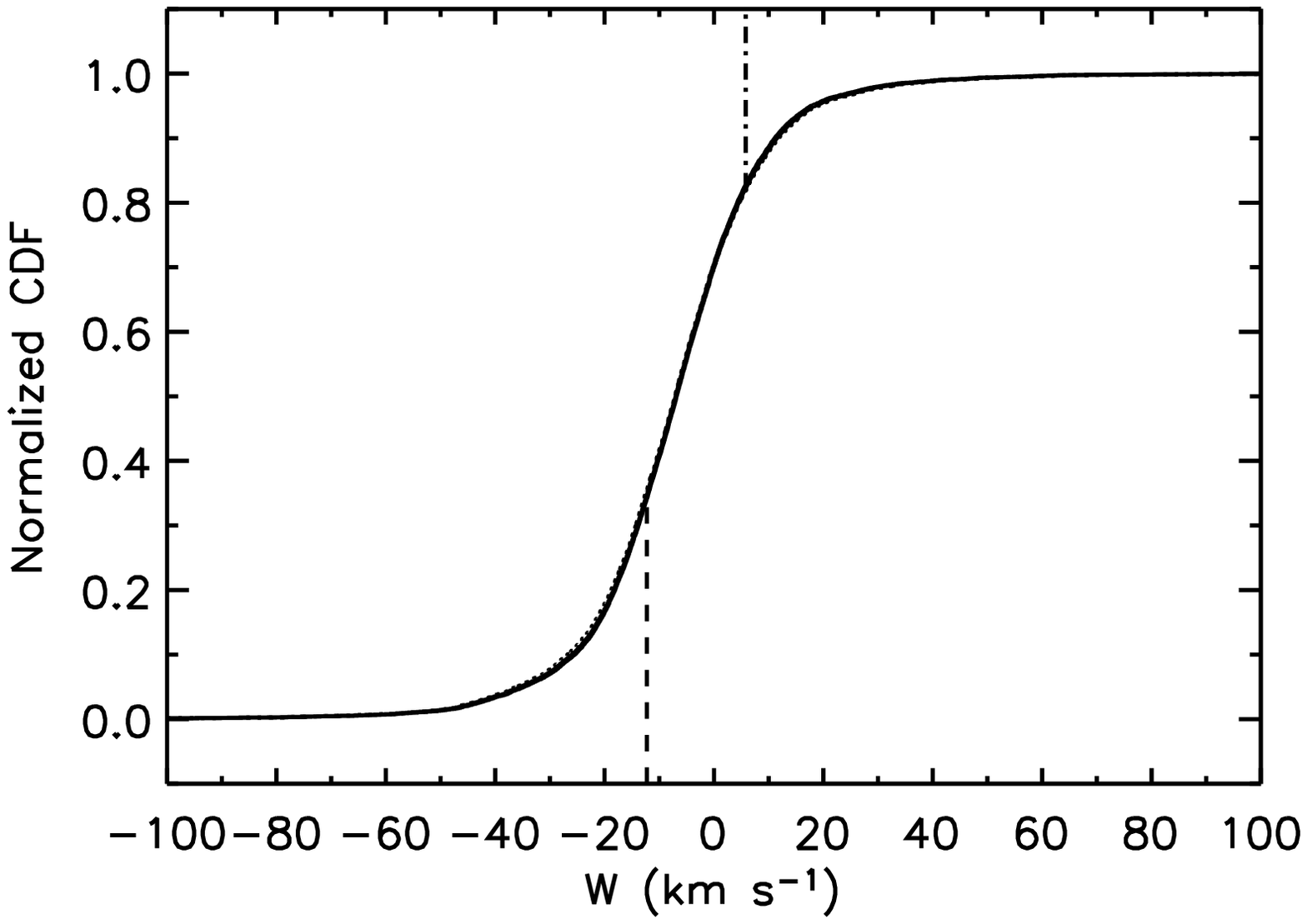,width=1.\columnwidth}
         \caption{{\it Top:} $W$ distribution of the CORAVEL dwarfs: all the 
        13 240 stars (solid histogram) and the 8589 single stars (dotted histogram).  The 3 km s$^{-1}$ velocity bin sizes are chosen to be twice the average space velocity error.  {\it Bottom:} Normalized CDF of all the 13 240 CORAVEL giants (solid line) and the 8589 single stars (dotted line) as a function of $W$, where the maximum differences between them are indicated by vertical lines: $D_{+}$ (dot-dashed line) and $D_{-}$ (dashed line).}  
                                  \label{fig:dwarf_w}
          \end{figure}

A double correlation peak may identify a spectroscopic binary
from a single RV measurement but an average of more than four Geneva-Copenhagen observations are available per CORAVEL dwarf, allowing N04
  to identify 3223 stars (out of 16 682, 19 per cent) as spectroscopic binaries of all
  kinds.  If the
velocities can be properly assigned to the two binary components, the
centre-of-mass velocities of double-lined spectroscopic binaries can be computed by the
method of \citet{wilson1941} without a full orbital solution.
N04 provide systemic velocities of double-lined
spectroscopic binaries
where possible, otherwise the raw average RV is given for CORAVEL dwarfs,
including single-lined spectroscopic binaries.  

The average scatter of individual RVs of single-lined
  spectroscopic binaries, which have orbital periods of years or less, is $\sim$15 km s$^{-1}$.  This scatter is an order of magnitude larger than the average error in each
  space velocity component of 1.5 km s$^{-1}$.  Therefore, the mean RV of a single-lined
  spectroscopic binary may be more representative of the instantaneous  radial component
of  the orbital  motion in the  binary system than its systemic
  line-of-sight motion in the Galaxy, resulting in inaccurate and even
  potentially misleading space velocities.  

Although any individual
  space velocity could be unreliable due to binarity, the statistical effect
  of binarity on the velocity distribution function of a large number of stars
  has not been empirically constrained before.  This is due to the sparsity of RV data available with sufficient temporal coverage to generate good
  binarity detection statistics.  The
  Geneva-Copenhagen survey is the largest, homogeneous RV survey
  with repeat measurements: 62 993 new RV observations of 13 464
  programme stars, representing $\sim$1000 nights of data.  Observations typically cover a time span of 1-3 years but some extend
  over a decade or more, allowing the majority of binaries to be identified.

The statistical effect of binarity on the velocity distribution function is
investigated for the first time here because it has fundamental implications
for our aims in this
paper, as well as the observational strategies and science of on-going RV surveys.  The vast majority of RAVE (and SEGUE) stars have never been spectroscopically observed
before and RAVE (and SEGUE) observes most of them only once.  Ideally
kinematic studies should use centre-of-mass velocities but when these are not available, it
is important to know how sensitive the analysis is to the kinematic affects of
binarity.

We investigate the kinematic effect of binarity by comparing the $W$
velocity distribution function of all the CORAVEL dwarfs, including binaries, to the
corresponding distribution function of single stars (see top of Fig. \ref{fig:dwarf_w}).  N04 provide a catalogue flag ($f_{b}$) that identifies confirmed
or suspected binaries, where the information can come from one or
several sources such as photometry, RV or astrometry.  They
identified 3537 visual binaries (out
of 16 682 stars, 21 per cent) and the majority of systems with periods below 1000 days should be
flagged. Single
stars, for which the derived space velocities are more reliable, are defined by
  N04 as having a null $f_{b}$ catalogue entry (8589
  stars out of the 13 240 with space velocities, 65 per cent).  Some CORAVEL dwarfs only have two observations.  Accordingly, some long-period and/or
low-amplitude binary stars will not be identified by the $f_{b}$ flag and will
thus be present in the sample of single stars.  Their effect should be
negligible compared to a bona fide sample of single stars due to the small amplitude
of their orbital
velocities.  
  
\citet{kepley2007} were able to use a parametric test (the Shapiro-Wilk
test for deviations from normality) to search for streams in their stellar halo sample
because the halo velocity distribution is approximated by a Gaussian. 
Both $W$
velocity distribution functions in the top of Fig. \ref{fig:dwarf_w}
approximately consist of the sum of three Gaussians.  These
represent the young thin disc, where $\sigma_{W} \sim$ 10-15 km s$^{-1}$,
the old thin disc, where $\sigma_{W} \sim$ 15-20 km s$^{-1}$, and the thick
disc, where $\sigma_{W} >$ 30 km s$^{-1}$ \citep{seabroke2007}.  The
shape of the resulting distribution is a Gaussian with positive kurtosis
(leptokurtic): it is more peaked and has heavier tails than a single Gaussian.  This means a non-parametric statistical comparison test is
required. 

A
standard distribution-free test is the Kolmogorov-Smirov (K-S) test.  The K-S
statistic is the maximum difference over all values of a single, independent
variable $x$ of two cumulative distribution functions (CDFs).  However, the
sensitivity of the K-S test is not independent of $x$: it tends to be most
sensitive around the median value and less sensitive at the extreme ends of
the distribution.  Identifying the points at $\pm\infty$ (wrapping the
$x$-axis around a circle) guarantees equal sensitivities
at all values of $x$.  This  is an invariant K-S test called the Kuiper test \citep{press1992}.
For comparing two different CDFs, the Kuiper statistic, defined as 

\begin{equation}
 D = D_{+} + D_{-},
  \label{equ:kuiper_v}
 \end{equation}

\noindent is the sum of the differences between the CDFs, where 

\begin{equation}
D_{+} = \max [S_{N_{1}}(x) - S_{N_{2}}(x)]
 \label{equ:kuiper_dp}
 \end{equation}
 
\noindent and 
 
 \begin{equation}
D_{-} = \max [S_{N_{2}}(x) - S_{N_{1}}(x)],
 \label{equ:kuiper_dm}
 \end{equation}
 
\noindent where $S_{N_{i}}(x)$ is the function giving the fraction of data
points $\le$$x$.  The statistical significance of $D$, $P(D >
\textrm{observed}) = Q(\lambda)$ is given by

\begin{equation}
Q = 2\sum_{j=1}^{\infty}(4j^{2}\lambda^{2} - 1)e^{-2j^{2}\lambda^{2}},
\label{equ:qkp}
\end{equation}

\noindent where $Q$ is a monotonic formula with variable $j$ for the asymptotic behaviour of the
statistic $D$, satisfying $Q(0) = 1$ and $Q(\infty) = 0$, where

\begin{equation}
\lambda = D\left(\sqrt{N_{e}} + 0.155 + \frac{0.24}{\sqrt{N_{e}}}\right)
 \label{equ:kuiper_p}
 \end{equation}
 
\noindent and 

\begin{equation}
N_{e} = \frac{N_{1}N_{2}}{N_{1} + N_{2}}.
 \label{equ:kuiper_ne}
 \end{equation}

To compare all the CORAVEL dwarfs ($N_{1}$ = 13 240) with the single-star
CORAVEL dwarfs ($N_{2}$ = 8589), $x = W$ in equations \ref{equ:kuiper_dp} and
\ref{equ:kuiper_dm}, giving $D_{+} = 0.004$, $D_{-} = 0.009$ (see bottom of
Fig. \ref{fig:dwarf_w}) and $D = 0.013$.  The sample size is sufficiently
large ($N_{e} = 5209.51$) to give the test enough power to detect significant
differences and reject the null hypothesis (that all the stars and the single
stars are drawn from the same parent population) if it is false.  $Q = 0.887$,
signifying the data fail to reject the null hypothesis.  This means that, statistically, stellar binarity
does not affect the $W$ distribution of the CORAVEL dwarfs significantly.  

\begin{figure}
\includegraphics[width=1.0\columnwidth]{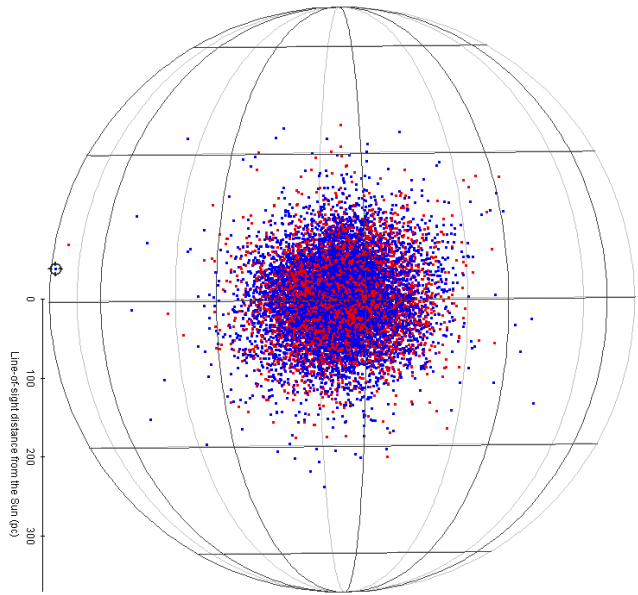}
\caption{Three dimensional sky distribution of all the 13 240 CORAVEL dwarfs with $W$
  velocities (red) and the 8589 single-star CORAVEL dwarfs (blue).  The spherical polar axes have a radius of 370 pc, centred on the
  Sun, chosen to include the most distant CORAVEL dwarf with a $W$
  velocity (363 pc, marked by the black cursor).  The viewer is in the Galactic plane, inside the
        solar circle at $l\sim30^{\circ}$ (the Galactic centre is over the
        viewer's left shoulder).  
The nearest line of constant $l$ is $l=30^{\circ}$ (with the majority
  of the stars behind it).  $l$ increases to the right (anticlockwise) with
        lines of constant $l$ every $30^{\circ}$.  All the lines of constant $l$ converge at the NGP ($b=+90^{\circ}$, top) and at
  the SGP ($b=-90^{\circ}$, bottom), with lines of constant $b$ every $30^{\circ}$.} 
\label{fig:gencop3d}
\end{figure}

Because binaries are, on average, apparently
  brighter than single stars, binaries are abundant in apparent magnitude
  limited samples like the Geneva-Copenhagen survey.  The Kuiper test suggests
  we can safely include binaries, nearly doubling the sample size of the CORAVEL
  dwarfs and increasing the power of the test. This means the RAVE RVs,
  which includes binaries, can be used on an equal footing with the CORAVEL
  data. Fig. \ref{fig:gencop3d} shows that including binaries does not increase the
  local volume of the solar neighbourhood sampled by the CORAVEL dwarfs.  
  
\subsection{Determining $W_{LSR}$ for the CORAVEL dwarfs}
\label{s:dwarf_lsr}

To search for streams, a Kuiper test could be used to compare
the $W$ distribution of the CORAVEL dwarfs with a Cauchy-Lorentz distribution
(heavy-tailed Gaussian).  Its peak location ($x_{0}$) and half-width at
half-maximum scale ($\gamma$) parameters could be theoretically determined according to
the Galactic volume sampled and the age-velocity dispersion relation of the
thin and thick discs.  However, the functional form of these relations remains
uncertain \citep{seabroke2007}.  In addition, the test hypothesis would
be whether the data is exactly a Cauchy-Lorentz distribution with the
specified $x_{0}$ and $\gamma$.  The hypothesis could be rejected due to the
data having a slightly different $x_{0}$ and $\gamma$ than those specified.
Thus, the test may not be very sensitive to the presence of streams.

A more straight-forward approach is to test the symmetry between the two
halves of the $W$ distribution.  In-falling streams on to the Milky Way disc will reveal themselves as
 overdensities in the distribution of $W$ with respect to the
 Galactic plane.  $W$ is positive towards the NGP and
 negative towards the SGP (symbolically denoted $+W$ and $-W$ respectively for brevity), so a stream falling onto
 the Galactic disc from the NGP (`above' the Galactic plane) would have $-W$
 and a stream falling onto the Galactic disc from the SGP (`below' the
 plane) would have $+W$.  A stream falling through the disc from the NGP to
 SGP would exhibit  $-W$ both above and below the plane and $+W$ both above
 and below the plane falling from the SGP to the NGP.  A symmetry test between
 the two sides of the $W$ distribution, if sensitive enough to the number of stream stars, will find the asymmetry caused by a single stream.
 
$W$ is measured relative to the Sun.  The top of Fig. \ref{fig:dwarf_w} shows that
the $W$ distribution is not centred on zero velocity.  The Kuiper test is
sensitive to the sign of the $W$ velocities (see bottom of
Fig. \ref{fig:dwarf_w}) so the absolute values of $-W$ velocities are
required to compare them to $+W$ to test the symmetry of the distribution.
The easiest method to implement this Kuiper symmetry test is to transform the
heliocentric reference frame to one centred on zero velocity -- the LSR reference frame, where $W^{\odot}_{LSR}$ in equation \ref{equ:wgsr} is the centre of the $W$ distribution in the top of Fig. \ref{fig:dwarf_w}.  $W^{\odot}_{LSR}$ needs to be determined to apply its correction to the $W$ distribution to convert it to a $W_{LSR}$ distribution.

\begin{figure}
\centering
        \psfig{figure=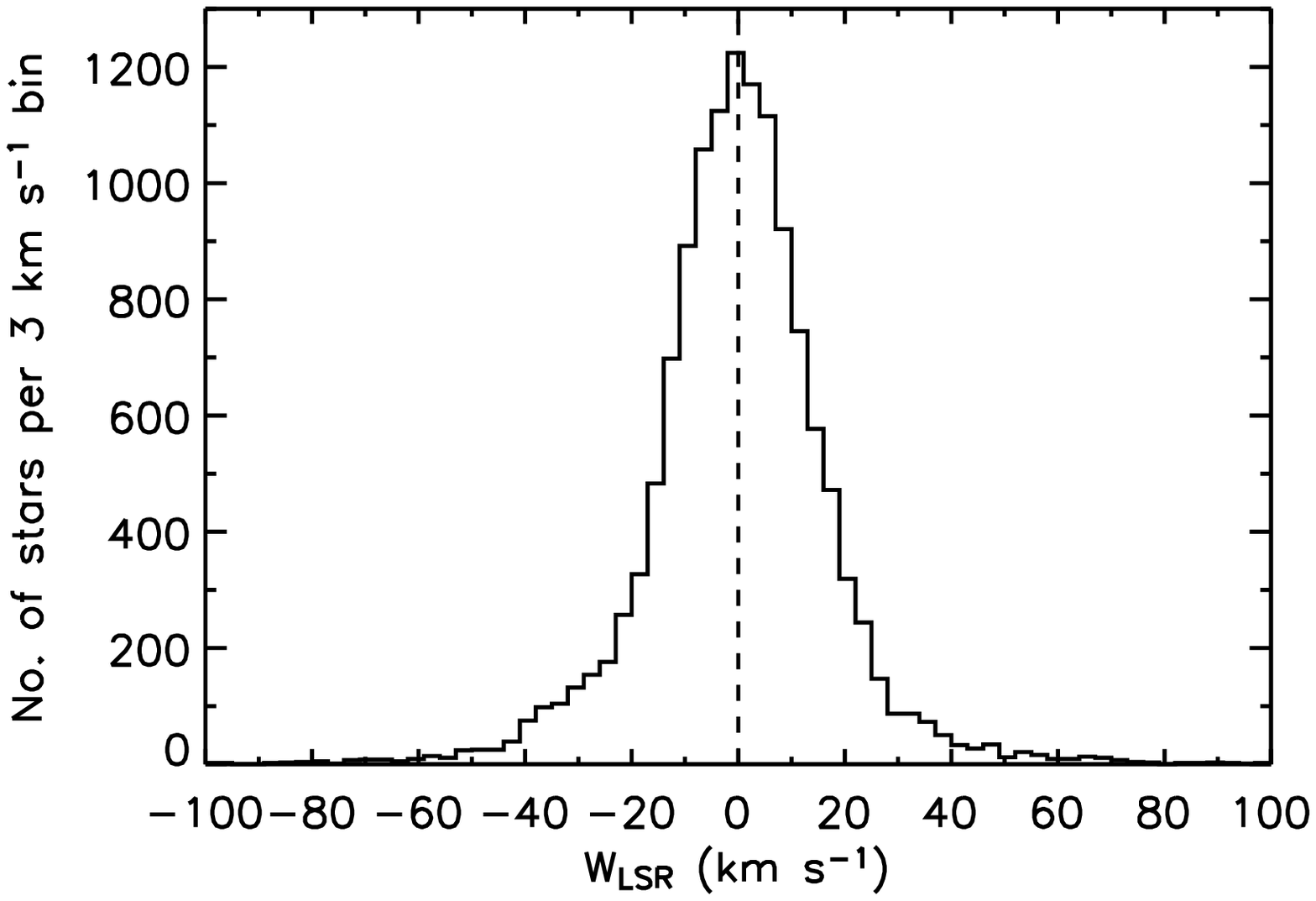,width=1.\columnwidth}
        \psfig{figure=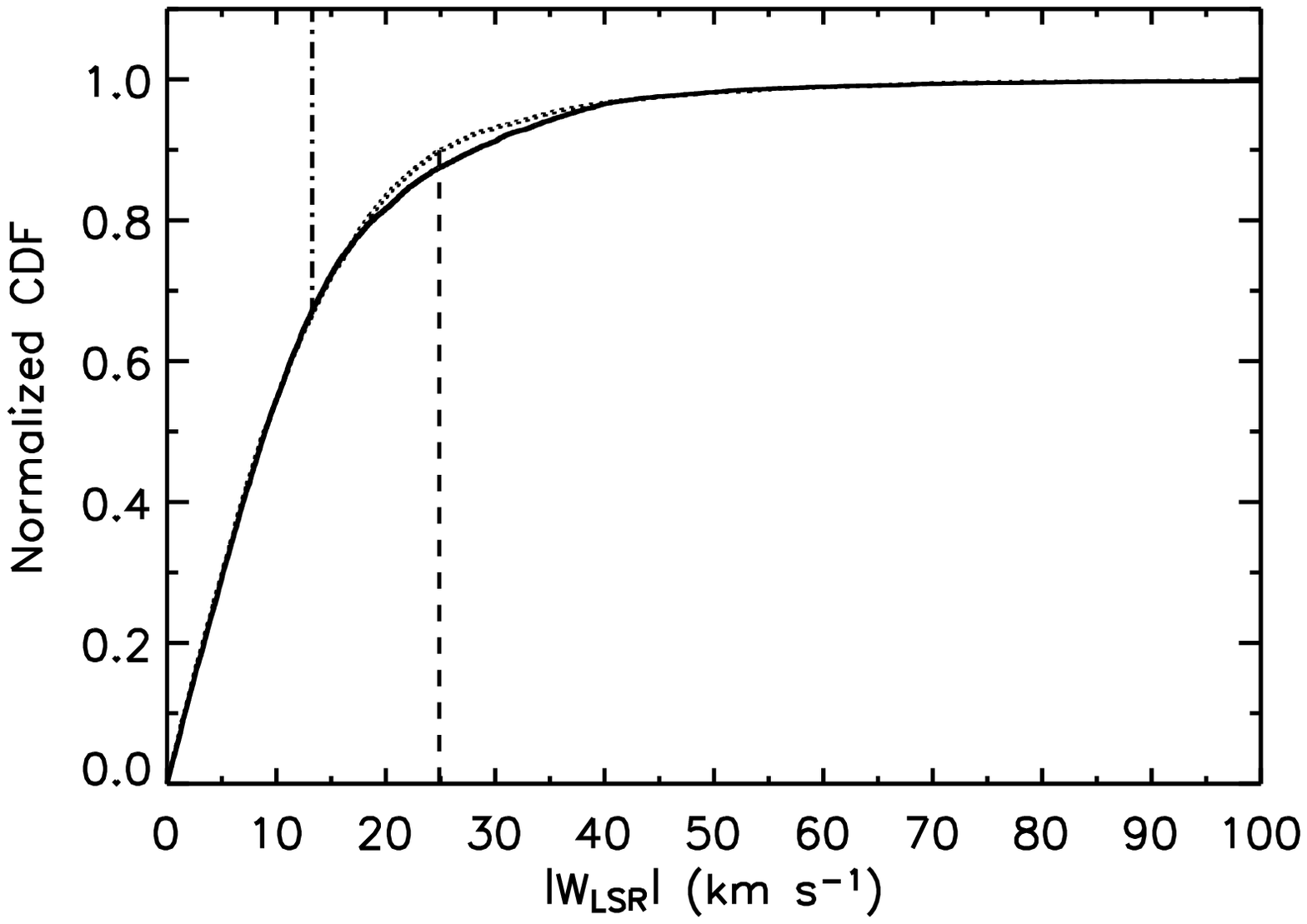,width=1.\columnwidth} 
                \caption{{\it Top:}  $W_{LSR}$ distribution of the CORAVEL
        dwarfs (solid histogram) either side of the $W_{LSR} = 0$ dividing line (dashed line).  
                  {\it Bottom:} Normalized CDF of $W_{LSR} < 0$ (solid line) and $W_{LSR} > 0$ (dotted line), showing $D_{-}$ (dashed line) and $D_{+}$ (dot-dashed line).}
                \label{fig:dwarf_kuiper_lsr}
\end{figure}

We determine $W^{\odot}_{LSR}$ by varying its value in the Kuiper symmetry test to find the best agreement between the two halves of the resulting $W_{LSR}$ distribution (minimum value of
$D$).  $D$ is calculated using equations \ref{equ:kuiper_dp} and
\ref{equ:kuiper_dm} but now $x$ = $|$$\pm$$W_{LSR}|$.  The maximum differences
between $S_{N_{W_{LSR} < 0}}(|$$-$$W_{LSR}|)$ and $S_{N_{W_{LSR} >
    0}}(+W_{LSR})$ are calculated over the range $0 < |$$\pm$$W_{LSR}| <
\infty$.  The top of Fig. \ref{fig:dwarf_kuiper_lsr} shows the $W_{LSR}$
distribution for the minimum value of $D$, where  $W^{\odot}_{LSR}$ = 7.0 km
s$^{-1}$ to the nearest 0.1 km s$^{-1}$.  Our value of $W^{\odot}_{LSR}$ also
agrees within the errors with the \citet{dehnen1998a} measurement:
$W^{\odot}_{LSR}$ = 7.17 $\pm$ 0.38 km s$^{-1}$ (see Appendix A for further discussion on the determination of $W^{\odot}_{LSR}$ and historical parallels).  

\begin{table*}
\begin{minipage}{170mm}
\centering
\label{t:dwarf_kuiper_centre}
  \caption{Results of the Kuiper symmetry test, applied to the CORAVEL dwarfs,
    to determine $W^{\odot}_{LSR}$ to the nearest 0.1 km s$^{-1}$, where $\Delta W^{\odot}_{LSR} = W^{\odot}_{LSR} - 7.0$ km s$^{-1}$.}
      \begin{tabular}{@{}ccccccccc@{}}
  \hline
$W^{\odot}_{LSR}$ & $\Delta W^{\odot}_{LSR}$ & $N_{W_{LSR} < 0}$ & $N_{W_{LSR} > 0}$ & $N_{e}$ & $D_{-}$ & $D_{+}$ & $D$ & $Q$\\
\hline
5.6 & $-1.4$ & 7176&        6064 &     3287 &0.0484 &  0.0021 &   0.0505&  $3\times10^{-6}$\\
7.0 & 0.0 &6583   &     6657  &    3310 &  0.0250 &  0.0064  &  0.0314&    0.0341\\
7.6 & 0.6 & 6334 &       6906 &     3304   &0.0242 &   0.0212 &   0.0454 & $6\times10^{-5}$\\
   \hline
\end{tabular}
\end{minipage}
\end{table*}

Having determined $W^{\odot}_{LSR}$, we continue to vary its value until a
4$\sigma$ statistically significant difference ($Q < 6 \times 10^{-5}$) is generated between the two halves of the resulting $W_{LSR}$ distribution to gauge how sensitive the Kuiper symmetry test is to the measured value of $W^{\odot}_{LSR}$.  Table 1 shows that $W^{\odot}_{LSR}$ can be placed $-1.4$ km s$^{-1}$ from our measured value of $W^{\odot}_{LSR}$ in the $-W_{LSR}$ distribution, whereas it can only be placed less than half that speed (+0.6 km s$^{-1}$) from our measured value of $W^{\odot}_{LSR}$ into the $+W_{LSR}$ distribution.  This is because the position of $W^{\odot}_{LSR}$ is sensitive to the slight asymmetry due to $D_{+}$ in the bottom of Fig. \ref{fig:dwarf_kuiper_lsr} at $|W_{LSR}| = 13$ km s$^{-1}$.  Although it may not appear so in the top of Fig. \ref{fig:dwarf_kuiper_lsr} because of the binning, there are more stars with $0 < W_{LSR} \le 13$ than $-13 < W_{LSR} \le 0$ km s$^{-1}$.  However, $S_{N_{W_{LSR} < 0}}(|$$-$$W_{LSR}|) > S_{N_{W_{LSR} > 0}}(+W_{LSR})$, allowing the line of symmetry ($W^{\odot}_{LSR}$) to be placed further into the $-W_{LSR}$ distribution than the $+W_{LSR}$ distribution.  If the position of $W^{\odot}_{LSR}$ was sensitive to the larger asymmetry due to $D_{-}$ in the bottom of Fig. \ref{fig:dwarf_kuiper_lsr} at $|W_{LSR}| = 25$ km s$^{-1}$, the reverse would be true i.e. even though there are more stars with $0 < W_{LSR} \le 25$ than $-25 < W_{LSR} \le 0$ km s$^{-1}$, $S_{N_{W_{LSR} > 0}}(+W_{LSR}) > S_{N_{W_{LSR} < 0}}(|$$-$$W_{LSR}|)$, which would allow the line of symmetry ($W^{\odot}_{LSR}$) to be placed further into the $+W_{LSR}$ distribution than the $-W_{LSR}$ distribution.  This demonstrates that $W^{\odot}_{LSR}$ is only sensitive to asymmetries at $|W_{LSR}| < \sigma_{W_{LSR}}$ (in this case $D_{+}$ rather than $D_{-}$), despite $D_{+} << D_{-}$.  

$W^{\odot}_{LSR}$ of the full sample of CORAVEL dwarfs coincides with the mode of the
$W_{LSR}$ distribution in the top of Fig. \ref{fig:dwarf_kuiper_lsr}.  This suggests
that there cannot be many CORAVEL dwarfs with $|W_{LSR}| < \sigma_{W_{LSR}}$ belonging
to vertically coherent streams with systemic $W_{LSR}$ velocities.  If there were, they would have biased the measurement of $W^{\odot}_{LSR}$ away
from the mode. 

A coherent stream with a systemic space velocity would be visible as an
overdensity in the CORAVEL dwarf $U$$-$$W$ and $V$$-$$W$ space diagrams.
These diagrams in fig. 20 in N04 do not exhibit any such vertical
substructure.  In fact, \citet{seabroke2007} show the CORAVEL dwarfs are well
phase-mixed vertically because of the sample's low scaleheight (see
Fig. \ref{fig:gencop3d}).  This suggests there are no vertically coherent
streams currently passing through the solar neighbourhood, in agreement with
the lack of strong clumping in Fig. \ref{fig:gencopj}.  

\subsection{Sensitivity of the $W_{LSR}$ symmetry of the CORAVEL dwarfs to tidal streams}
\label{s:dwarf_stream}

In this section, 
we investigate the sensitivity of the $W_{LSR}$ symmetry of the CORAVEL dwarfs to tidal streams at $|W_{LSR}| \ge \sigma_{W_{LSR}}$.
 The previous section suggests the sample of CORAVEL dwarfs is free from
 vertically coherent streams.  This means its distribution can be used as a
 smooth background against which we can investigate the effect of a tidal
 stream passing vertically through the solar neighbourhood.  We do this by
 simulating a stream, pseudo-randomly generating a Gaussian with $\sigma_{s} =
 10$ km s$^{-1}$, and adding it to the sample.  This dispersion was chosen
 because the best studied tidal stream in the literature, the  Sgr stream, is
 found to be dynamically cold by two different groups:  \citet{majewski2004}
 used medium-resolution spectroscopy to find an intrinsic dispersion of
 $\sigma = 10.4 \pm 1.3$ km s$^{-1}$ (after removing random errors of
 $\sim$5.3 km s$^{-1}$); and \citet{monaco2007} fitted a Gaussian of  $\sigma
 = 8.3 \pm 0.9$ km s$^{-1}$ to their sample of high-resolution
 spectroscopically derived RVs.  

To investigate how sensitive the Kuiper symmetry test is to the choice of
 $\sigma_{s}$, we repeated the above experiment with $\sigma_{s} = 3$ and 17.3
 km s$^{-1}$.  These dispersions are physically motivated: H99a state that the
 intrinsic dispersion for a tidal stream from a Large Magellanic Cloud-type
 progenitor is $\sim$3-5 km s$^{-1}$ after a Hubble time; and
 \citet{duffau2006} discovered the Virgo stellar stream with $\sigma = 17.3$
 km s$^{-1}$.   We find that the number of stars required to generate a
 4$\sigma$ tidal stream detection is relatively insensitive to $\sigma_{s}$.  
 The conservation of phase-space density results in $\sigma$ decreasing as
 1/time at each point along a stream (H99a).  Hence, the ability of the test
 to detect a stream is not biased by the dynamical age of the stream.

The simulated tidal stream is added to the sample of CORAVEL dwarfs at
$W = \pm$(1, 2, 3, $\sim$13)$\sigma_{W_{LSR}}$, where $\sigma_{W_{LSR}} \approx$ 18
km s$^{-1}$, to test the sensitivity of the $W_{LSR}$ symmetry to the presence
of each stream.  Approximately $-13$$\sigma_{W_{LSR}}$ corresponds to $W_{LSR}
\approx$ $-242$ km s$^{-1}$ calculated by \citet{freese2004} using the eight
clump stars from \citet{chiba1998} in the H99b stream (the CORAVEL
dwarfs with $W_{LSR}$ nearest to this value are at $-286$ and $-179$ km s$^{-1}$).  

\begin{table*}
 \begin{minipage}{170mm}
 \centering
 \label{tab:dwarf_kuiper}
   \caption{Results of the Kuiper symmetry test, applied to the CORAVEL dwarfs, to determine the
     approximate number of tidal stream stars ($N_{s}$) required, in
     pseudo-randomly generated $\sigma_{s} = 10$ km s$^{-1}$ Gaussians, placed
     at $\pm$(1, 2, 3, $\sim$13)$\sigma_{W_{LSR}}$ in the $W_{LSR}$ distribution,
     to cause the test to reject the null hypothesis at 4$\sigma$ ($Q < 6 \times 10^{-5}$).  \% = $N_{s}/(N_{s} +
     N_{total})$ where $N_{total}$ = 13 240. $\rho_{s}$ is the stellar density
     of each stream, in $N \times 10^{6}$ stars kpc$^{-3}$,
     calculated by dividing the $N_{s}$ of each stream by the volume within which the CORAVEL
     dwarfs are complete, approximated by a sphere with radius $\sim$40 pc
     ($\sim$3$\times 10^{-4}$ kpc$^{-3}$).}
     \begin{tabular}{@{}rrrrccccccc@{}}
   \hline
 $\sigma_{W_{LSR}}$ & $N_{s}$ & \% & $\rho_{s}$ & $N_{W_{LSR} < 0}$ & $N_{W_{LSR} > 0}$ &$N_{e}$ & $D_{-}$ & $D_{+}$ & $D$ & $Q$\\
 \hline
 $-1$ &  600 & 4.3 & 2.2 & 7165   &    6675  &    3456   &0.0414  & 0.0033  & 0.0447 & $5\times10^{-6}$\\
 1 & 500 & 3.6 & 1.9 &6601&        7139&      3430&   0.0163 &   0.0339&    0.0502& $2\times10^{-6}$\\
 $-2$ &  200 & 1.5 &0.7 & 6783        &6657      &3360   &0.0469  &0.0010   &0.0479& $1\times10^{-5}$\\
 2 & 400 & 2.9 & 1.5 &6583        &7057      &3406  &0.0026    &0.0458    &0.0484& $7\times10^{-6}$\\
  $-3$ & 200 & 1.5 &0.7&6783 &       6657&      3360   &0.0508 & 0.0004  &  0.0512& $1\times10^{-6}$\\
 3 & 400 & 2.9 & 1.5&6583 &       7057&      3406  &0.0026 &   0.0504 &   0.0530&$3\times10^{-7}$\\
 $\sim$$-13$ & 200 & 1.5&0.7 & 6783  &      6657  &    3360  & 0.0508 & 0.0004 &    0.0512& $1\times10^{-6}$\\
 $\sim$13 & 300 &2.2& 1.1 &6583 &       6957  &    3382  &0.0036 &  0.0445 &   0.0481 & $9\times10^{-6}$\\
 \hline
 \end{tabular}
 \end{minipage}
 \end{table*}

\begin{figure}
\centering
               \psfig{figure=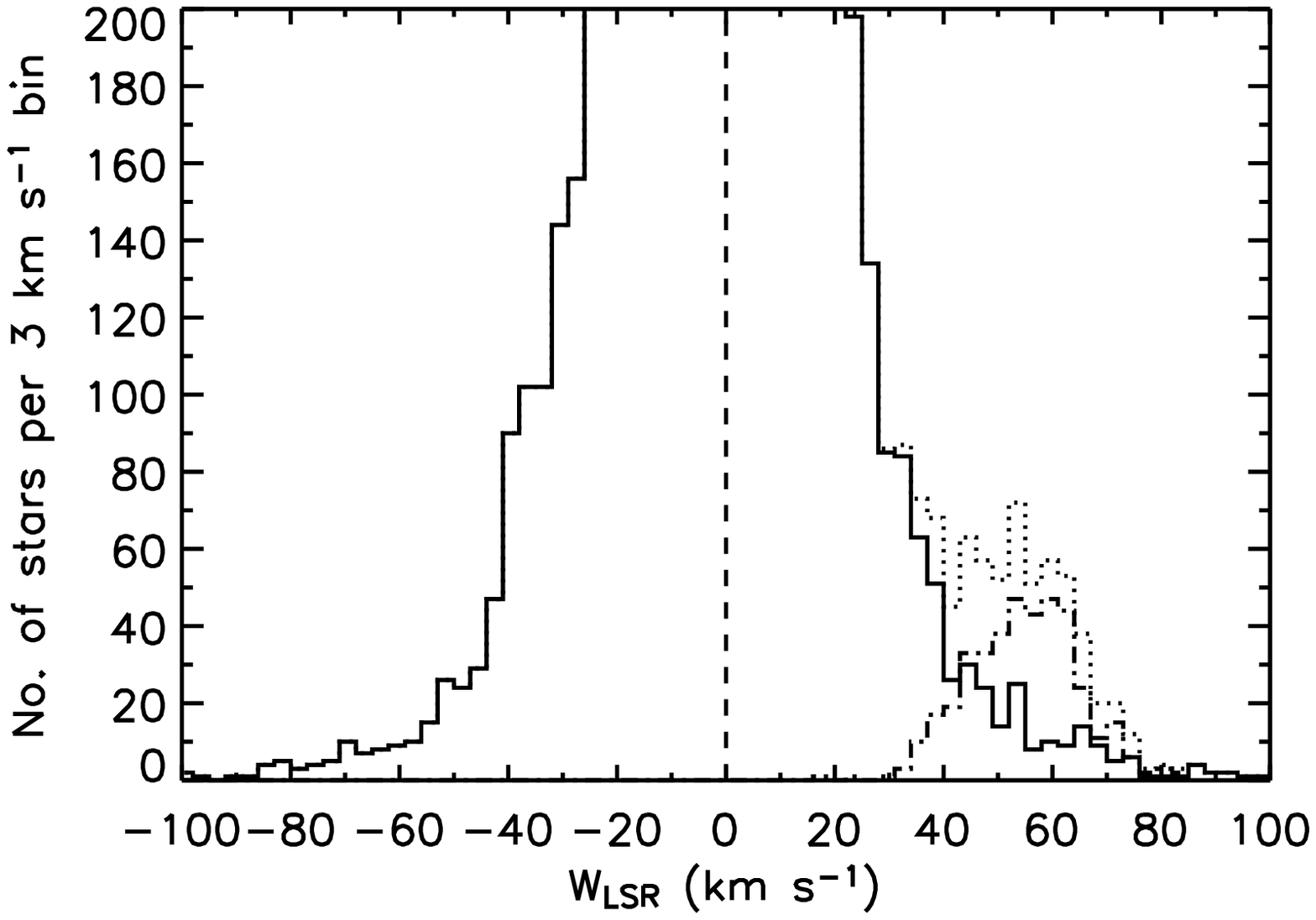,width=1.\columnwidth}
        \psfig{figure=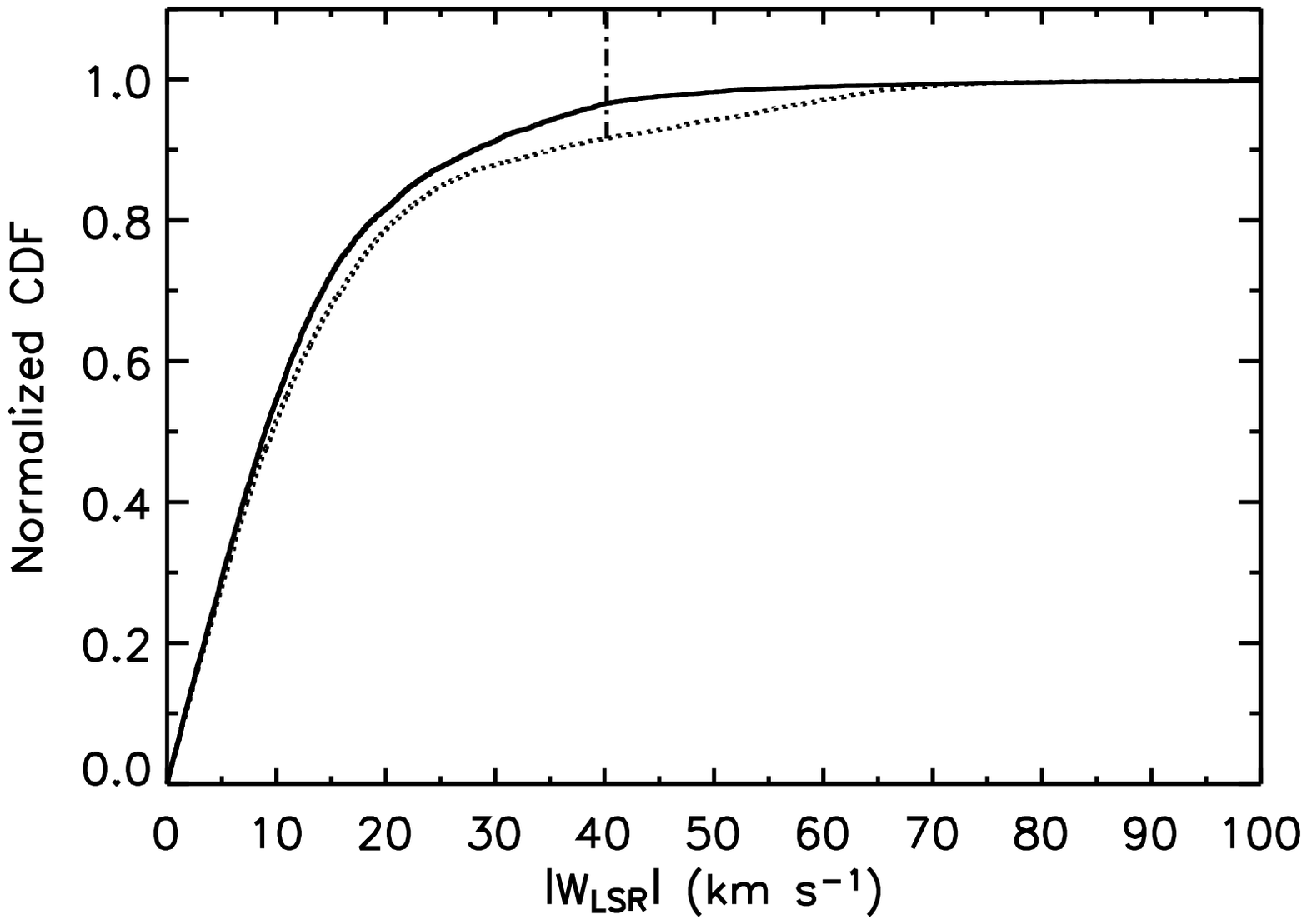,width=1.\columnwidth}  
     \caption{{\it Top:} $W_{LSR}$ distribution of the CORAVEL dwarfs (solid
               histogram) either side of the $W_{LSR} = 0$ dividing line
               (dashed line).  The pseudo-randomly generated Gaussian
               (representing a $\sigma = 10$ km s$^{-1}$ tidal stream)
               included in the data set (dotted histogram) that caused the
               Kuiper symmetry test to reject the null hypothesis at 4$\sigma$
               when placed at 3$\sigma_{W_{LSR}}$ is also plotted separately (dot-dashed line).  {\it Bottom:} Normalized CDF of $W_{LSR} < 0$ (solid line) and $W_{LSR} > 0$ (dotted line), showing  $D_{+}$ (dot-dashed line).  $D_{-}$ is not visible in this plot at $|W_{LSR}| \approx 0$.}
                \label{fig:dwarf_kuiper_gausp3}
\end{figure}

At each position, the random population of the simulated tidal stream is
increased until it causes the Kuiper symmetry test to reject the null
hypothesis (that $W_{LSR} < 0$ and $W_{LSR} > 0$ are drawn from the same parent population) at the 4$\sigma$ statistical significance level.  The test
results in Table 2 are indicative because they represent a single realization
of a pseudo-random Gaussian added to the sample and we have not conducted a
full Monte Carlo simulation.  Nevertheless, the indication that more tidal
stream stars are required to generate a 4$\sigma$ tidal stream detection at 2
and 3$\sigma_{W_{LSR}}$ than at $-2$ and $-3$$\sigma_{W_{LSR}}$ seems
plausible due to the asymmetry already in the $W_{LSR}$ distribution: $D_{-}$
in the bottom of Fig. \ref{fig:dwarf_kuiper_lsr} at $|W_{LSR}| = 25$ km
s$^{-1}$.  Placing streams in the $-W_{LSR}$ distribution, further increases
$D_{-}$ and decreases $D_{+}$.  The reverse is true for placing streams in the
$+W_{LSR}$ distribution but even more stream stars are required to cancel out
the intrinsic $D_{-}$ and increase $D_{+}$ to the 4$\sigma$ statistical
significance level (see Fig. \ref{fig:dwarf_kuiper_gausp3}).  

The test becomes more sensitive (less stream
stars required to generate a $4\sigma$ detection), the further a stream is from the centre of the
distribution.  However, Table 2 shows that the number of stream stars required for a $4\sigma$
detection stays
approximately constant (saturates) at $\le$$-3$$\sigma_{W_{LSR}}$.  This is because
the positions of $D_{-}$ and $D_{+}$ have become fixed and their values
saturated, whereas $D_{+}$ is still sensitive to perturbation at $W_{LSR} > 3\sigma_{W_{LSR}}$ by hundreds of stream stars.

H99a demonstrate that phase-mixing allows different tidal streams from a
single disrupted object to be observed with opposite motion in the vertical
direction (e.g. the two H99b streams).  If two such streams contain similar
numbers of stars, the Kuiper symmetry test may not find them because their
presence may not create enough of an asymmetry between the two sides of the
$W$ distribution.  H99a predict inside a 1 kpc$^{3}$ volume centred on the
Sun, there should be 300-500 tidal streams containing a few hundred stars each
if the whole stellar halo was built up by disrupting satellites.  However,
neither the CORAVEL nor RAVE surveys provide complete coverage of this volume
and so are very incomplete with respect to halo stars.  Thus, if these streams exist in our samples, they will most likely contain fewer stars than the detection thresholds of the Kuiper symmetry test in Table 2.  This means the test is not sensitive to these streams, regardless of their possible symmetries.  The number of stars required for a detection suggests the test is better suited to detect less phase-mixed, more coherent streams in configuration space, like the Sgr streams that contain hundreds of stars. 

\citet{seabroke2007} estimated that are $\sim$1300 CORAVEL dwarfs in the Hercules dynamical stream.  This substructure corresponds to 10 per cent of the sample and is easily visible in the N04 $U$$-$$V$ space diagram (their fig. 20).   F05 show in their sample of CORAVEL giants that the Hercules stream has $\sigma_{U} = 28.35 \pm 1.68$ and $\sigma_{V} = 9.31 \pm 1.22$ km s$^{-1}$.    More than halving $\sigma_{U}$ would also more than halve the number of Hercules stars, bringing them into agreement with the numbers in Table 2 but its nucleus would still be discernible as an overdensity in the N04 $U$$-$$V$ space diagram.   

The high number of stream stars needed to cause a statistically
significant asymmetry would most likely be visible as substructure in
the N04 $U$$-$$W$ and $V$$-$$W$ space diagrams and strong clumping in
Fig. \ref{fig:gencopj}.
Given this and the symmetry of the $W_{LSR}$ distribution, we conclude that it is unlikely that the sample of CORAVEL dwarfs
contains any  vertically coherent streams consisting of $\gtrsim$200
stars.  Veltz et al. (2007) find the scaleheight of the thin disc is
$\sim$225 pc so the CORAVEL dwarfs only sample a very local volume of the thin disc (N04 showed the local thick disc consists of $<$10 per cent of the sample).  

F-G turnoff stars are a good indicator of stellar number density. The
only part of the Sgr stream close enough to the Sun where this quantity can be
reliably measured is 50 kpc from the Sgr dSph (SDSS Stripe 82).  \citet{freese2005} estimate the full-width at half-maximum of
the Sgr stream in Stripe 82 is 4-8 kpc, which is consistent with $\sigma \approx$ 2 kpc,
derived from 2MASS M giants in the stream \citep{majewski2003}.
\citet{freese2005} use this stream width to estimate its stellar density: 210-740 stars kpc$^{-3}$.  
They use Blue Horizontal Branch stars in Stripe 82 and in another part of the
stream, also 50 kpc from the Sgr
dSph (Stripe 10), to estimate how stellar density varies as a function of
position along the stream and find the variation is a factor of 1-2, implying
the stellar density of the Sgr stream in the solar neighbourhood is in the
range of 210-1480 stars kpc$^{-3}$.  An upper limit on the stream density of
1.6 $\times 10^{4} M_{\odot}$ kpc$^{-3}$ is derived from the models 
shown in \citet{martinez-delgado2007}, assuming all the mass in the Sgr dSph
is in the form of stars.  
 
The CORAVEL dwarfs are volume complete
out to $\sim$40 pc, which corresponds to a spherical volume of $\sim$3 $\times
10^{-4}$ kpc$^{3}$.  Therefore, even the upper bound on the number of Sgr
stream stars expected to be passing through this volume is $<$1.
Fig. \ref{fig:gencop3d} illustrates that there are a significant number of stars out to $\sim$100-200
pc so if the Sgr stream is passing through the spherical CORAVEL dwarf volume (radius $\sim$200
pc), only of order unity CORAVEL dwarfs are likely to be Sgr stars.
Table 2 shows that the Kuiper symmetry test is not sensitive to such a small number of
stream stars so if there is subtle clumping in Fig. \ref{fig:gencopj} the test
would not find such weak streams.  

From fig. 1 of \citet{fuchs2006}, we estimate that at $Z \sim$ 2 kpc, the VOD minus
the vertical density profile at the position of the Sun of a 
smooth Galaxy model is 1.2 $\times 10^{5}$ stars kpc$^{-3}$ in the 13-h CADIS field.  Therefore, if this structure exists in
the solar neighbourhood with this density, there will not be many more
than 30-40 CORAVEL dwarfs belonging to it.  Table 2 shows that the
Kuiper symmetry test is not sensitive to such a small number of stars.  The
table illustrates that even the smallest
number of stars required for a stream detection in this sample only puts a very weak
constraint on the stellar density of a coherent structure in the solar
neighbourhood with vertically
coherent kinematics.

\section{CORAVEL giants}
\label{s:giants}

\subsection{Derivation of the space velocities of the CORAVEL giants}
\label{s:giants_dev}

The F05 sample of CORAVEL giants is the intersection of several data sets: (i)
spectral types K and M in the {\it Hipparcos} catalogue; (ii) proper motions
from the {\it Tycho-2} catalogue; (iii) CORAVEL RVs for stars in list (i) in the Northern hemisphere ($\delta > 0^{\circ}$), observed with the Swiss 1-m telescope at Observatoire
de Haute-Provence, France.  The
{\it Hipparcos} parallaxes were used to construct a crude Hertzsprung-Russell (H-R) diagram
from which dwarfs were identified and removed from the sample.  Stars
with peculiar spectra, such as T Tau stars, Mira variables and S stars were
eliminated using diagnostics based on RV variability and CORAVEL
cross-correlation profiles, combined with literature searches.  Binaries for
which no centre-of-mass RV could be estimated were removed using the F05
binarity flag ($B$).  A star was excluded from their (and our) kinematic analysis if it is a spectroscopic binary ($B$ = 0) or a visual binary with no orbit available ($B$ = 5) or an uncertain case, either a spectroscopic binary or supergiant ($B$ = 3) or for a visual binary ($B$ = 8), leaving 6030 stars (5311 K and 719 M giants).   

Because the CORAVEL giants are at distances approximately an order of
magnitude greater than the CORAVEL dwarfs, the $\sigma_{\pi}/\pi$ distribution
for the giants is correspondingly less accurate than that for the dwarfs.
This means, unlike for the dwarfs, simple inversion of the parallaxes of the
giants results in biased estimates of their distances for the $>$3000 giants
with $\sigma_{\pi}/\pi >$ 20 per cent \citep{brown1997}.  To derive bias-free
distances for all the CORAVEL giants, F05 used the Bayesian Luri-Mennessier
maximum-likelihood method \citep{luri1996} to model the sample in order to
assign each star to a kinematic `base' group.  Given the maximum-likelihood
estimator of its assigned group parameters and its observed values,  F05
obtain the marginal probability density law for the distance of each star from
the global probability density function.  These distances were used with the
CORAVEL RVs and {\it Tycho-2} proper motions to derive the space velocities.

\subsection{Orbital angular momenta of the CORAVEL giants}

\begin{figure}
        \psfig{figure=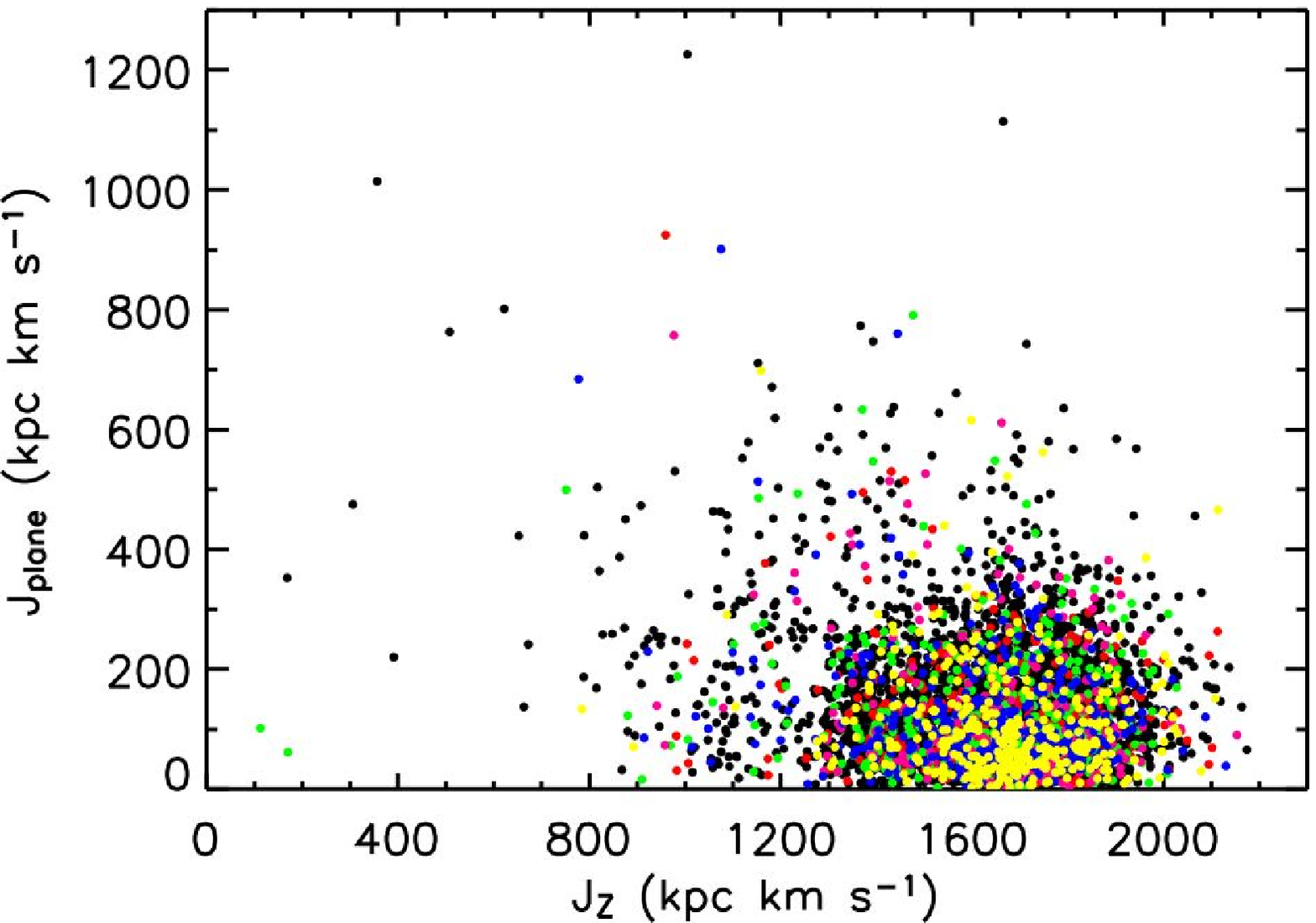,width=1.\columnwidth} 
        \psfig{figure=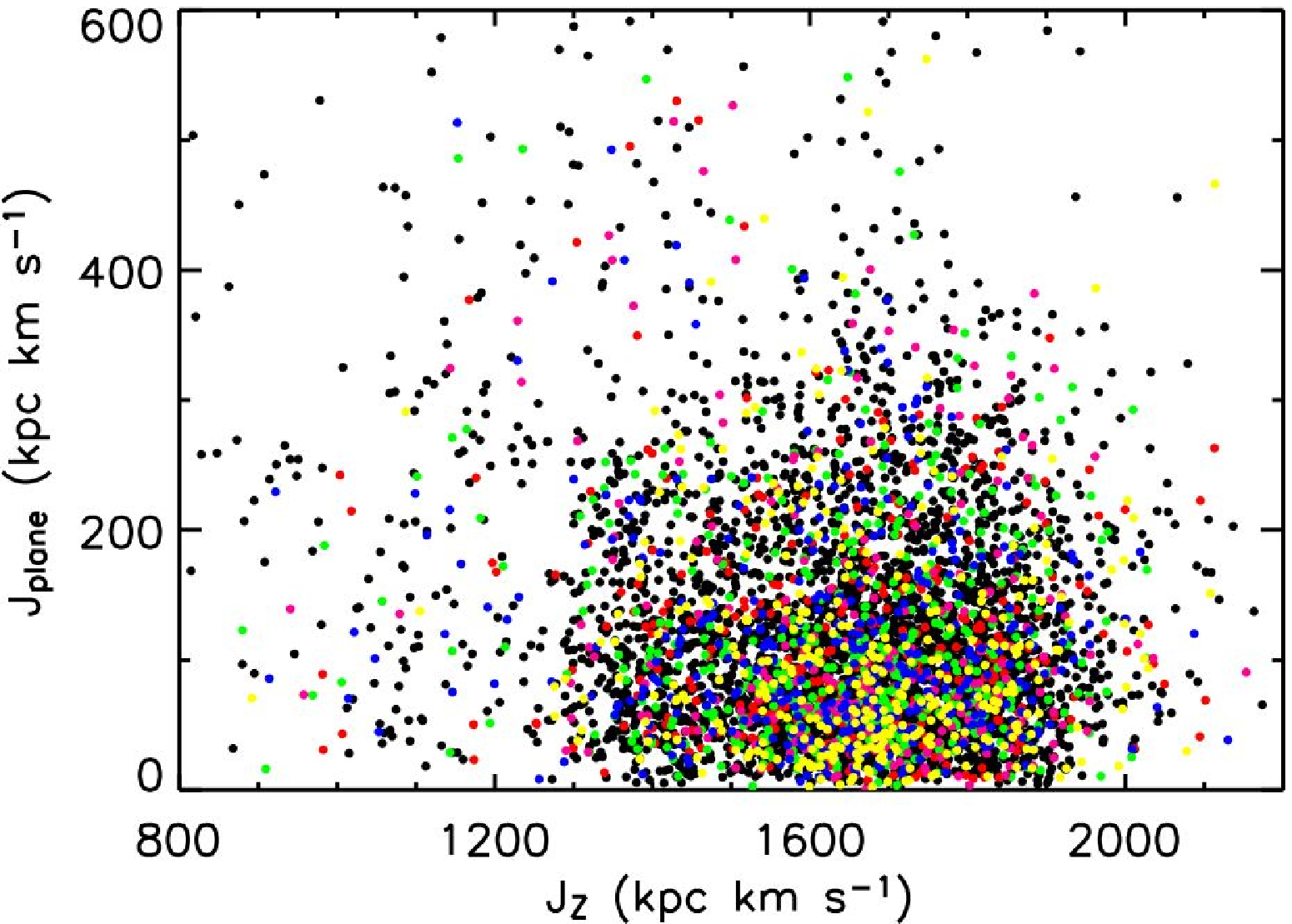,width=1.\columnwidth}
         \caption{Distribution of CORAVEL giants in the plane of orbital angular momentum components, colour-coded according to the F05 maximum
        likelihood base group
        assignment to known kinematic features of the solar neighbourhood: smooth
  background (black), Sirius moving group (magenta), young
  kinematics (yellow), Hyades-Pleiades supercluster (red),
  Hercules stream (green) and high-velocity stars (blue).  Note the different
        scale between the top plot above and in Fig. \ref{fig:gencopj}.}  
         \label{fig:famaeyj}
          \end{figure}
          
Fig. \ref{fig:famaeyj} shows that, despite probing a larger volume than the
CORAVEL dwarfs, the giants, unlike the dwarfs, do not include a symmetric (no
net rotation) halo sample.  This figure includes all the sample and thus none
of the CORAVEL giants are members of the H99b streams.  The F05 dynamical
streams are part of the smooth distribution in Fig. \ref{fig:famaeyj} because
all the stars exhibit a similar range of  thin disc $J_{Z}$ values.  The lack
of strong clumping again suggests there are not any coherent tidal streams in the sample.

\subsection{Determining $W_{LSR}$ for the CORAVEL giants}
\label{s:giant_lsr} 

\begin{figure}
\centering
        \psfig{figure=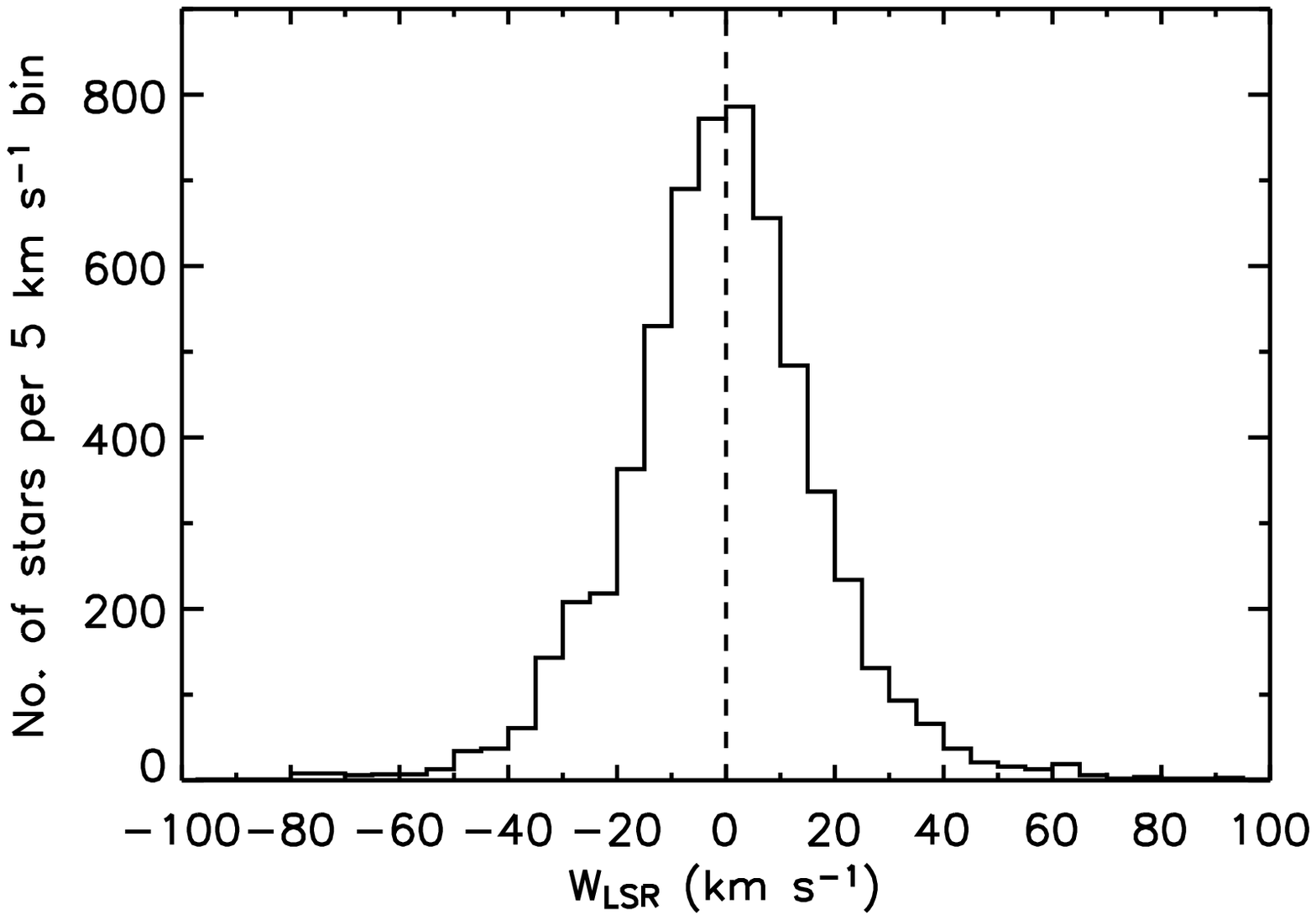,width=1.\columnwidth}
        \psfig{figure=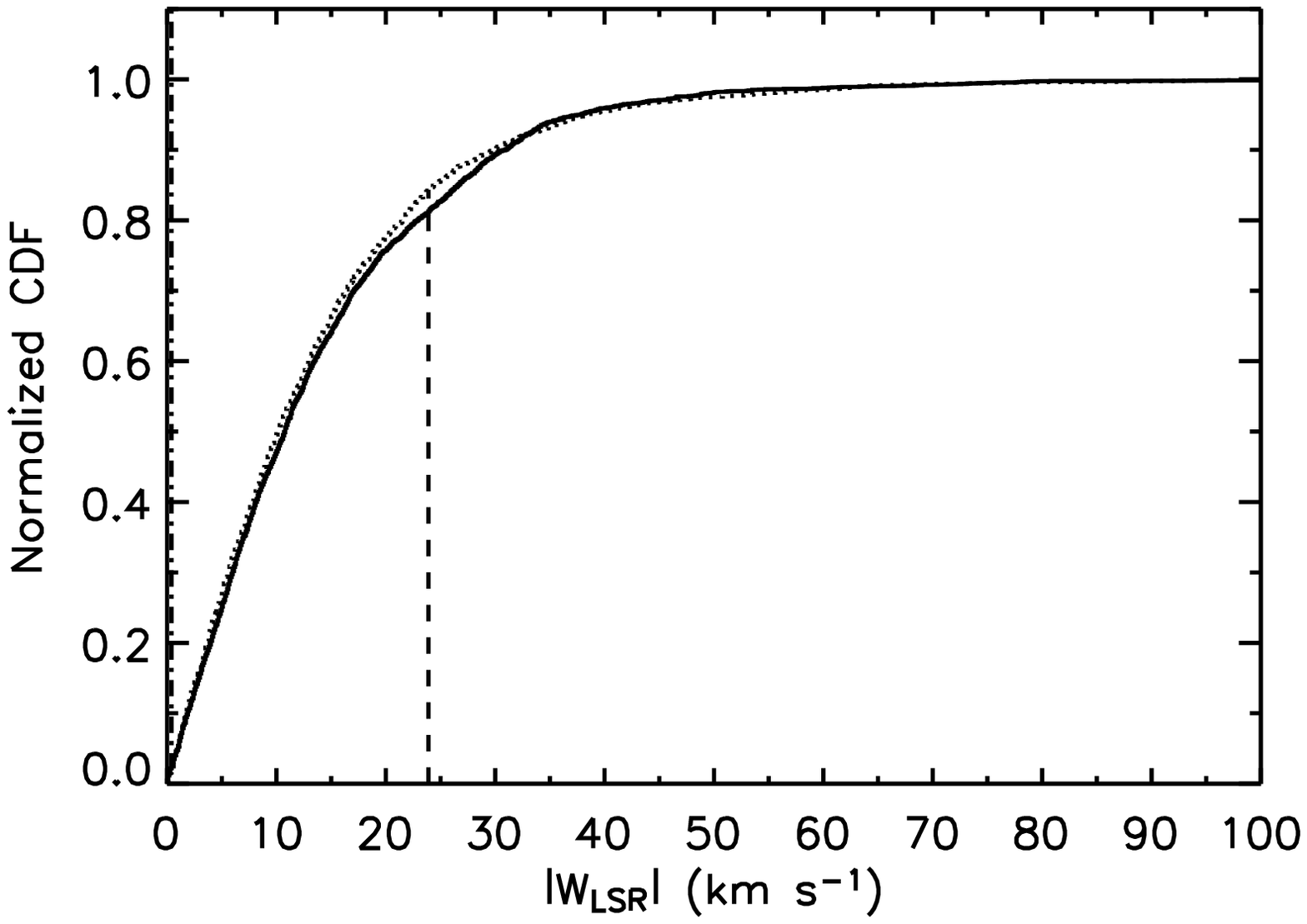,width=1.\columnwidth} 
              \caption{{\it Top:} $W_{LSR}$ distribution of the CORAVEL giants
        (solid histogram) either side of the $W_{LSR} = 0$ dividing line (dashed line).  The 5 km s$^{-1}$ velocity bin sizes are chosen to be twice the average space velocity error.    
                  {\it Bottom:} Normalized CDF of $W_{LSR} < 0$ (solid line) and $W_{LSR} > 0$ (dotted line), showing $D_{-}$ (dashed line) and $D_{+}$ (dot-dashed line, just visible at $|W_{LSR}| \approx 0$).}
                \label{fig:giant_kuiper_lsr}
\end{figure}

This section repeats the technique applied to the CORAVEL dwarfs in Section
\ref{s:dwarf_lsr} to the CORAVEL giants.  The top of
Fig. \ref{fig:giant_kuiper_lsr} shows the $W_{LSR}$ distribution for the
minimum value of $D$, where  $W^{\odot}_{LSR}$ = 7.0 km s$^{-1}$ to the
nearest 0.1 km s$^{-1}$.  This is the same value as found for the CORAVEL
dwarfs.  In the reverse situation to the dwarfs, $W^{\odot}_{LSR}$ for the
giants can be placed further in the $+W_{LSR}$ distribution than the
$-W_{LSR}$ distribution.  The placement is twice as far as the maximum
$W^{\odot}_{LSR}$ position for the dwarfs.  This shows the position of
$W^{\odot}_{LSR}$ for the giants is now sensitive to the slight asymmetry due to $D_{-}$ in the bottom of Fig. \ref{fig:giant_kuiper_lsr} at $|W_{LSR}| = 24$ km s$^{-1}$ because $D_{+}$ is at the centre of the distribution.  

 $W^{\odot}_{LSR}$ is in agreement with the mode of the $W_{LSR}$ distribution
 in Fig. \ref{fig:giant_kuiper_lsr}.  Again, this suggests that there cannot
 be many stars belonging to vertically coherent streams with systemic
 $W_{LSR}$ velocities.  If there were, they would have biased the measurement
 of $W^{\odot}_{LSR}$ away from the mode.   Table 3 shows the value of $D_{-}$
 for the giants is greater than for the dwarfs and so the value of $D$ is also
 larger for the giants than for the dwarfs.  In spite of this, the $Q$ value
 for the giants is less significant than for the dwarfs because the $N_{e}$
 value for the giants is half the value for the dwarfs.  Hence, the $D_{-}$
 asymmetry does not bias the
 $W^{\odot}_{LSR}$ value for the giants.
 
 \begin{table*}
\begin{minipage}{170mm}
\centering
\label{tab:giant_kuiper_centre}
  \caption{Results of the Kuiper symmetry test, applied to the CORAVEL
     giants, to determine $W^{\odot}_{LSR}$ to the nearest 0.1 km s$^{-1}$, where $\Delta W^{\odot}_{LSR} = W^{\odot}_{LSR} - 7.0$ km s$^{-1}$.}
      \begin{tabular}{@{}rrccccccc@{}}
  \hline
$W^{\odot}_{LSR}$ & $\Delta W^{\odot}_{LSR}$ & $N_{W_{LSR} < 0}$ & $N_{W_{LSR} > 0}$ & $N_{e}$ & $D_{-}$ & $D_{+}$ & $D$ & $Q$\\
\hline
5.4  & $-1.6$ & 3397 &       2633 &     1483   &0.0609 &  0.0082  &  0.0691& $3\times10^{-5}$\\
7.0 & 0.0 & 3127        &2903      &1505   &0.0316  &0.0081    &0.0397&0.1415\\
10.4 & 3.4 & 2589     &   3441 &     1477  & 0.0016 &   0.0677 &   0.0693& $3\times10^{-5}$\\
   \hline
\end{tabular}
\end{minipage}
\end{table*}

\subsection{Sensitivity of $W_{LSR}$ symmetry of the CORAVEL giants to tidal streams}
\label{s:giant_stream}

 \begin{table*}
 \begin{minipage}{170mm}
 \centering
 \label{tab:giant_kuiper}
   \caption{Results of the Kuiper symmetry test, applied to the CORAVEL
     giants, to determine the
     approximate number of tidal stream stars ($N_{s}$), required in
     pseudo-randomly generated $\sigma_{s} = 10$ km s$^{-1}$ Gaussians, placed
     at $\pm$(1, 2, 3, $\sim$13)$\sigma_{W_{LSR}}$ in the $W_{LSR}$ distribution,
     to cause the test to reject the null hypothesis at 4$\sigma$ ($Q < 6 \times 10^{-5}$). \% =
     $N_{s}/(N_{s} + N_{total})$ where $N_{total}$ = 6030. $\rho_{s}$ is the stellar density of each stream, in $N$ stars kpc$^{-3}$,
     calculated by dividing the $N_{s}$ of each stream by the volume within which the CORAVEL
     giants are complete, approximated by a hemisphere with radius $\sim$0.29 kpc ($\sim$0.05 kpc$^{-3}$).}
     \begin{tabular}{@{}rrrrccccccc@{}}
   \hline
 $\sigma_{W_{LSR}}$ & $N_{s}$ & \% & $\rho_{s}$ & $N_{W_{LSR} < 0}$ & $N_{W_{LSR} > 0}$ & $N_{e}$ & $D_{-}$ & $D_{+}$ & $D$ & $Q$\\
 \hline
 $-1$ &  300 &4.7 & 6000&3417     &   2913  &    1572   &0.0604  & 0.0084  &  0.0688& $1\times10^{-5}$\\
 1 & 800 &11.7&16 000& 3153     &   3677  &    1697  &0.0039  &  0.0665  &  0.0704& $2\times10^{-6}$\\
 $-2$ &  200 &3.2&4000&3327     &   2903    &  1550   &0.0772 &  0.0064  &  0.0836& $3\times10^{-8}$\\
 2 & 300 &4.7& 6000&3127     &   3203   &   1582  &0.0052  &  0.0687  &  0.0739& $2\times10^{-6}$\\
 $-3$ & 200 &3.2&4000&  3327     &   2903    &  1550   &0.0804 &  0.0064 &   0.0868& $5\times10^{-9}$\\
 3 & 200 &3.2&4000&3127      &  3103    &  1557   &0.0119  &  0.0678   & 0.0797& $2\times10^{-7}$\\
  $\sim$$-$13 & 200 &3.2&4000& 3327 &  2903 & 1550 & 0.0804 &   0.0064 &  0.0868 & $5\times10^{-9}$\\
   $\sim$13 & 200 &3.2&4000&3127     &   3103 &     1557   &0.0119 &   0.0697 &    0.0816 &  $7\times10^{-8}$\\
 \hline
 \end{tabular}
 \end{minipage}
 \end{table*}

This section repeats the technique applied to the CORAVEL dwarfs in Section
\ref{s:dwarf_stream} to the CORAVEL giants, except the dispersion of the
giants is slightly larger: $\sigma_{W_{LSR}} \approx$ 19 km s$^{-1}$.  Table 4
shows that, because $D_{+}$ is at the centre of the distribution, many more
stream stars are required to cause $D_{+}$ to generate a 4$\sigma$ stream
detection at $W_{LSR}$ = 1$\sigma_{W_{LSR}}$ than at $-$1$\sigma_{W_{LSR}}$.
As before, the test becomes more sensitive (less stream
stars required to generate a $4\sigma$ detection), the further a stream is from the centre of the
distribution.  However, Table 4 shows that, again, the number of stream stars required for a $4\sigma$
detection stays
approximately constant (saturates) at $\le$$-3$$\sigma_{W_{LSR}}$ for the same
reason as before.

\begin{figure}
\centering
               \psfig{figure=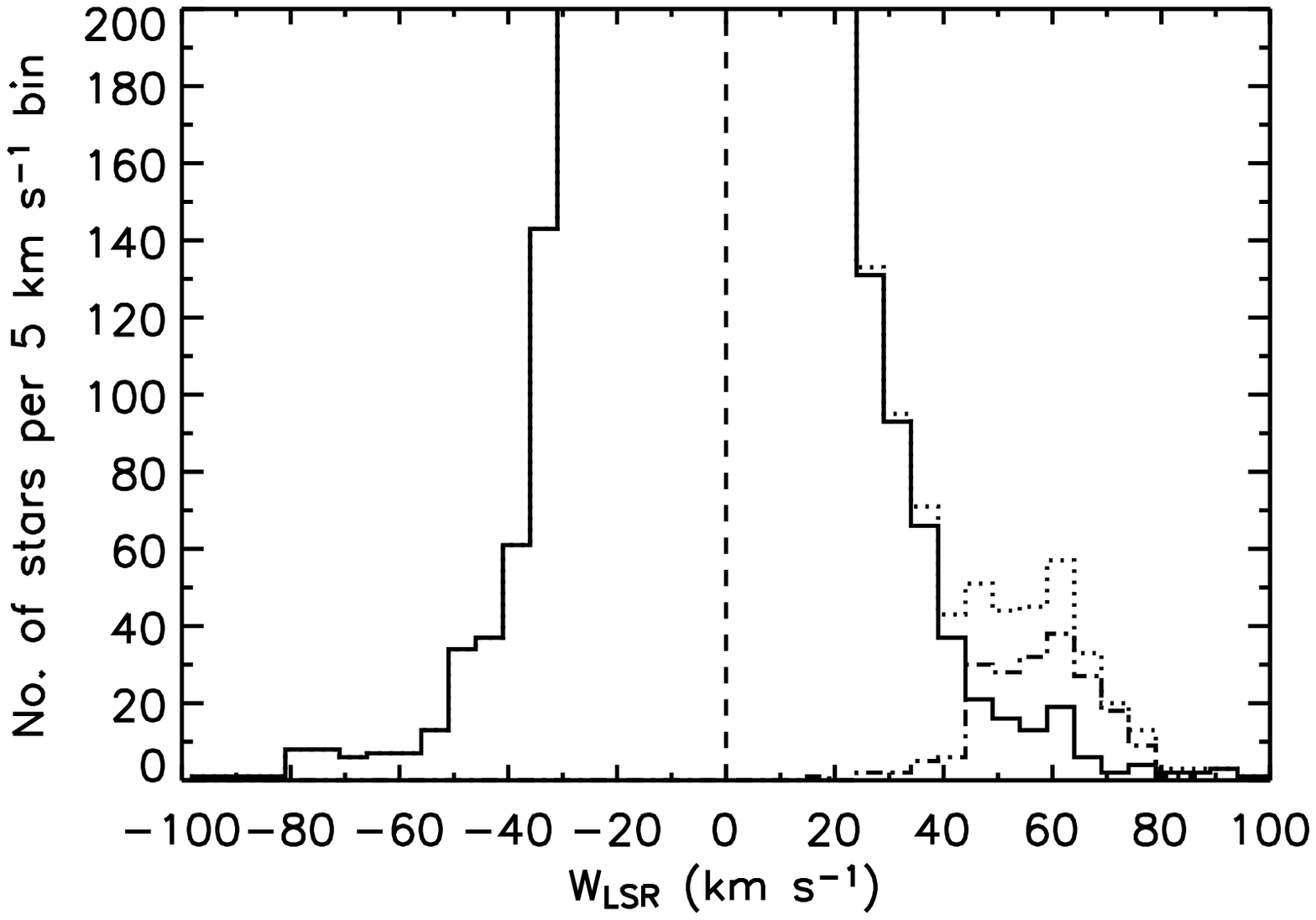,width=1.\columnwidth}
        \psfig{figure=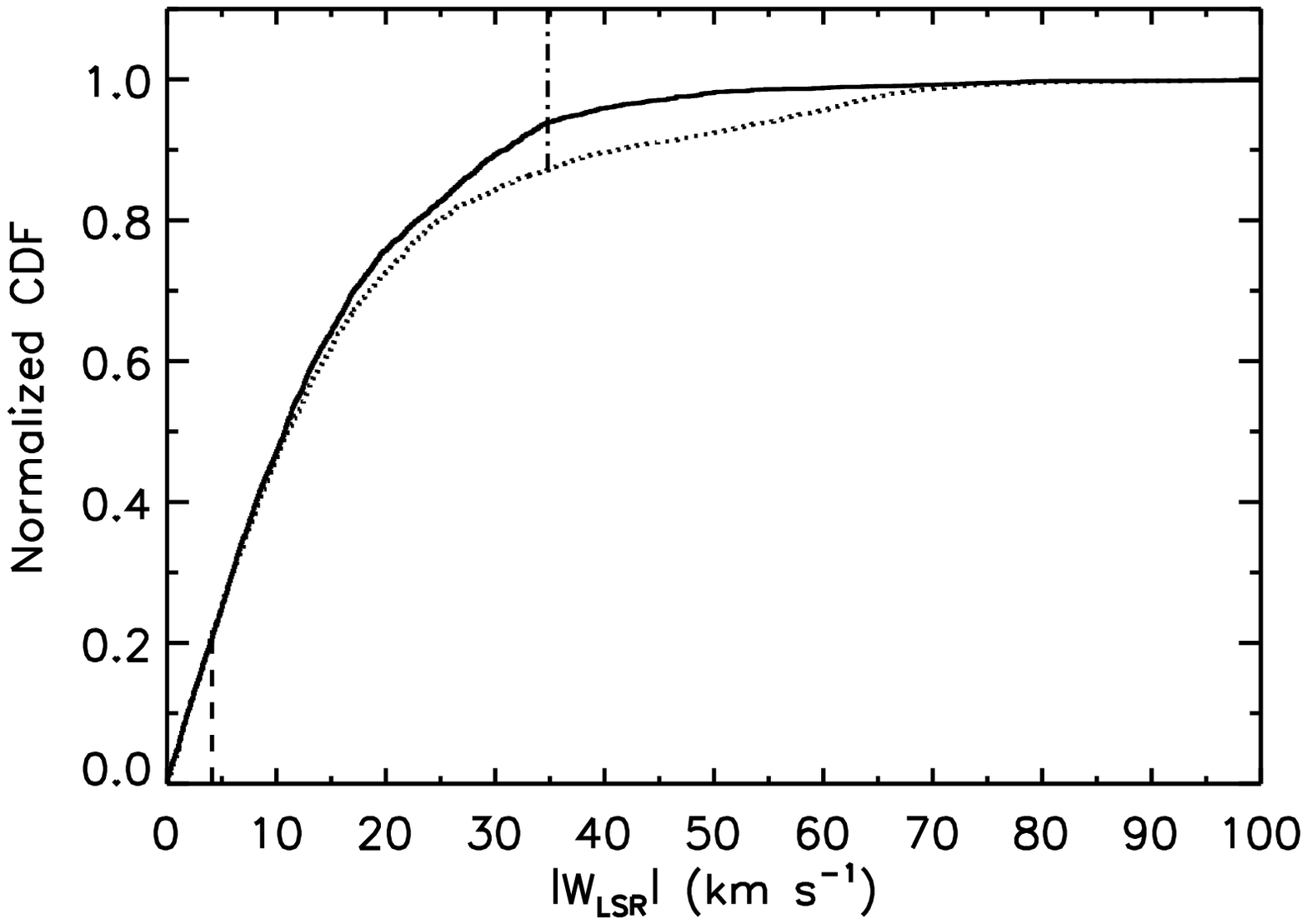,width=1.\columnwidth}  
\caption{{\it Top:} $W_{LSR}$ distribution of the CORAVEL giants (solid histogram)
               either side of the $W_{LSR} = 0$ dividing line (dashed line).
               The pseudo-randomly generated Gaussian (representing a $\sigma
               = 10$ km s$^{-1}$ tidal stream) included in the data set
               (dotted histogram) that caused the Kuiper symmetry test to
               reject the null hypothesis at 4$\sigma$ when placed at 3$\sigma_{W_{LSR}}$ is also plotted separately (dot-dashed line).  {\it Bottom:} Normalized CDF of $W_{LSR} < 0$ (solid line) and $W_{LSR} > 0$ (dotted line), showing $D_{-}$ (dashed line) and $D_{+}$ (dot-dashed line).}
                \label{fig:giant_kuiper_gausp}
\end{figure}

\begin{figure}
\centering
        \psfig{figure=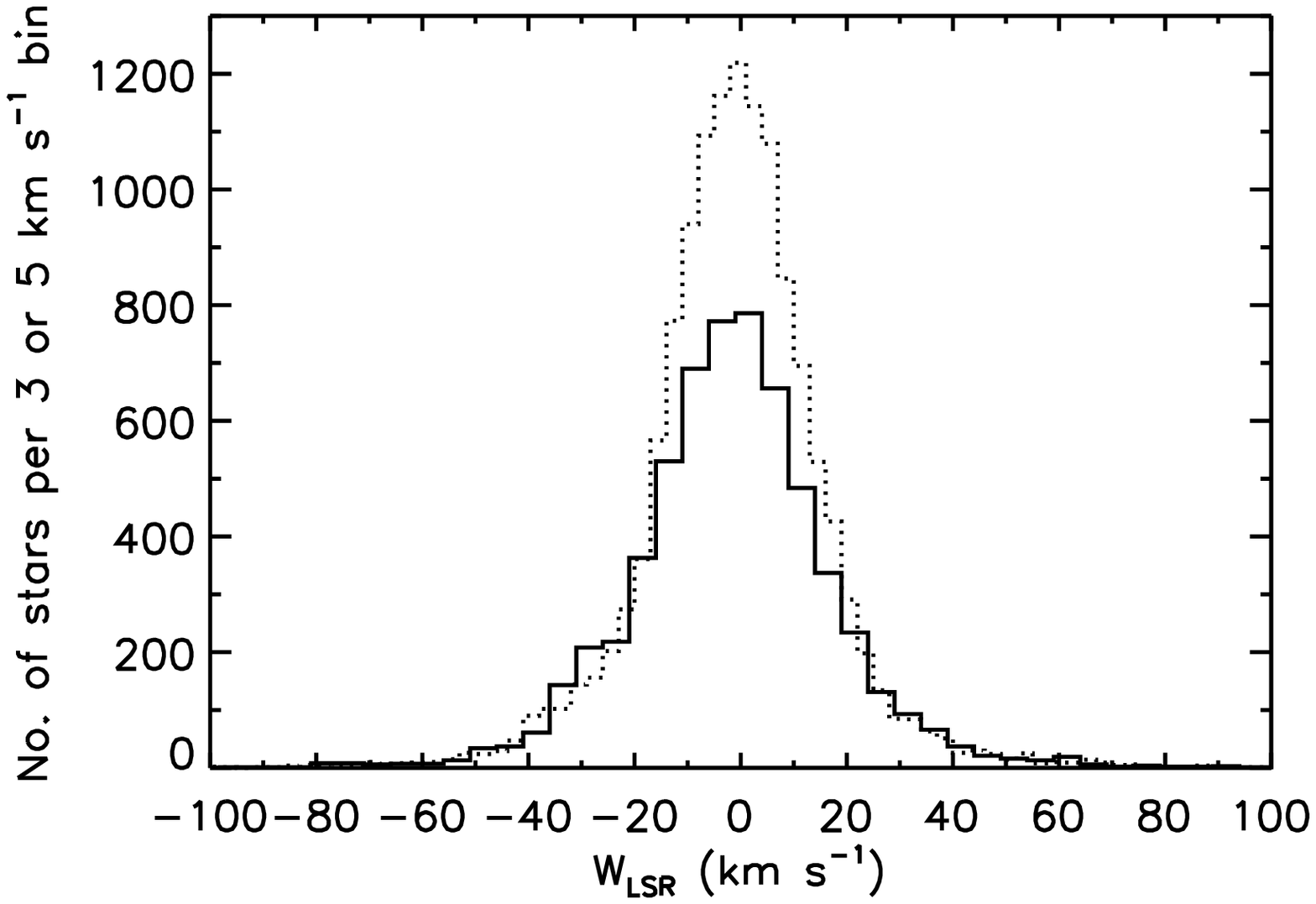,width=1.\columnwidth}
        \psfig{figure=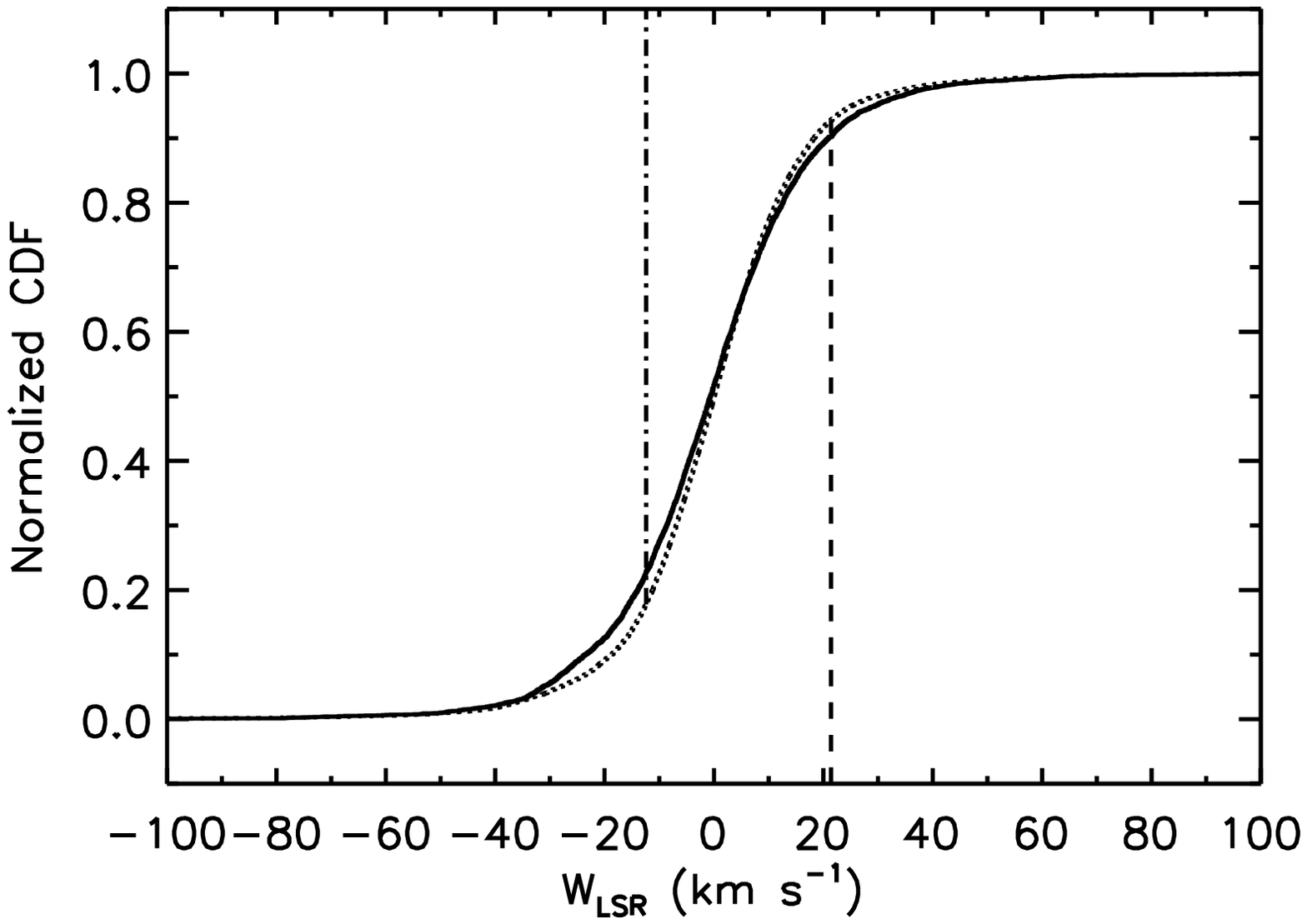,width=1.\columnwidth}
         \caption{{\it Top:} $W_{LSR}$ distribution of the CORAVEL survey: 6030 giants (solid histogram) and 13 240 dwarfs (dotted histogram).  {\it Bottom:} Normalized CDF of the giants ($N_{1}$, solid line) and the dwarfs ($N_{2}$, dotted line) as a function of $W_{LSR}$, where the maximum differences between them are indicated by vertical lines: $D_{+}$ (= 0.0509, dot-dashed line) and $D_{-}$ (= 0.0257, dashed line), where $N_{e}$ = 4143.08, $D$ = 0.0766 and $Q = 1\times10^{-19}$.}  
                                  \label{fig:giant_w}
          \end{figure}

\begin{figure}
\centering
        \psfig{figure=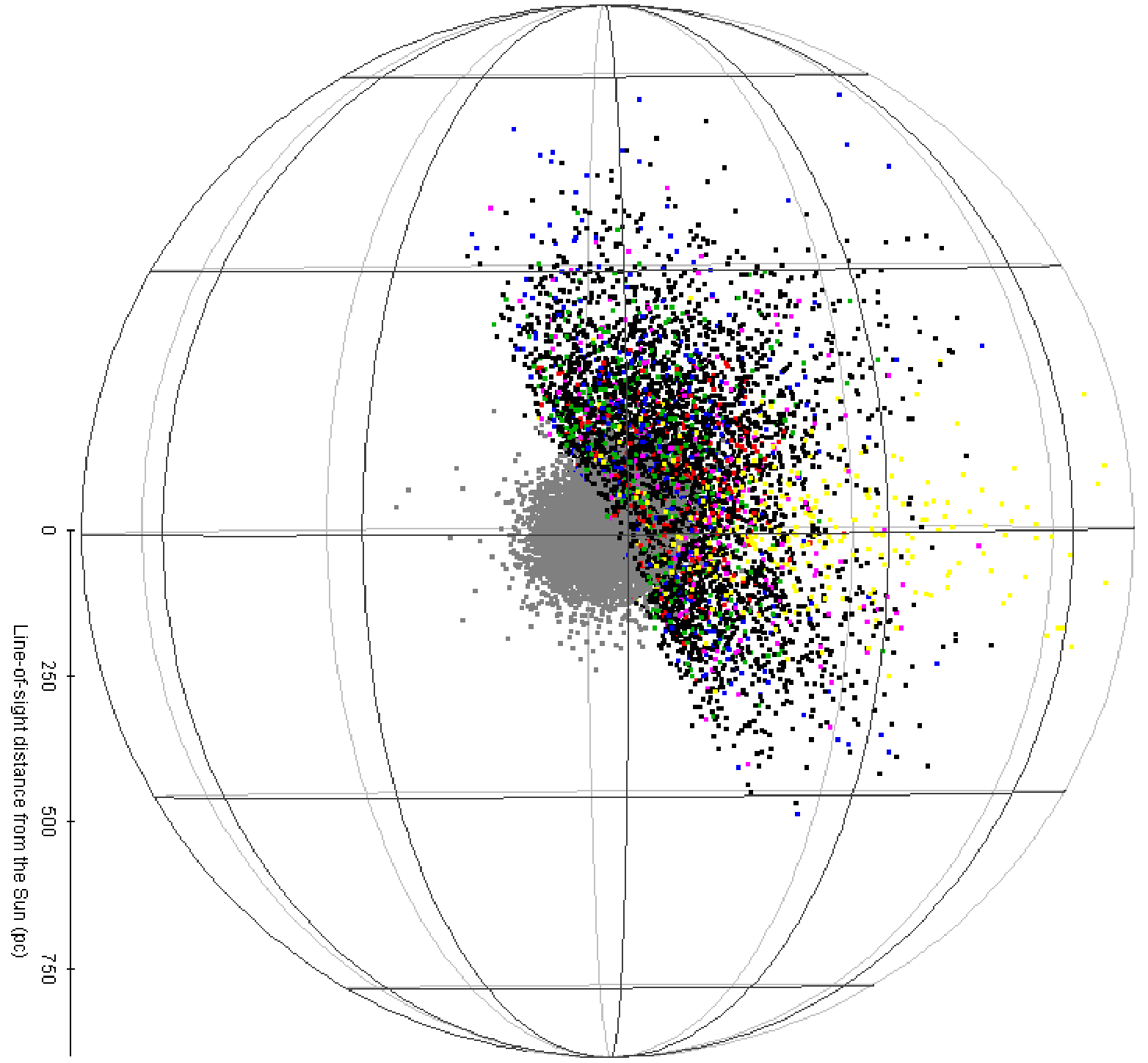,width=1.0\columnwidth}
 \caption{Three dimensional Galactic sky distribution of all the 13 240
        CORAVEL dwarfs with $W$ space
  velocities 
(grey) and the 6030 CORAVEL giants
        with $W$ space velocities, including spectroscopic binaries with
        centre-of-mass RVs, colour-coded the same as in Fig. \ref{fig:famaeyj}.
  The spherical polar axes have a radius of 900 pc, centred on the
  Sun. 
         The viewer's angle is the same as in Fig. \ref{fig:gencop3d}: chosen to
        illustrate the $63^{\circ}$ between the Earth's equator and the
        Galactic plane (longest horizontal line), emphasizing that the CORAVEL
  giants are only visible from the Earth's Northern
        hemisphere.}
                \label{fig:famaey3d}
\end{figure}

The number of stream stars required in the dwarf
sample are also needed in the giant sample at $W_{LSR}$ = 3$\sigma_{W_{LSR}}$ to generate
a 4$\sigma$ detection (see Fig. \ref{fig:giant_kuiper_gausp}).  This is
because the top of Fig. \ref{fig:giant_w} shows that both samples include
similar numbers of stars at $W_{LSR} =$ $\pm$3$\sigma_{W_{LSR}}$, even though the dwarf sample is twice as large as the giant one.   The larger volume sampled by the giants (see Fig. \ref{fig:famaey3d}) increases the numbers of giants in the tails of its $W_{LSR}$ distribution.  It is this greater volume, and correspondingly different stellar population mix, that makes the giant distribution differ from the dwarf one as seen in the bottom of Fig. \ref{fig:giant_w}.

Fig. \ref{fig:famaey3d} shows that the nine stars at $\sim$4.55 kpc towards
the NGP, proposed as a candidate halo stream by \citet{majewski1992}, are not
in the sample of CORAVEL giants. \citet{majewski1994} measured the RVs of six out
of the nine stars and H99a calculated their mean vertical velocity to be
$\langle W\rangle = -76 \pm 18$ km s$^{-1}$ and $\sigma_{W} = 35 \pm 24$ km
s$^{-1}$.  There are giants in the sample with these velocities but they do
not share the angular momenta of these stars computed by H99a as $J_{Z} <<
0$.  This is far from the range in Fig. \ref{fig:famaeyj}, suggesting none of
the CORAVEL giants are members of this stream.  We do not see any evidence of the more general halo streaming
from the NGP towards the Galactic plane
reported by \citet{majewski1996} and \citet{kinman2007}. \citet{kinman2007}
speculated their streaming could be connected with the VOD. 

  The method briefly described in Section \ref{s:giants_dev} allowed F05 to
        search for and characterize kinematic substructure in their sample.  Their method identified three kinematic base groups:
        the Sirius moving group, giants with young kinematics and the
        Hyades-Pleiades supercluster.  F05 found
        these groups have dispersions similar to cold streams
        ($\sigma_{U,V,W} \sim$ 5-10 km s$^{-1}$).  The smooth background
        contains more than an order of magnitude more stars than in each
        group.  The
        groups are barely visible as overdensities against the background (see \citealt{seabroke2007}
        fig. 3). Each contains similar numbers of stars
        (see table 2 in F05) to the numbers in Table 4 required for our Kuiper
        symmetry test to detect vertically coherent streams.  F05 did not report any vertical
        substructure.  This agrees with our Kuiper symmetry test results and the
        CORAVEL giants being well phase-mixed vertically (see
        \citealt{seabroke2007} figs 4 and 5).  This suggests the $D_{-}$
        asymmetry in Fig. \ref{fig:giant_kuiper_lsr} is not caused by a
        vertically coherent stream.

Approximately 80 per cent of the CORAVEL giants are Hipparcos `survey' stars.  Therefore, the sample is complete for the K and
M giants brighter than $V = 7.3 + 1.1 \vert$$\sin b\vert$
\citep{udry1997}. This magnitude threshold translates into distances of 290 pc
in the galactic plane and 480 pc in the direction of the galactic pole for a
typical $M_V=0$ giant.  These values are, however, very sensitive to the
adopted absolute magnitude.  They become 45 pc for a subgiant with $M_V =
+4$ in the galactic plane, and 2.9 kpc for a supergiant with $M_V = -5$
(e.g. the most luminous giants and supergiants colour-coded yellow in
Fig. \ref{fig:famaey3d}).  We consider the volume within which the CORAVEL giants
are complete to be approximated by a hemisphere with radius $\sim$0.29 kpc ($\sim$0.05 kpc$^{-3}$).

From the \citet{freese2005} arguments, we estimate that
        10-80 Sgr stream stars could be within the volume where the CORAVEL
        giants are complete.  A caveat against interpreting the $D_{-}$
        asymmetry as not due to a stream is that the
        numbers of stars found by F05 to be part of substructure are much larger than the expected number of Sgr and
        \citet{majewski1992} halo stream stars.  This implies that both the F05 method and the Kuiper symmetry test cannot
        rule out the presence of these streams in the volume sampled by CORAVEL giants.

The larger volume sampled by the giants than the dwarfs can, however, rule out
the presence of the VOD, if their stellar densities estimated from the
\citet{fuchs2006} CADIS fields are realistic. If this is the case, they would consist of 6000-8000 CORAVEL
giants, which the Kuiper test can strongly rule out.  Table 4 shows that the
test can rule out these streams down to a density of $\sim$4000 stars
kpc$^{-3}$ (depending on kinematics).

Thick disc and halo stars make up $<$10 per cent of the sample (high-velocity stars, colour-coded blue in Fig. \ref{fig:famaey3d}).  Therefore, although the CORAVEL giants have extended our search volume to greater vertical distances than the thin disc scaleheight ($\sim$220 pc), it is still dominated by the thin disc.  RAVE is the only RV survey that vertically samples significant numbers of thick disc stars (as well as old thin disc stars) at the solar position.

\section{RAVE}
\label{s:rave}

\subsection{RAVE stellar populations}
\label{s:ravepop}

\begin{figure}
               \psfig{figure=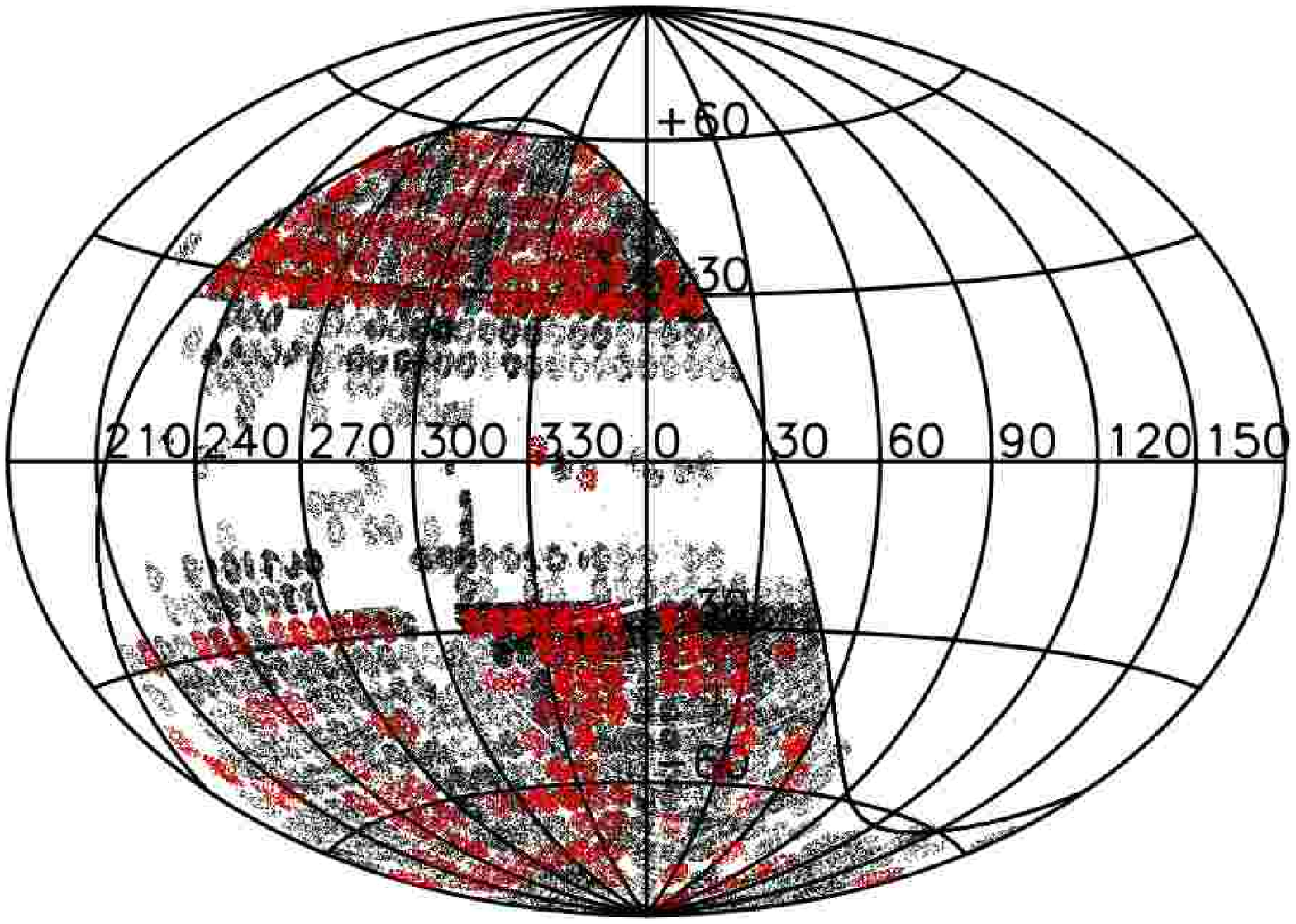,width=1.0\columnwidth}
\caption{Aitoff projection of the Galactic co-ordinates (in degrees) of the 151 856 RVs in the internal RAVE data release (black dots) and the 24 748 publicly available RVs from the first  data release (red dots).  The vast majority of RAVE data is below the plotted celestial equatorial plane ($\delta = 0\hbox {$^\circ $}$).  The pattern is due to the survey field centres of the first and second input catalogues being defined on $5.7^{\circ}$ and $5.0^{\circ}$ grid spacing respectively (the field of view of the 6dF
field plates is $5.7^{\circ}$).}
\label{fig:ravesky}
\end{figure}

To date (2007 June 26), 220 070 RAVE spectra have been amassed from 196 131
stars.   This paper utilizes an internal RAVE data release
containing 151 856 RVs, which includes 24 748 RVs from the first public data release\footnote{The first public data release of the RAVE catalogue is available and can be retrieved or queried from the RAVE Collaboration website: http://www.rave-survey.org.} \citep{steinmetz2006paper}, a
similar number of currently unpublished RVs from the forthcoming second public data
release (Zwitter et al. 2007, in preparation) and future data releases (see Fig. \ref{fig:ravesky}).

\begin{figure}
               \psfig{figure=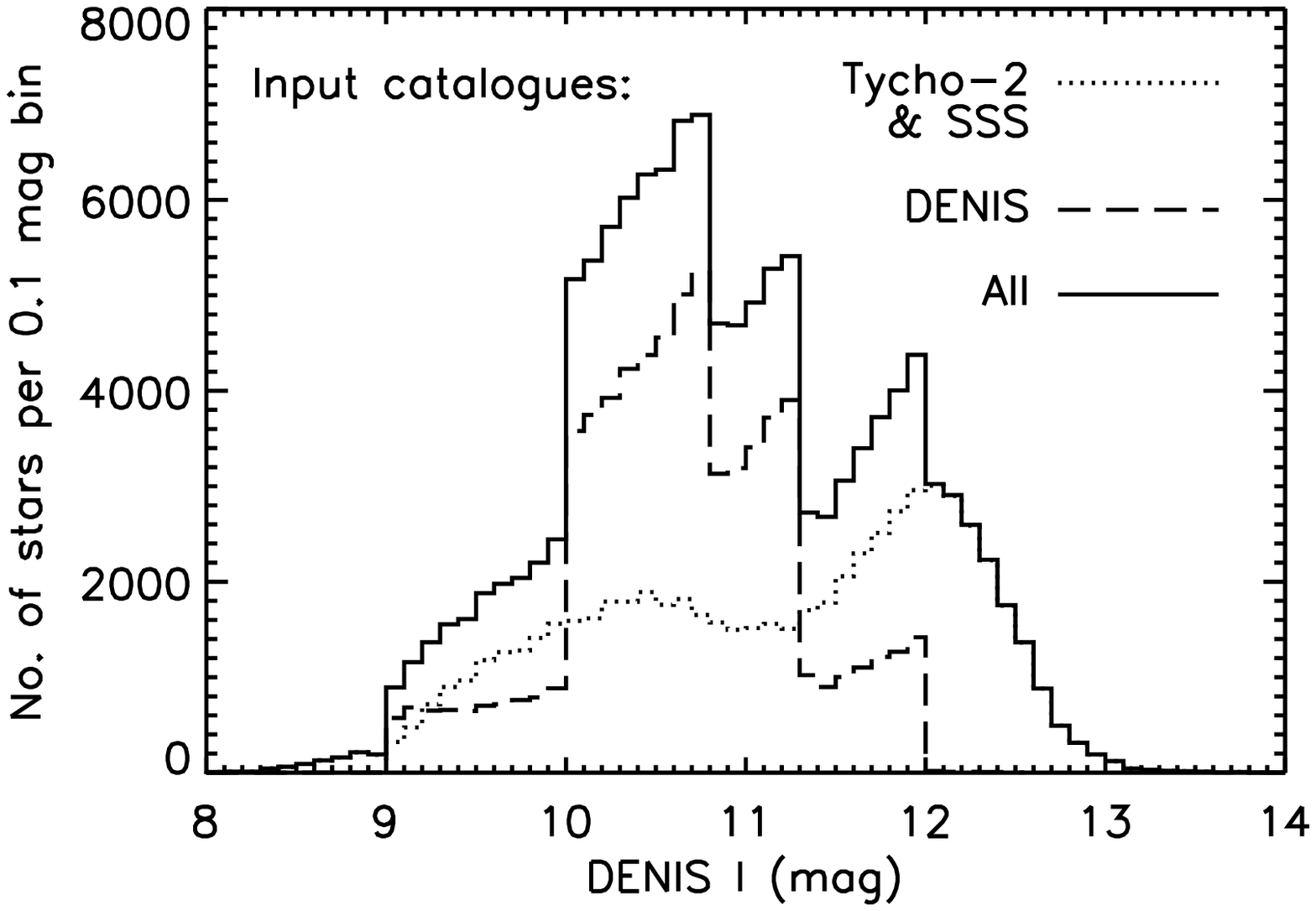,width=1.0\columnwidth}
\caption{DEep Near-Infra Red Survey
  (DENIS, \citealt{epchtein1999paper}) $I$-band selection functions
  of 131 632 RAVE RVs, observed with the two RAVE input catalogues.  {\it Tycho-2} and
  the SuperCOSMOS Sky Survey (SSS, \citealt{hambly2001a}) were used to
  generate the first and second RAVE data releases (see
  \citealt{steinmetz2006paper} for details) and contribute 64 581 RVs to the plot (49 per cent).  DENIS was used to define the current
  input catalogue, which is still being observed and will generate all future
  data releases and contributes 67 051 RVs to the plot (51 per cent). (20 224 RAVE RVs are not included in the plot because of incomplete DENIS sky coverage.)}
\label{fig:ravei}
\end{figure}

Unlike the CORAVEL surveys that targeted specific stellar populations (F-G dwarfs and K-M giants), RAVE does not employ any colour selection or spectral type cuts.  This means RAVE observes all the stellar populations within the apparent $I$ magnitude-limited selection functions of its input catalogues (see Fig. \ref{fig:ravei}).  \citet{steinmetz2006paper} fig. 5 shows that this selection function is much fainter than the CORAVEL (and {\it Hipparcos}) surveys.  Therefore, there is only a few per cent serendipitous overlap of RAVE stars with {\it Hipparcos} parallaxes (with large errors) and/or Str{\" o}mgren photometry at the bright end of RAVE.  Hence, the distance determination methods described in Section \ref{s:dwarfs_dev} are not available to RAVE stars and the method described in Section \ref{s:giants_dev} is beyond the scope of this study.  RAVE stars cannot be placed in a H-R diagram because sufficiently accurate photometrically derived distances are not yet available (discussed in more detail below).  

\begin{figure}
               \psfig{figure=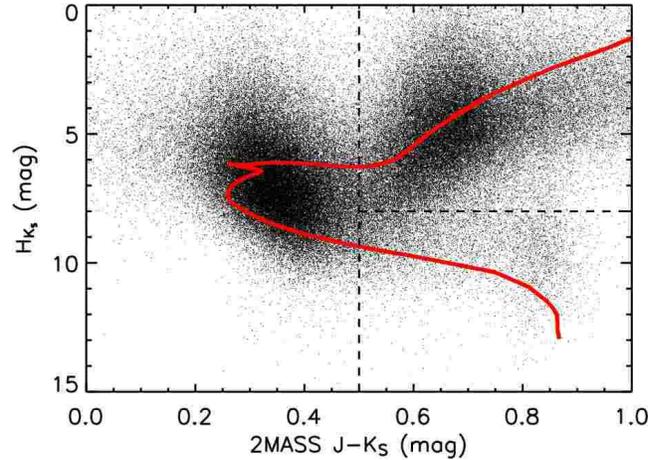,width=1.0\columnwidth}
\caption{Reduced proper motion diagram of 149 660 RAVE RVs (2196 either do not have 2MASS photometry or measured proper motions).  The solar-scaled Padova 2MASS isochrone
  (\citealt{bonatto2004}, red line), chosen to be representative of thin
  disc stellar populations (2.5 Gyr old, initial metal fraction = 0.019,
  [Fe/H] $\approx -0.02$ dex), has been given thin disc kinematics ($v_{T}
  \approx$ 30 km s$^{-1}$ in equation \ref{equ:rpm}) in order to interpret
  RAVE's stellar populations (approximately partitioned by the dashed lines).}  
\label{fig:raverpm}
\end{figure}

Instead, we illustrate RAVE's different stellar populations using a reduced proper motion diagram.  The concept of reduced proper motion ($H$) was first used by E. Hertzsprung \citep{luyten1968} in the absence of absolute magnitude.   Just as parallax fixes the absolute magnitude exactly, $H$ determines the absolute magnitude approximately.  When this is plotted against colour, it produces a statistical H-R diagram (kinematic colour-apparent magnitude diagram).   $H$ combines the observable properties of proper motion ($\mu$, in arcsec yr$^{-1}$) and apparent magnitude [the most comprehensive photometric coverage of observed RAVE stars is the 2MASS Point Source Catalogue \citep{cutri2003paper} so here we choose 2MASS $K_{S}$] in the equation 

\begin{equation}
H_{K_{S}} = K_{S} + 5\log\mu + 5 = M_{K_{S}} + 5\log v_{T} - 3.379,        
\label{equ:rpm}
\end{equation} 

\noindent which also expresses $H_{K_{S}}$ in terms of the more fundamental
properties of absolute magnitude ($M_{K_{S}}$) and tangential velocity ($v_{T}$ in km s$^{-1}$).  Fig.  \ref{fig:raverpm} shows that the $H_{K_{S}}$$-$($J-K_{S}$) plane can divide the observed RAVE stars into three distinct stellar populations.  

The largest stellar population in the RAVE sample has $J-K_{S} < 0.5$ mag ($\sim$51 per cent).  These stars are mainly thin disc F-G dwarfs [represented on the isochrone by the Main Sequence (MS) at $H_{K_{S}}$ = 8-9 mag].  Although the subgiant branch (represented on the isochrone by the horizontal branch along $H_{K_{S}}$ = 6 mag)  is
 a short phase of stellar evolution (see the Hertzsprung Gap at $J-K_{S}$ =
 0.5 mag in Fig. \ref{fig:raverpm}), there will be a non-negligible number of
 subgiants in the RAVE sample.  $H_{K_{S}}$ is unable to kinematically
 separate the subgiants from the F-G dwarfs because their values of absolute magnitude are too
 similar for their proper motions to resolve the overlapping kinematics.  This results
 from the overlapping distance ranges of the two populations, due to the large
 spread in apparent magnitude in Fig. \ref{fig:ravei}.
 
 The two populations with  $J-K_{S} > 0.5$ mag have sufficiently different absolute magnitudes for their proper motions to resolve the very different kinematics resulting from the very different distance ranges of the two populations.  The fainter population ($\sim$5 per cent) are thin disc K-M dwarfs (represented on the isochrone by the MS at $H_{K_{S}}$ = 9-14 mag).   The brighter population ($\sim$44 per cent) are K-M giants (represented on the isochrone by the Red Giant Branch at $H_{K_{S}}$ = 1-4 mag).  Although the RAVE giant sample is still dominated by the thin disc, their intrinsic brightnesses probe distances that sample a statistically significant number of thick disc stars and a non-negligible number of inner halo stars.
 
 \subsection{Distances to RAVE stars}
 
Fig. \ref{fig:raverpm} shows that $H_{K_{S}}$ bifurcation can be used to
kinematically select K-M dwarfs, without spectroscopically derived log $g$, to
derive photometric distances.  Their atmospheres are dominated by saturated molecular
bands so their luminosity is not very sensitive to metallicity.  This means
isochrones for this population are very similar for different metallicities
and ages, leading to small photometric distance errors.  RAVE K0 dwarfs probe
$\sim$50-250 pc from the Sun.  This volume has already been sampled by the
CORAVEL dwarfs (see Fig. \ref{fig:gencop3d}) and shown to be devoid of tidal
streams  in Section \ref{s:dwarfs}.  Therefore there is no advantage to
searching this volume again with RAVE's less accurate $W$ space velocities.

The largest uncertainty in deriving photometric distances to the RAVE F-G
dwarfs is stellar age.  Absolute magnitude is most sensitive to age close to
the MS turnoff.  Unlike the CORAVEL dwarfs, RAVE does not have an independent
source of trigonometric parallax from which distances can be inferred.  This enables stars to be placed in the H-R diagram and allows age to be estimated from a star's evolution
away from the zero-age MS.   In principle, a chromospheric activity indicator
is present in RAVE spectra in the form of Ca II emission.  This feature
declines with time (as stellar rotation decreases with age) but decays into
invisibility at about the age of the Sun.  Ranking age using this diagnostic
is fraught with systematic errors.  RAVE F-G dwarfs probe similar distances
($\sim$200-400 pc) to the CORAVEL giants but are complementary in their
non-overlapping sampled volumes (CORAVEL giants are Northern celestial
hemisphere only and RAVE is Southern celestial hemisphere only -- see
Fig. \ref{fig:famaey3d}).  Therefore, there is potential advantage to deriving
RAVE F-G dwarf space velocities.  However, the associated errors
will not yield a reliable $W$ distribution to search for asymmetries.

Absolute magnitude is sensitive to [$\alpha$/Fe] as well as [Fe/H].  RAVE K-M
giants include thick disc and halo stars, which have, compared to the thin
disc, enhanced [$\alpha$/Fe].  Metallicity uncertainty dominates
 the uncertainty in deriving photometric distances to the RAVE K-M giants.  

Stars in the first RAVE data release were observed without a blue light blocking filter (OG531).  Their spectra were contaminated with second order light, which means stellar parameters are not currently available for these stars (see \citealt{steinmetz2006paper} for details).  Stars in all future RAVE data releases have all been observed with an OG531 filter so their spectra only contain first order light.  Stellar parameters can be derived for these spectra from the best-match synthetic spectrum.   At the time of writing, the accuracy of these parameters is being finalized ready for the second data release (Zwitter et al. 2007, in preparation).  The log $g$ is likely to be sufficiently accurate to resolve the $J-K_{S} > 0.5$ mag sample into dwarfs and giants. However, the log $g$ errors may not be small enough to resolve the $J-K_{S} < 0.5$ mag sample into dwarfs  and subgiants.  

RAVE K-M giants probe greater distances ($\sim$0.7-3 kpc) than the CORAVEL
giants.  Therefore, there is great advantage to deriving the space
velocities of RAVE K-M giants because it is unexplored phase space on Galactic scales in between
the very local thin disc and inner halo at the solar position.  However, tangential velocity errors scale with distance
from the Sun.  The absolute magnitudes of nearby {\it Hipparcos} red clump
giants show no correlation with [Fe/H].  They have a mean $M_{K} = -1.61 \pm
0.03$ and $\sigma_{M_{K}} \approx 0.22$ mag \citep{alves2000}, which
translates to a distance error of $\sim$11 per cent.  At 1 kpc, this distance error converts $\Delta \mu$ = 2
mas yr$^{-1}$ to $\Delta v_{T}$ = 10 km s$^{-1}$ (Veltz et al. 2007)
and ergo $\Delta W >$ 10 km s$^{-1}$.  The red clump RAVE giants have the
smallest distance error of all the RAVE giants but their values of $\Delta
W$ are still too large to resolve any dynamically cold, vertical tidal streams
falling through the sample.  
The availability of RAVE giant metallicities would therefore not improve the
situation.  Outside the red clump, a RAVE K giant at a distance of 1 kpc with proper motion errors at a more typical level of 3.5 mas yr$^{-1}$ and a distance error
$>$15 per cent, due to the uncertainty in [Fe/H], results in $\Delta v_{T} >$
20 km s$^{-1}$ and ergo $\Delta W >$ 20 km s$^{-1}$.  

\subsection{RAVE radial velocity determination and accuracy}
\label{s:raverv}

\begin{figure}
               \psfig{figure=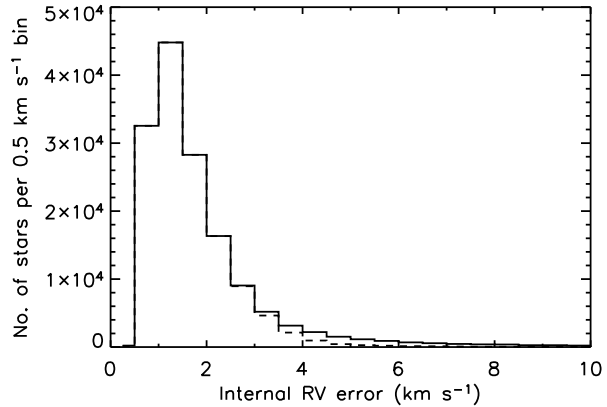,width=1.0\columnwidth}
\caption{Distribution of internal RV errors for all 151 856 RAVE RVs (all values of $r$, solid line) and the 140 228 RVs with $r > 15$ (dashed line).}
\label{fig:raveerv}
\end{figure}

The RAVE RV pipeline matches each RAVE spectrum with a
theoretical spectrum from the \citet{zwitter2004} and \citet{munari2005a}
libraries.  These libraries contain thousands of stellar templates, whereas
CORAVEL uses only one template.  The RAVE RV pipeline uses the standard cross-correlation procedure
\citep{tonry1979}, implemented in the IRAF package XCSAO \citep{kurtz1992}, to derive RVs and internal RV errors.  Fig. \ref{fig:raveerv} shows the mode of the internal RV error
distribution is $\sim$1 km s$^{-1}$.  A RV zero-point error of $\sim$1 km s$^{-1}$ makes the mode of the external RV error distribution $\sim$2 km s$^{-1}$.  

We use the Tonry-Davis cross-correlation coefficient ($r$,
\citealt{tonry1979}) to impose a quality cut.  $r$ is a measure of template
match to spectra.  Excluding RVs with $r < 15$, removes poor-quality spectra
and a small number of hot dwarfs due to template-mismatches.  This is seen in
the proportion of RAVE RVs with $J-K_{S} < 0.5$ mag and $r > 15$ decreasing
from $\sim$51 to 50 per cent.   Fig. \ref{fig:raveerv} shows this cut
decreases the amplitude of the tail of the internal RV error distribution,
which has sufficient accuracy to resolve dynamically cold streams.

RAVE RV accuracy compares favourably with the CORAVEL accuracy because the two strongest Calcium triplet absorption lines in the RAVE spectroscopic region (8410-8795 \AA) are powerful RV diagnostics at RAVE's medium resolution ($R \sim 7500$).  \citet{steinmetz2006paper} cross-matched RAVE first data release targets with the CORAVEL dwarfs.  Thirteen matches were
found but two of them were classified by N04 as binaries.
The remaining eleven single targets show good agreement,
with a mean difference of 1.4 $\pm$ 0.4 km s$^{-1}$ and $\sigma$ = 1.4 km s$^{-1}$.  

  \subsection{RAVE pseudo-$W$ without distances}
    \label{s:pseudow}
    
\begin{figure*}
\begin{minipage}{180mm}
\centering
        \psfig{figure=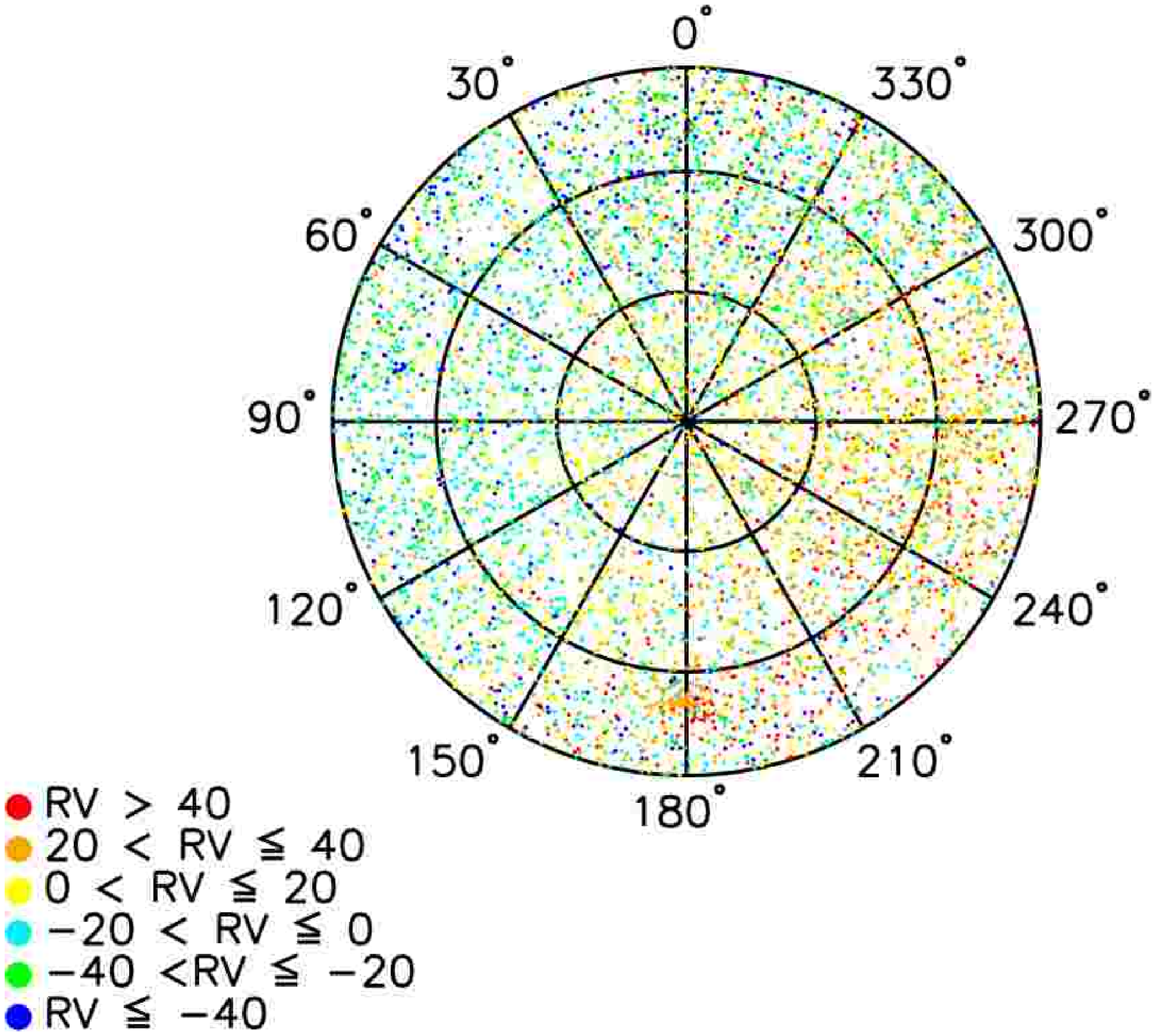,width=0.4\columnwidth}
        \psfig{figure=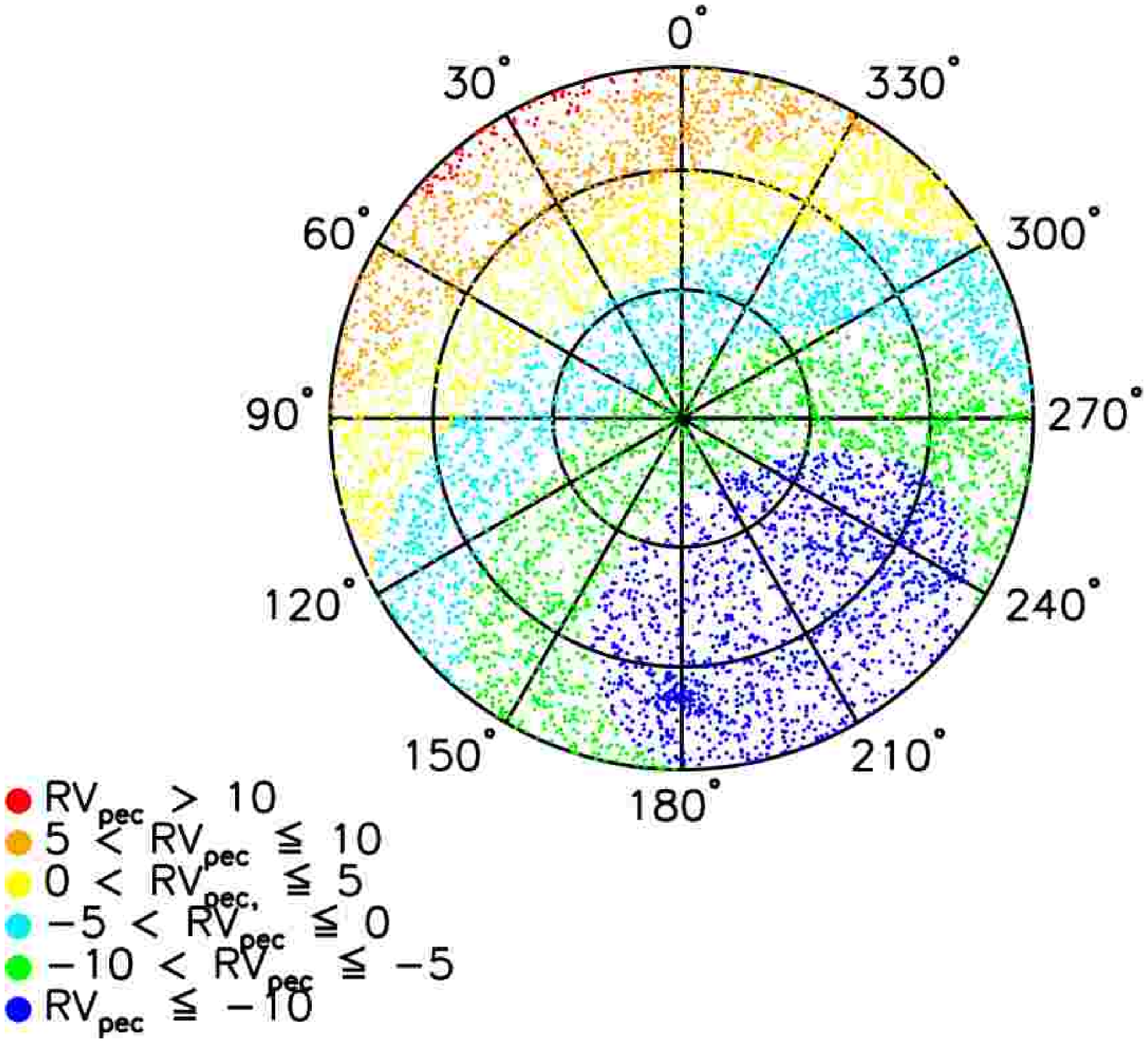,width=0.4\columnwidth}
        \psfig{figure=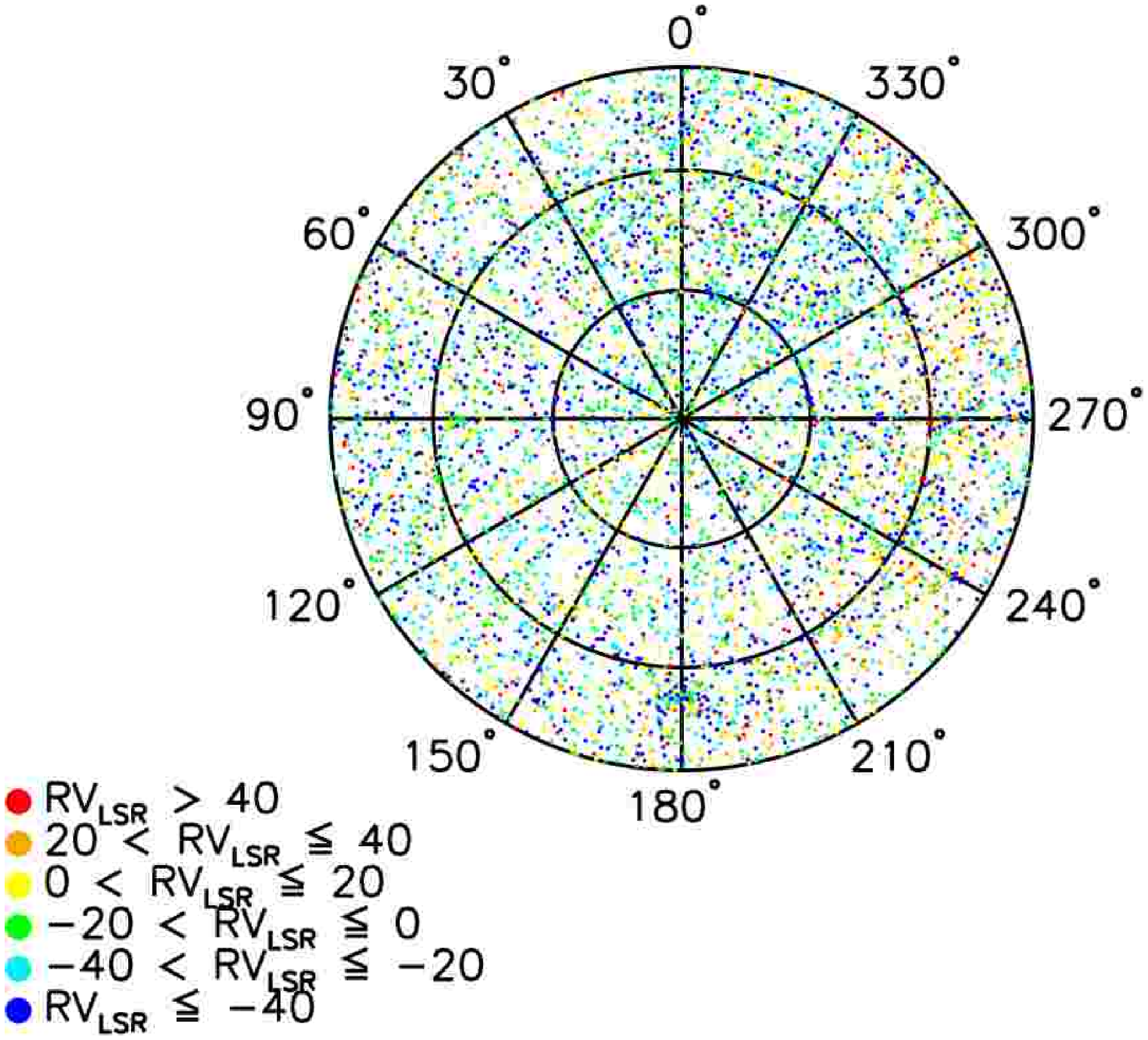,width=0.4\columnwidth}
        \psfig{figure=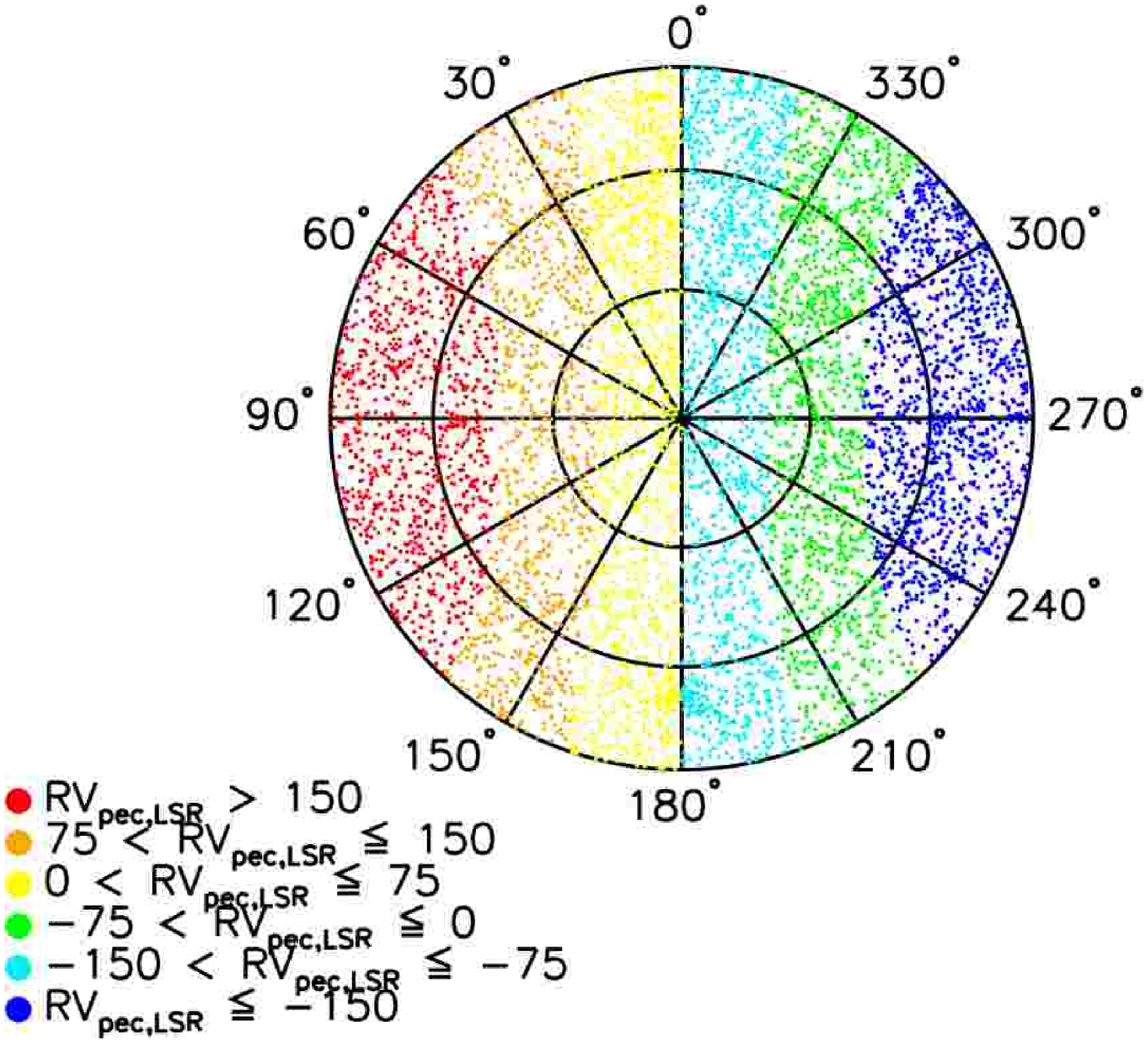,width=0.4\columnwidth}
        \psfig{figure=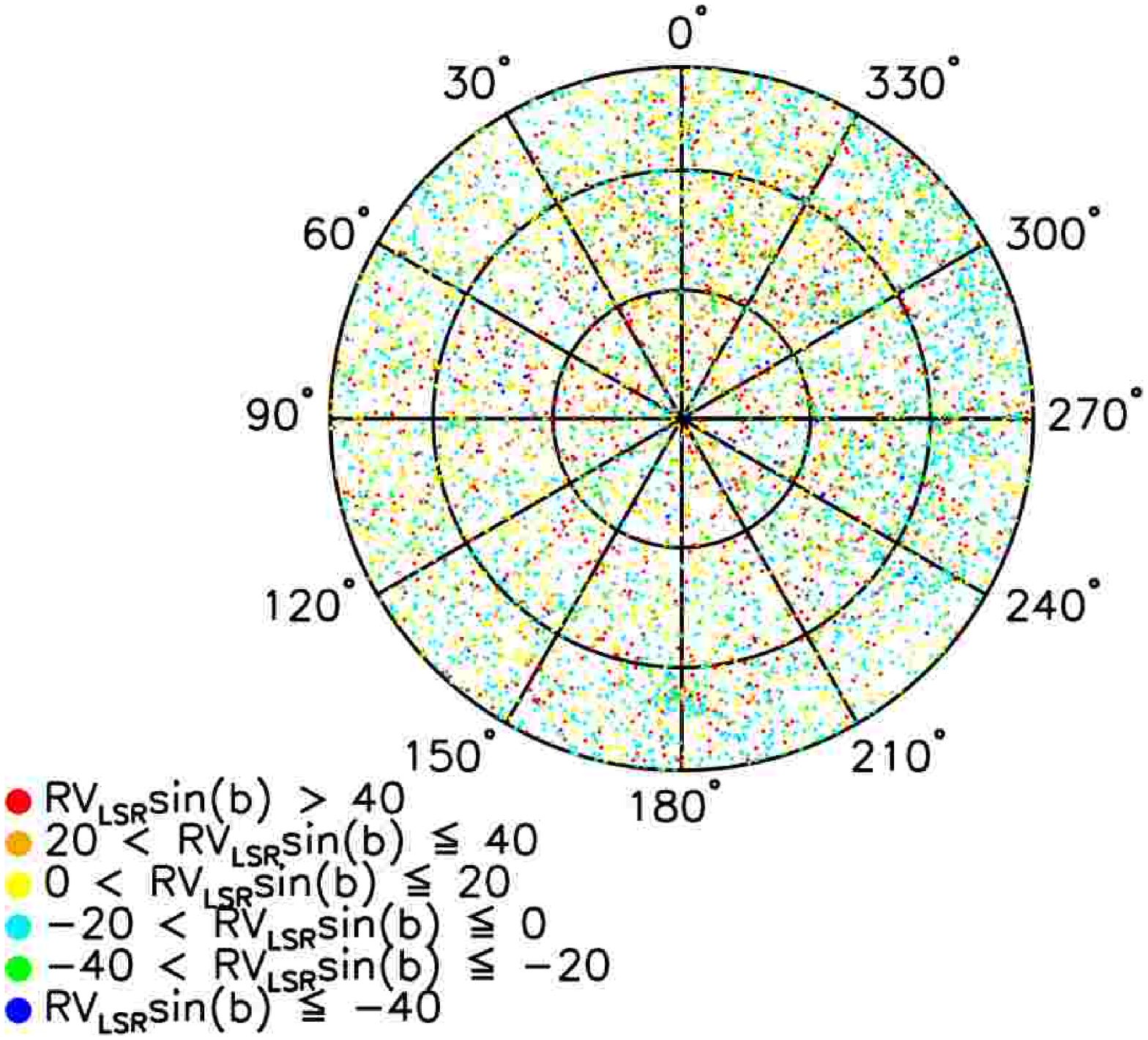,width=0.4\columnwidth}
        \psfig{figure=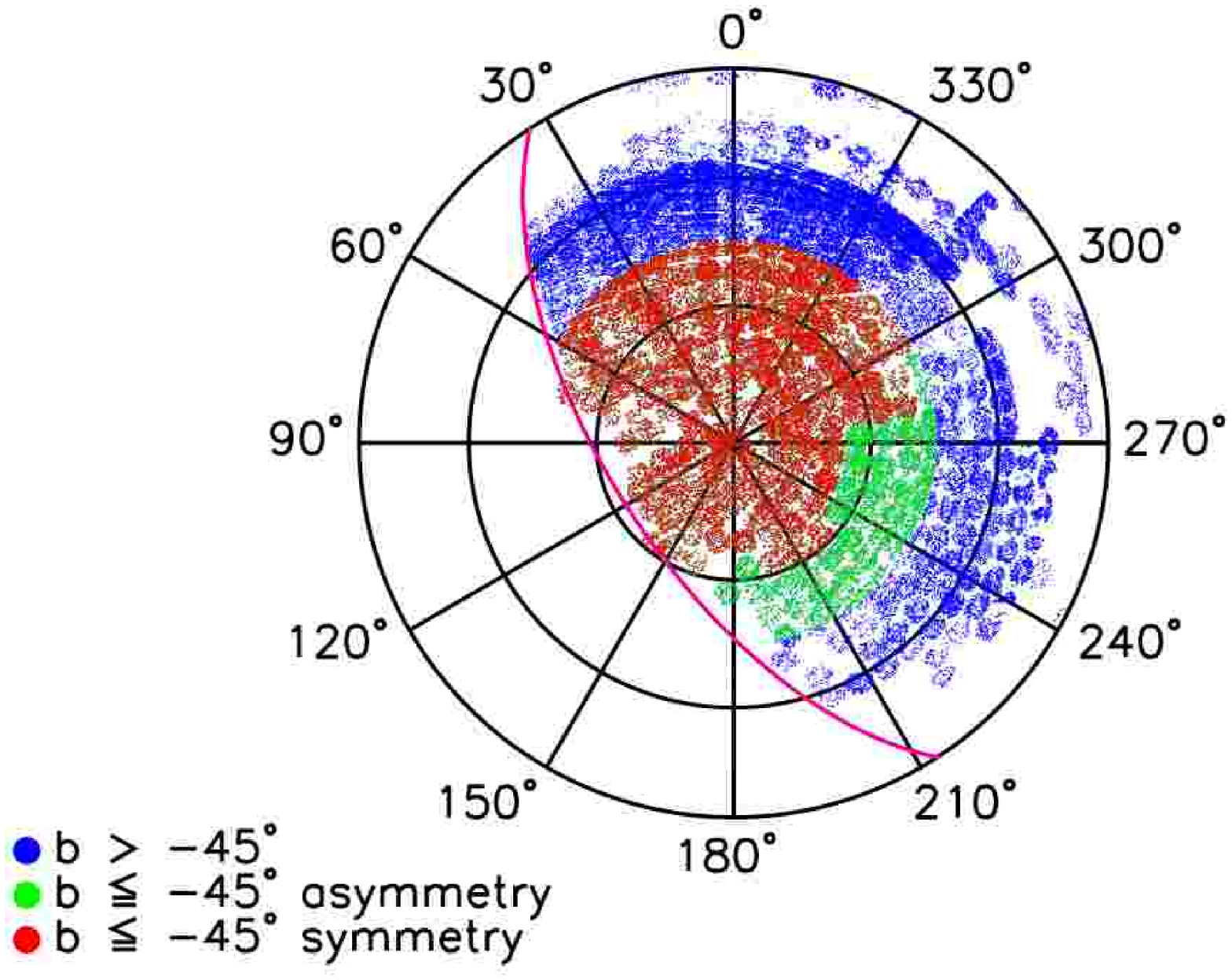,width=0.4\columnwidth}
        \vspace{0.5cm}
       \caption{Lambert polar (equal area corresponds to equal solid angle on
        the sky) projections of the Southern Galactic hemisphere, where
        the SGP ($b = -90^{\circ}$) is at the centre and concentric circles correspond to
        constant $b$ at $-60^{\circ}$, $-30^{\circ}$ and 0$^{\circ}$ (Galactic
        equator), around which the radial lines of constant $l$ are
        labelled.  CORAVEL dwarfs are plotted with their RVs in km s$^{-1}$ colour-coded in
        different reference frames according to the legend of each plot:
        heliocentric RV (top left); peculiar RV of the Sun relative to the LSR
        (solar motion, RV$_{pec}$, top right, see Appendix A for more details);
        RV with respect to the LSR (RV$_{LSR}$, middle left); peculiar RV of the LSR relative to the
        GSR (RV$_{pec,GSR}$, middle right); $W^{los}$ (RV$_{LSR}\sin b$,
        bottom left).  The overdensity at $l \sim 180^{\circ}$, $b \sim
        -20^{\circ}$ is the Hyades open cluster.  {\it Bottom right}: RAVE
        stars colour-coded according to RAVE field symmetry about
        $l = 180^{\circ}$.  The magenta curve traces the celestial equatorial
        plane ($\delta = 0\hbox {$^\circ $}$).}
        \label{fig:gencopsgc}
\end{minipage}
\end{figure*}

A star's heliocentric RV cannot be decomposed into the star's constituent $U$, $V$ and $W$ space velocities because, by definition, RV does not include any tangential contribution.  However, a star's RV can be resolved into its line-of-sight (los) components of these space velocities:

\begin{equation}
\textrm{RV} = \sqrt{(U^{los})^{2} + (V^{los})^{2} + (W^{los})^{2}}.
\label{equ:rv}
\end{equation} 
 
\noindent We remove the dependence of Galactic latitude on RV by only resolving the components in equation \ref{equ:rv} in a single dimension in the Galactic plane,

\begin{equation}
 UV^{los} = \sqrt{(U^{los})^{2} + (V^{los})^{2}} = \textrm{RV}\cos b,
 \label{equ:UVlos}
\end{equation} 

\noindent and the dimension perpendicular to the Galactic plane,

\begin{equation}
W^{los} = \textrm{RV}\sin b.
\label{equ:Wlos}
\end{equation} 

\noindent The above equations show that for a stellar los along $b =
-45^{\circ}$, RV consists of equal contributions from $UV^{los}$ and $W^{los}$
but for stars with $b < -45^{\circ}$, as $b$ decreases the contribution of
$W^{los}$ to RV increases and RV becomes less and less sensitive to
$UV^{los}$.  If a star's los is exactly towards the SGP ($b = -90^{\circ}$),
$W^{los} = W$ because here is there is no tangential component of $W$.  For
los with $-90^{\circ} < b < -45^{\circ}$, $W^{los} < W$ but $W^{los}$ is a useful proxy for $W$. Therefore, the issues discussed in the previous section can be circumvented by using the accuracy of the RAVE RVs
 towards the SGP with $b < -45^{\circ}$.
 
The top two plots in Fig. \ref{fig:gencopsgc} show that for differing los, RV in equation \ref{equ:Wlos} is sensitive to differing amounts of solar motion with respect to the LSR.  In order to compare $W^{los}$ along different los in the same reference frame, RV in equation \ref{equ:Wlos} is replaced by RV$_{LSR}$, which is RV corrected for the 
 solar motion, decomposed into its cardinal directions along the los relative to the LSR:
 
\begin{eqnarray}
\textrm{RV}_{LSR}  &=& \textrm{RV}+U^{\odot}_{LSR}\cos l\cos b+V^{\odot}_{LSR}\sin l\cos b \nonumber\\
                     & & +W^{\odot}_{LSR}\sin b.   
\label{equ:RVlsr}
\end{eqnarray} 

\noindent The middle left plot in Fig. \ref{fig:gencopsgc} shows that the LSR reference frame removes the peculiar components of the solar RV (top right plot) from the RV los (top left plot).  However, like RV, RV$_{LSR}$ is still dependent on the los.  The middle right plot in Fig. \ref{fig:gencopsgc} shows that RV$_{LSR}$ is sensitive to differing amounts of the peculiar RV of the LSR relative to the GSR,
        
 \begin{equation}
\textrm{RV}_{pec,GSR} = V_{rot}\sin l\cos b.
\label{equ:pecgsr}
\end{equation} 

\noindent The net Galactic rotation from $l=270^{\circ}$ to $l=90^{\circ}$ is
just visible in the middle left plot of Fig. \ref{fig:gencopsgc} towards ($l=90^{\circ}$) and against
($l=270^{\circ}$) the Galactic rotation directions: the $0^{\circ} < l <
180^{\circ}$ CORAVEL dwarfs have slightly more stars with RV$_{LSR} < 0$ than
RV$_{LSR} > 0$ and vice versa for $180^{\circ}  < l < 360^{\circ}$ (RV$_{LSR}$
exhibits the pattern in equation \ref{equ:pecgsr} and the middle right plot in
Fig. \ref{fig:gencopsgc} because it mirrors RV$_{GSR}$).  Although F05 showed that the presence of
 dynamical streams causes there to be a net radial motion in the solar neighbourhood (see Appendix A), this is in phase space and 
 independent of $l$.  Thus the $U^{los}$ contribution to RV$_{LSR}$ is isotropic as illustrated
 by the symmetry in RV$_{LSR}$ between the regions near the cardinal $U$ directions ($l
 = 0^{\circ}$ and $l = 180^{\circ}$) in the middle left plot in Fig. \ref{fig:gencopsgc}.

It is not appropriate to  correct RV$_{LSR}$ for Galactic rotation because we
are resolving RV$_{LSR}$ perpendicular to the Galactic plane and the
correction does not affect $W^{los}$ because $W = 0$ in equation
\ref{equ:pecgsr}.  However, because this correction is not applied to RAVE
stars towards the SGP ($b < -45^{\circ}$), the small component of RV$_{LSR}$
sensitive to Galactic rotation ($V^{los}$) propagates through equation
\ref{equ:Wlos}.  The effect of $V^{los}$ is just visible in the bottom left
plot in Fig. \ref{fig:gencopsgc}, where sin($b < 0^{\circ}$) reverses the sign of the effect i.e. the $0^{\circ} < l < 180^{\circ}$ CORAVEL dwarfs have slightly more stars with $W^{los} > 0$ than $W^{los} < 0$ and vice versa for $180^{\circ}  < l < 360^{\circ}$.  A Kuiper symmetry test on the $W^{los}$ distribution of $b < -45^{\circ}$ CORAVEL dwarfs is similar to the results in Section \ref{s:dwarf_lsr} because the $W^{los}$ sample is symmetric about $l = 180^{\circ}$, so that the positive and negative $V^{los}$ contributions cancel out. 

A Kuiper symmetry test on the $W^{los}_{LSR}$
distribution of RAVE stars with $b < -45^{\circ}$ and $r > 15$ strongly rejects the null hypothesis (that $W^{los}_{LSR} > 0$ and
$W^{los}_{LSR} < 0$ are drawn from the same parent population) at $Q \sim 1 \times 10^{-9}$, where $W^{los,\odot}_{LSR}$ = 4.1 km s$^{-1}$ to the nearest 0.1 km s$^{-1}$.
The bottom right plot in Fig. \ref{fig:gencopsgc} shows that, because the RAVE
sample is Southern celestial hemisphere only, it does not symmetrically sample the SGP out to $b = -45^{\circ}$.  This plot reveals the largest part of the $b < -45^{\circ}$ asymmetry at $l > 180^{\circ}$ (colour-coded green) is not observed by RAVE at $l < 180^{\circ}$ because it is in the Northern hemisphere.  The contribution of $V^{los}$ to RV$_{LSR}$ in this region of $l > 180^{\circ}$ cannot cancel with its $l < 180^{\circ}$ mirrored region so the Kuiper symmetry test finds the difference in the $W^{los}_{LSR}$ distribution.  

To rectify this, we only sample $b < -45^{\circ}$ RAVE fields that are
symmetrically positioned either side of $l =180^{\circ}$ (28~868 RVs
colour-coded red in the bottom right plot of Fig. \ref{fig:gencopsgc}).  Because RAVE observes some fields
more than once (normally different stars each visit), differing stellar
densities are visible in this plot.  In the Galactic sky-symmetric sample,
there are more $r > 15$ RVs with $l > 180^{\circ}$ (16 235) than $l <
180^{\circ}$ (12 633). Nevertheless, the next section shows that the Kuiper
symmetry test fails to reject the null hypothesis, showing that our Galactic
sky-symmetric sample can be used to search for streams without any further processing.  Its selection function is almost identical to Fig. \ref{fig:ravei}, apart from nearly all the $9 < I < 10$ mag stars come from  the first input catalogue rather than the second.  The stellar population proportions of the subsample are different to the whole RAVE sample.  F-G dwarfs have increased most to $\sim$61 per cent, K-M dwarfs have increased to $\sim$7 per cent and K-M giants have decreased to $\sim$32 per cent.  The internal RV error distribution of the subsample is the same as the whole sample with $r > 15$.

\subsection{Determining RAVE $W^{los}_{LSR}$}

\begin{figure}
\centering
        \psfig{figure=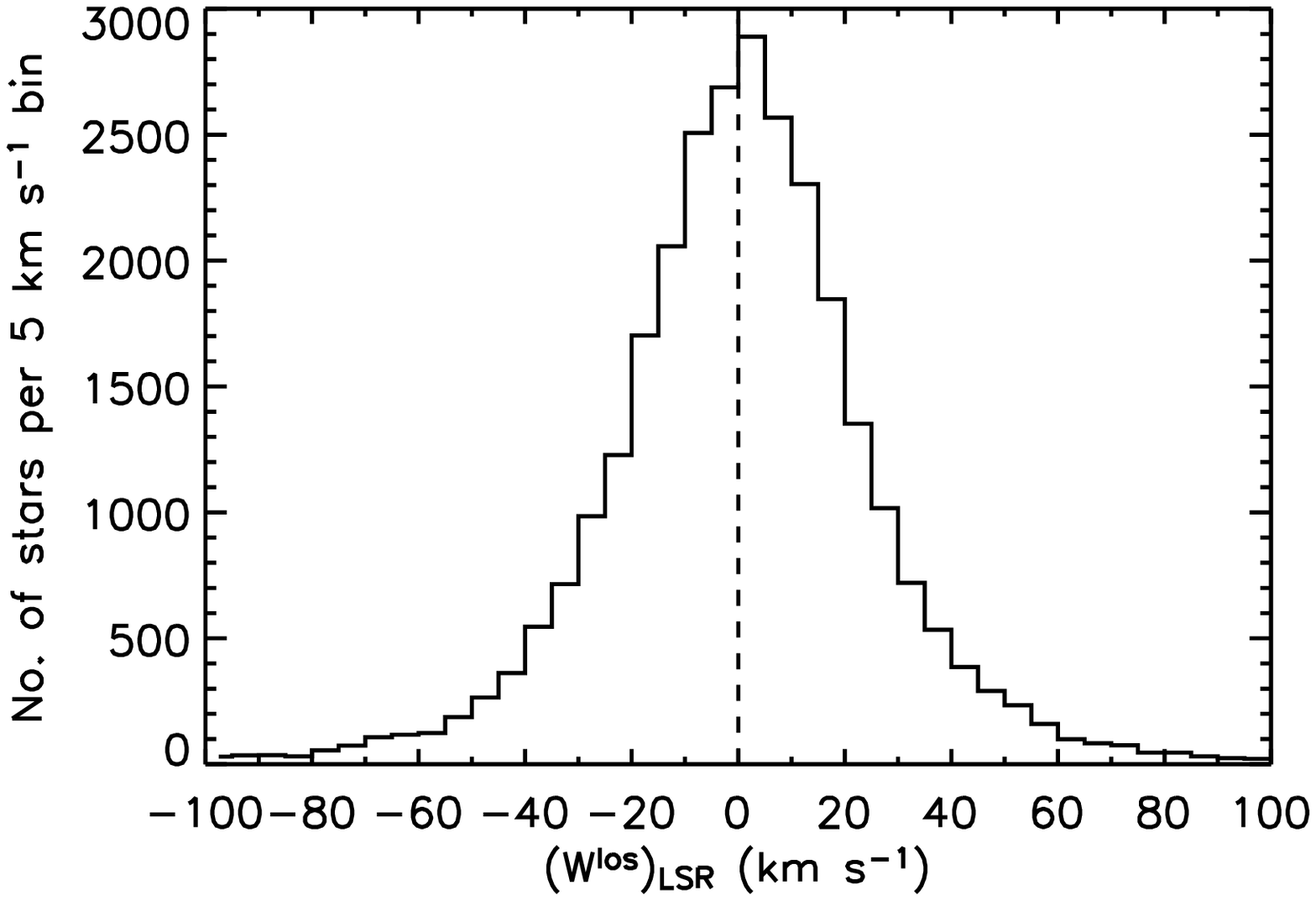,width=1.\columnwidth}
        \psfig{figure=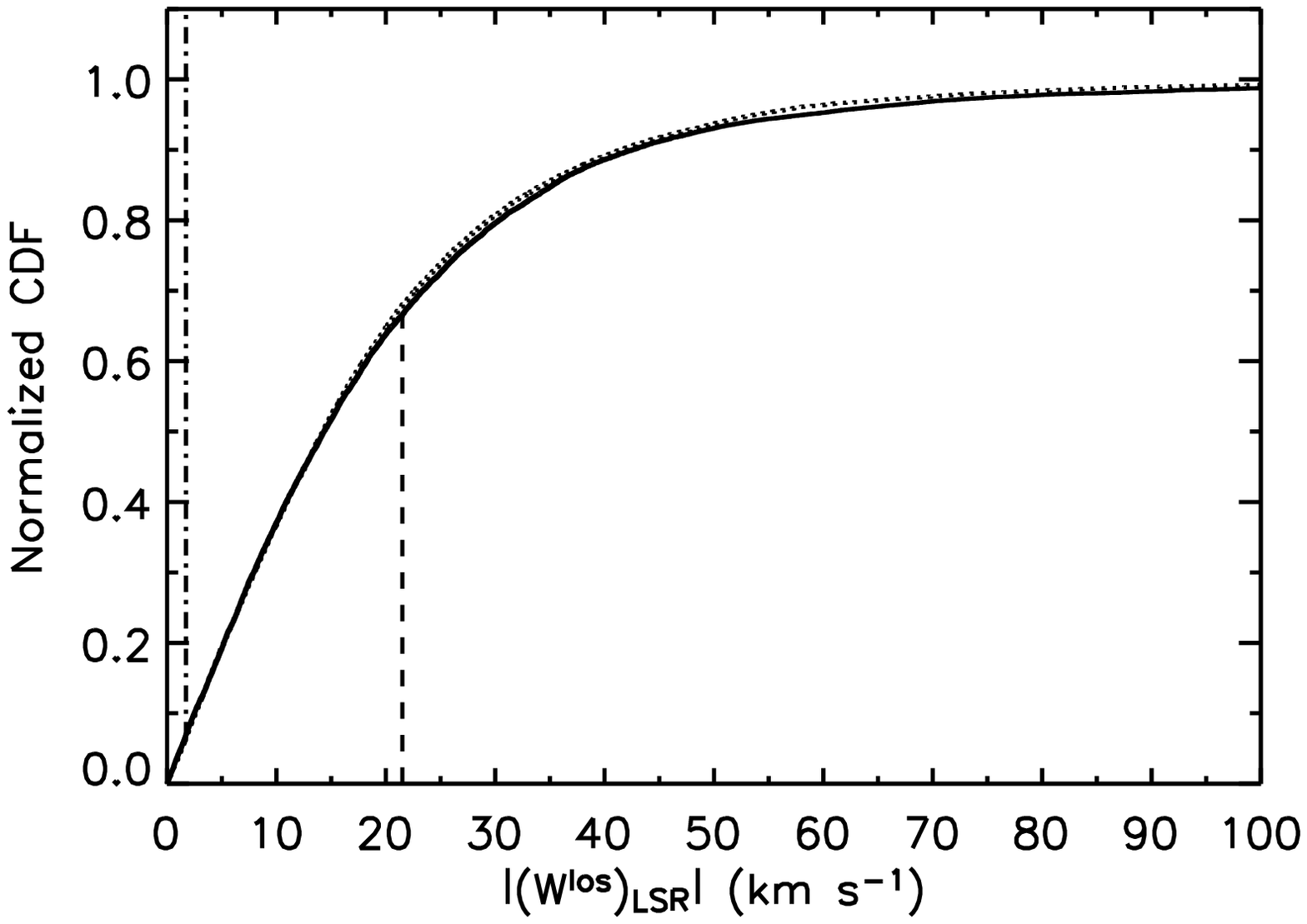,width=1.\columnwidth} 
\caption{{\it Top:} $W^{los}_{LSR}$ distribution of RAVE stars with $b < -45^{\circ}$ 
        (solid histogram) either side of the $W^{los}_{LSR} = 0$ dividing
        line (dashed lines).  The 5 km s$^{-1}$ velocity bin sizes are chosen
        to be more than twice the mode of the RAVE external RV error ($\sim$2 km s$^{-1}$).
                  {\it Bottom:} Normalized CDF of $W^{los}_{LSR} < 0$ (solid lines) and $W^{los}_{LSR} > 0$ (dotted lines), showing $D_{-}$ (dashed lines) and $D_{+}$ (dot-dashed lines).}
                \label{fig:RAVE_kuiper_lsr}
\end{figure}

We repeated the Kuiper test in Section \ref{s:binary}, comparing the $b <
 -45^{\circ}$ RV$_{LSR}\sin b$ distribution of all the CORAVEL dwarfs with the
 single dwarfs and found them to be more similar than the $W_{LSR}$ test ($Q$
 = 0.9998 but $N_{e}$ is only 779).  This demonstrates that all the RAVE (and SEGUE) RVs can also be used to look for tidal streams without worrying about the affects of binarity.  This section repeats the technique applied to the CORAVEL stars in Sections
 \ref{s:dwarf_lsr} and \ref{s:giant_lsr} to RAVE stars.  
  
The top of Fig. \ref{fig:RAVE_kuiper_lsr} shows the $W^{los}_{LSR}$ distribution for the minimum value of $D$, where  $W^{los,\odot}_{LSR}$ = 1.7 km s$^{-1}$ to the nearest 0.1 km s$^{-1}$.  The difference between this value and the $W^{\odot}_{LSR}$ = 7.0 km s$^{-1}$ values derived from the CORAVEL dwarfs and giants highlights that $W^{los}_{LSR} \not\equiv W_{LSR}$ but merely a proxy.   The $D_{-}$ and $D_{+}$ positions in the bottom plot of Fig. \ref{fig:RAVE_kuiper_lsr} are similar to the CORAVEL positions in Figs \ref{fig:dwarf_kuiper_lsr} and \ref{fig:giant_kuiper_lsr}, suggesting $W^{los}_{LSR}$ does approximate $W_{LSR}$.  $W^{los,\odot}_{LSR}$ is nearly in agreement with the mode of the $W^{los}_{LSR}$ distribution in Fig. \ref{fig:RAVE_kuiper_lsr}.  Again, this suggests that there cannot
 be many stars belonging to vertically coherent streams with systemic
 $W_{LSR}$ velocities. Table 5 shows that the RAVE $N_{e}$ value is more than double the value for the CORAVEL samples.  The RAVE $Q$ value is double the value for the CORAVEL dwarfs but less than the value for the CORAVEL giants.  The RAVE $\Delta W^{los,\odot}_{LSR}$ values are more symmetric than the CORAVEL values.
 
\begin{table*}
\begin{minipage}{170mm}
\centering
\label{tab:RAVE_kuiper_centre}
  \caption{Results of the Kuiper symmetry test, applied to our RAVE sample, to determine $W^{los,\odot}_{LSR}$ to the nearest 0.1 km s$^{-1}$, where $\Delta W^{los,\odot}_{LSR} = W^{los,\odot}_{LSR} - 1.7$ km s$^{-1}$.}
      \begin{tabular}{@{}rrccccccc@{}}
  \hline
$W^{los,\odot}_{LSR}$  & $\Delta W^{los,\odot}_{LSR}$  & $N_{W^{los}_{LSR} < 0}$ & $N_{W^{los}_{LSR} > 0}$ & $N_{e}$ & $D_{-}$ & $D_{+}$ & $D$ & $Q$\\
\hline
0.4 & $-1.3$ & 14 742     &  14 126    &  7214 &   0.0318&  0.0005 & 0.0323 & $1 \times 10^{-5}$\\
1.7 & 0.0 & 14 026    &   14 842   &   7211 & 0.0146  & 0.0053 &    0.0199 & 0.0658\\
2.7 & 1.0 &  13 455  &     15 413    &  7184 &  0.0119 &    0.0196 &   0.0315 &  $3 \times 10^{-5}$\\
   \hline
\end{tabular}
\end{minipage}
\end{table*}

\begin{table*}
 \begin{minipage}{170mm}
 \centering
 \label{tab:RAVE_kuiper}
   \caption{Results of the Kuiper symmetry test, applied to our RAVE sample, to determine the approximate
     number of tidal stream stars ($N_{s}$), required in pseudo-randomly
     generated $\sigma_{s} = 10$ km s$^{-1}$ Gaussians, placed at
     $\pm$(1, 2, 3, $\sim$7)$\sigma_{W^{los}_{LSR}}$ in the $W^{los}_{LSR}$
     distribution, to cause the test to reject the null hypothesis at
     4$\sigma$ ($Q < 6 \times 10^{-5}$).   \% =
     $N_{s}/(N_{s} + N_{total})$ where $N_{total}$ = 28 868. $\rho_{s}$ is the
     range of stellar density of each stream, in $N$ stars kpc$^{-3}$,
     calculated by dividing the $N_{s}$ of each stream by our RAVE sample volume ($\sim$8 kpc$^{-3}$, see
     text for details), modulated by the 5-15 per cent RAVE completeness,
     which increases $\rho_{s}$ by a factor of 20-6.7 respectively.}
     \begin{tabular}{@{}rrrrccccccc@{}}
   \hline
 $\sigma_{W^{los}_{LSR}}$ & $N_{s}$ & \% & $\rho_{s}$ & $N_{W^{los}_{LSR} < 0}$ & $N_{W^{los}_{LSR} > 0}$ & $N_{e}$ & $D_{-}$ & $D_{+}$ & $D$ & $Q$\\
 \hline
      $-1$ & 500 & 1.7 & 400-1300 & 14 525 & 14 843 & 7341 & 0.0303 & 0.0031 & 0.0334 & $5\times10^{-6}$\\
            1 & 600 & 2.0 & 500-1500 & 14 027 & 15 441 & 7350 & 0.0135 & 0.0185 & 0.0320 & $2\times10^{-5}$\\
      $-2$ & 300 & 1.0 & 250-750 & 14 326 & 14 842 & 7290 & 0.0296 & 0.0039 & 0.0335 & $5\times10^{-6}$\\
            2 & 500 & 1.7 & 400-1300 & 14 026 & 15 342 & 7327 & 0.0074 & 0.0235 & 0.0309 & $4\times10^{-5}$\\
      $-3$ & 300 & 1.0 & 250-750 & 14 326 & 14 842 & 7290 & 0.0318 & 0.0039 & 0.0357 & $6\times10^{-7}$\\ 
                  3 & 600 & 2.0 & 500-1500 & 14 026 & 15 442 & 7350 & 0.0041 & 0.0310 & 0.0351 & $9\times10^{-7}$\\ 
$\sim$$-$8 & 300 & 1.0 & 250-750 & 14 326 & 14 842 & 7289 & 0.0319 & 0.0039 & 0.0358 & $5\times10^{-7}$\\ 
   $\sim$8 & 500 & 1.7 & 400-1300 & 14 026 & 15 342 & 7327 & 0.0002 & 0.0321 & 0.0323 & $1\times10^{-5}$\\
     \hline
 \end{tabular}
 \end{minipage}
 \end{table*}

\subsection{Sensitivity of RAVE $W^{los}_{LSR}$ symmetry to tidal streams}
\label{s:rave_stream}

\begin{figure*}
\centering
\begin{minipage}{180mm}
        \psfig{figure=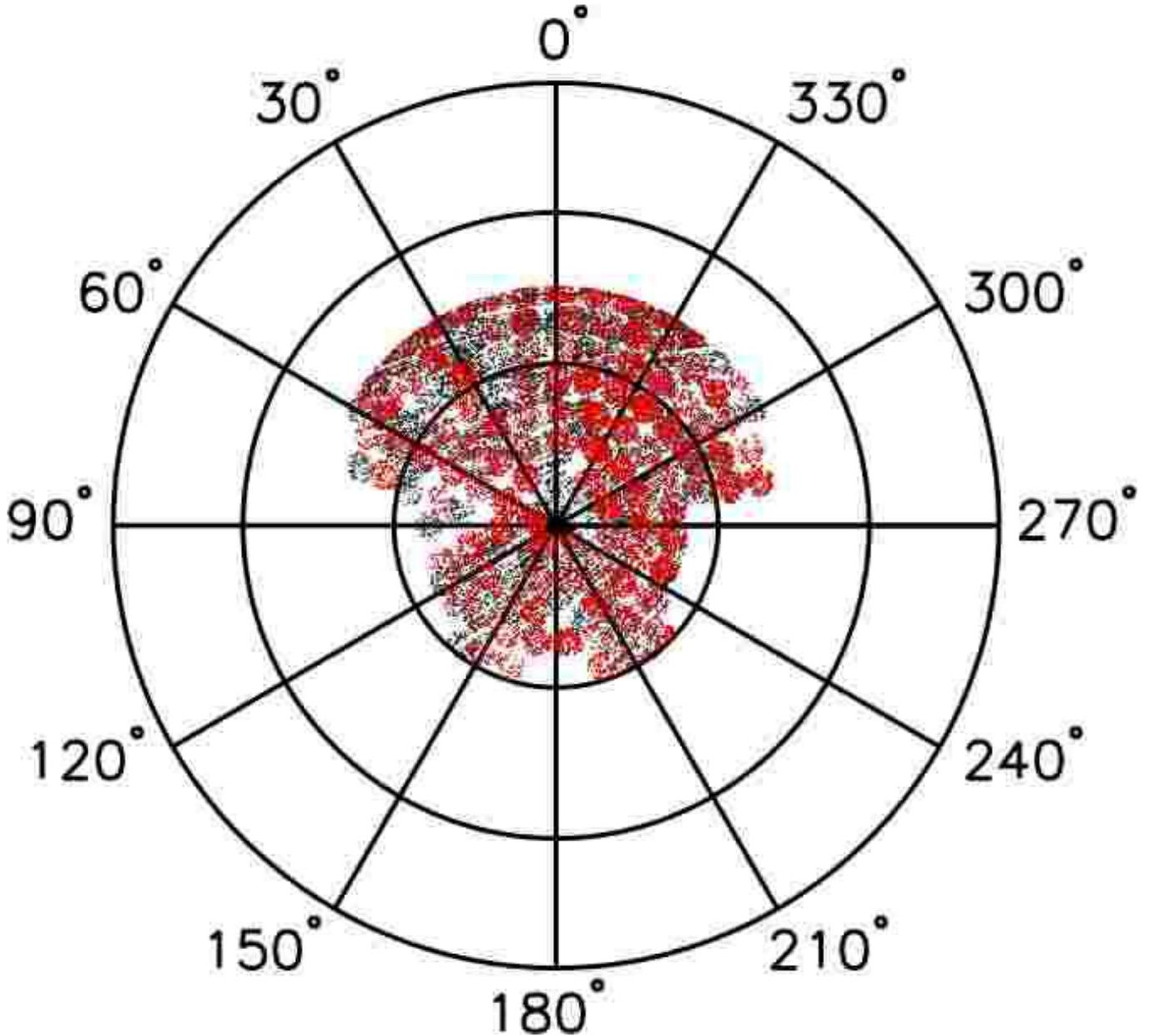,width=1.0\columnwidth}
          \caption{Same as the bottom right plot of Fig. \ref{fig:gencopsgc}
            except only the Galactic sky-symmetric RAVE sample is plotted and
        is colour-coded according to input catalogue: 5.7$^{\circ}$ grid
        spacing (red) and 5.0$^{\circ}$ grid
        spacing (black).  The blank strips (aligned with the declination axis)
        in between the black dots are due
        to the incomplete sky coverage of DENIS data.}
        \label{fig:fields}
\end{minipage}
\end{figure*}

To estimate the volume probed by our RAVE sample, we consider that RAVE stars fainter
than $I$ = 12 mag contribute a negligible number of stars to the sample (see
Fig. \ref{fig:ravei}).  This sets a distance limit for a typical RAVE giant ($M_{I} =
-0.25$ mag) at $\sim$3 kpc.  We approximate the volume of each RAVE field as a
cone of height 3 kpc subtending the RAVE field diameter of 5.7$^{\circ}$.  Our sample is complicated by the inclusion of stars from two input catalogues with the co-ordinates of its field centres based on grids offset from each other.  Fig. \ref{fig:fields} shows that the first RAVE input catalogue (colour-coded red) consists of contiguous fields on a $5.7^{\circ}$ grid.  The second RAVE input catalogue (colour-coded black) is based on a $5.0^{\circ}$ grid and so the 5.7$^{\circ}$ field of view 6dF fields overlap each other, as well as overlapping the fields of the first input catalogue.  We have counted the number of contiguous (red) fields in Fig. \ref{fig:fields} and estimated, to the nearest half field, the additional sky covered by the non-overlapping parts of the other fields (black).  The sample consists of $\sim$113 fields, giving a volume
of $\sim$8 kpc$^{3}$.  

\citet{steinmetz2006paper} estimated the RAVE
completeness of the old {\it Tycho-2} and SSS input catalogue in their fig. 4 to be $\sim$15
per cent compared to DENIS at the bright end of the selection function ($I
\sim$ 10 mag).  This drops to $\sim$5
per cent at the faint end ($I
\sim$ 12 mag).  Subtle colour biases exist in the
old input catalogue but more than half our sample comes from the unbiased DENIS
input catalogue.  This should reduce the effect of the biases and increase the
completeness levels but we use the old input catalogue levels as a
conservative estimate.  

This section repeats the technique applied to the CORAVEL stars in Sections
\ref{s:dwarf_stream} and \ref{s:giant_stream} to the RAVE stars, except the
RAVE dispersion is much larger: $\sigma_{W^{los}_{LSR}} \approx$ 29 km
s$^{-1}$ and thus now $\pm$8$\sigma_{W^{los}_{LSR}}$ approximately corresponds to the $\pm$$W^{los}_{LSR}$ values of the H99b
streams. 

Table 6
shows that the positions of $D_{-}$ and $D_{+}$ appear to cause the $+W^{los}_{LSR}$ distribution to be insensitive to the position of the stream and the number of its members.  As before, the test becomes more sensitive (less stream
stars required to generate a $4\sigma$ detection), the further a stream is from the centre of the
distribution in the $-W^{los}_{LSR}$ distribution.  However, Table 6 shows that, again, the number of stream stars required for a $4\sigma$
detection stays
approximately constant (saturates) at $W_{LSR} \le -3$$\sigma_{W_{LSR}}$.

\begin{figure}
\centering
               \psfig{figure=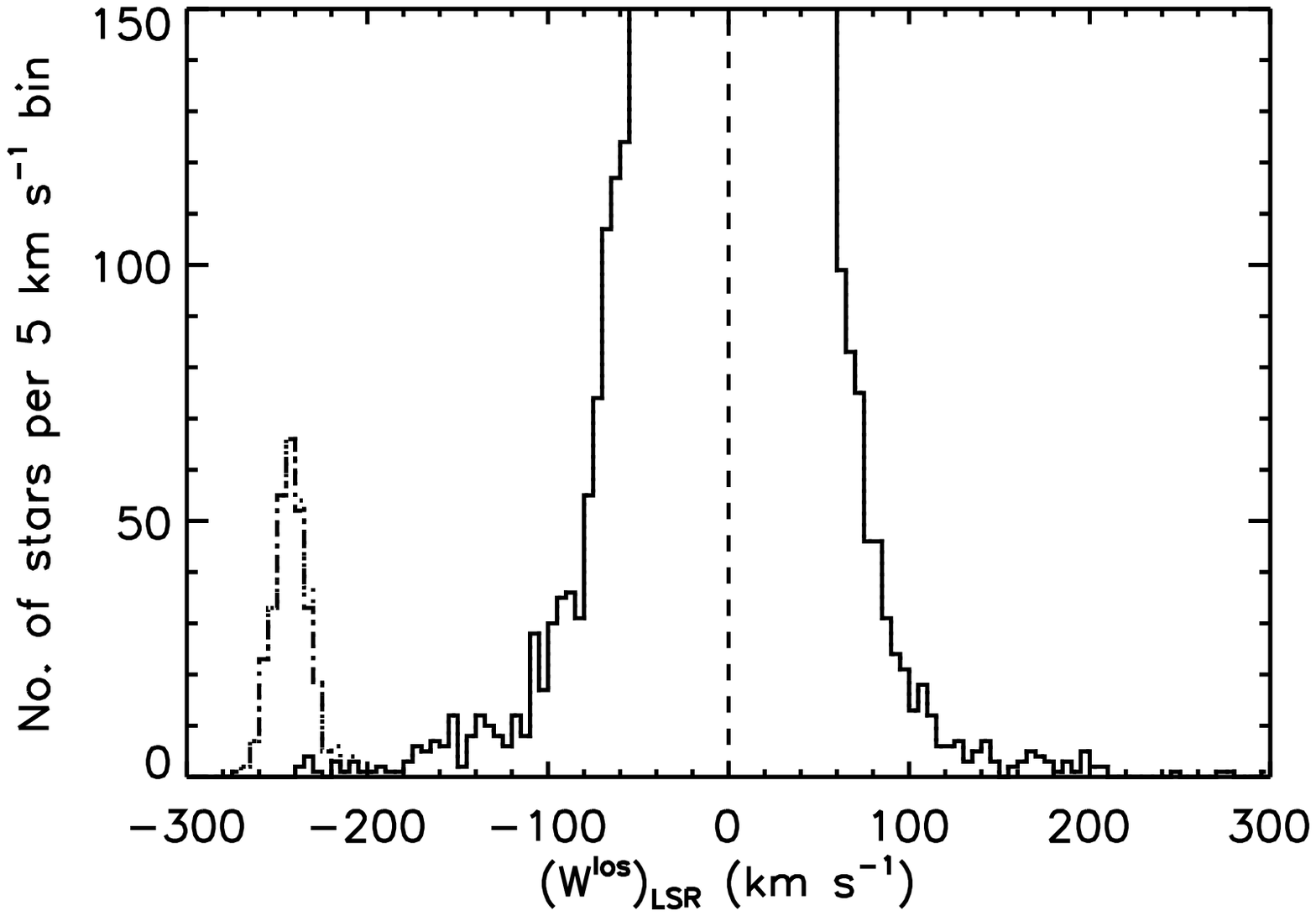,width=1.\columnwidth}
        \psfig{figure=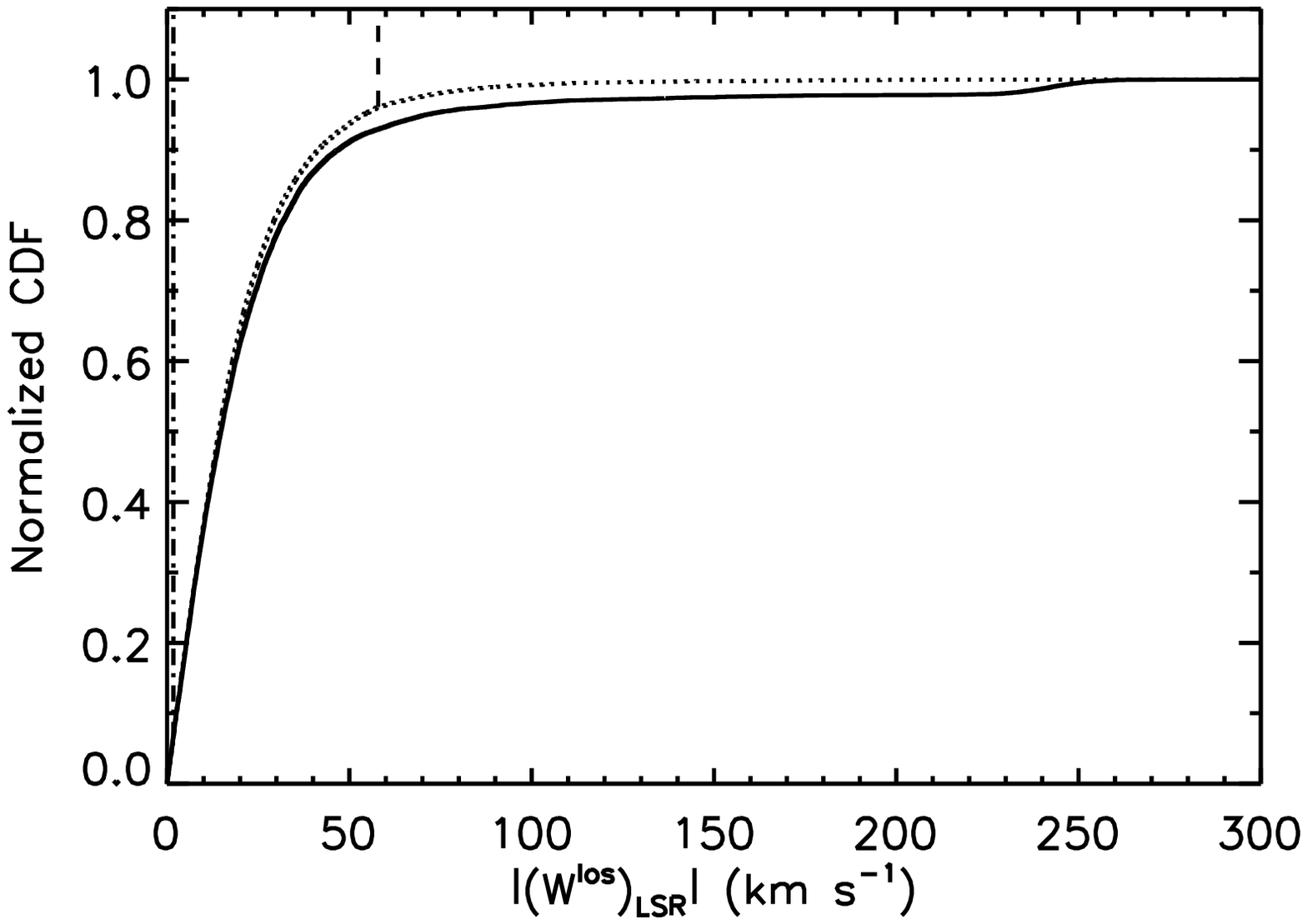,width=1.\columnwidth}  
\caption{{\it Top:} RAVE $W^{los}_{LSR}$ distribution (solid histogram) either
               side of the $W^{los}_{LSR} = 0$ dividing line (dashed line).
               The randomly-generated Gaussian (representing a $\sigma = 10$ km
               s$^{-1}$ tidal stream) included in the data set (dotted
               histogram) that cause the Kuiper symmetry test to reject the
               null hypothesis at 4$\sigma$ when placed at $\sim$$-$8$\sigma_{W_{LSR}}$ is also plotted separately (dot-dashed line).  {\it Bottom:} Normalized CDF of $W^{los}_{LSR} < 0$ (solid line) and $W^{los}_{LSR} > 0$ (dotted line), showing $D_{-}$ (dashed line) and $D_{+}$ (dot-dashed line, just visible at $|W^{los}_{LSR}| \approx 0$).}
                \label{fig:RAVE_kuiper_gausSgr}
\end{figure}

Fig. \ref{fig:RAVE_kuiper_gausSgr} shows that
there are a few RAVE stars with similar $W^{los}_{LSR}$ as both the $+W_{LSR}$
($250 < W_{LSR} < 350$ km s$^{-1}$)  and
$-W_{LSR}$ ($-250 < W_{LSR} < -200$ km s$^{-1}$) H99b streams.  Without distances to these stars to check their
orbital angular momenta, they cannot be confirmed as members of the H99b
streams.  Regardless of the symmetry between the two streams, Table 6 shows
that there are far too few of these stars for the Kuiper symmetry test to
detect them as a stream.  

Assuming 5 per cent RAVE completeness, the lower bound of the \citet{freese2005} estimates of the
Sgr stream stellar density predict that $\sim$80 RAVE stars could be Sgr
stream stars.  This increases to $\sim$590 stars using the upper bound.
Assuming a less conservative RAVE completeness level of 15 per cent, predicts
between $\sim$250-1780 Sgr stars.  The number of VOD stars that could be in RAVE is $\gtrsim$48 000.  Therefore, the superior
volume of RAVE rules out the possibilities that the Sgr stream and VOD cross the
Galactic disc through
the solar neighbourhood.  It also puts much tighter constraints than the
CORAVEL surveys on the density
of any stream that could be passing through the solar neighbourhood, ruling
out any streams with similar densities to the Sgr stream. 

\subsection{Comparing RAVE and CORAVEL $W^{(los)}_{LSR}$ and sampled volumes}

\begin{figure}
\centering
        \psfig{figure=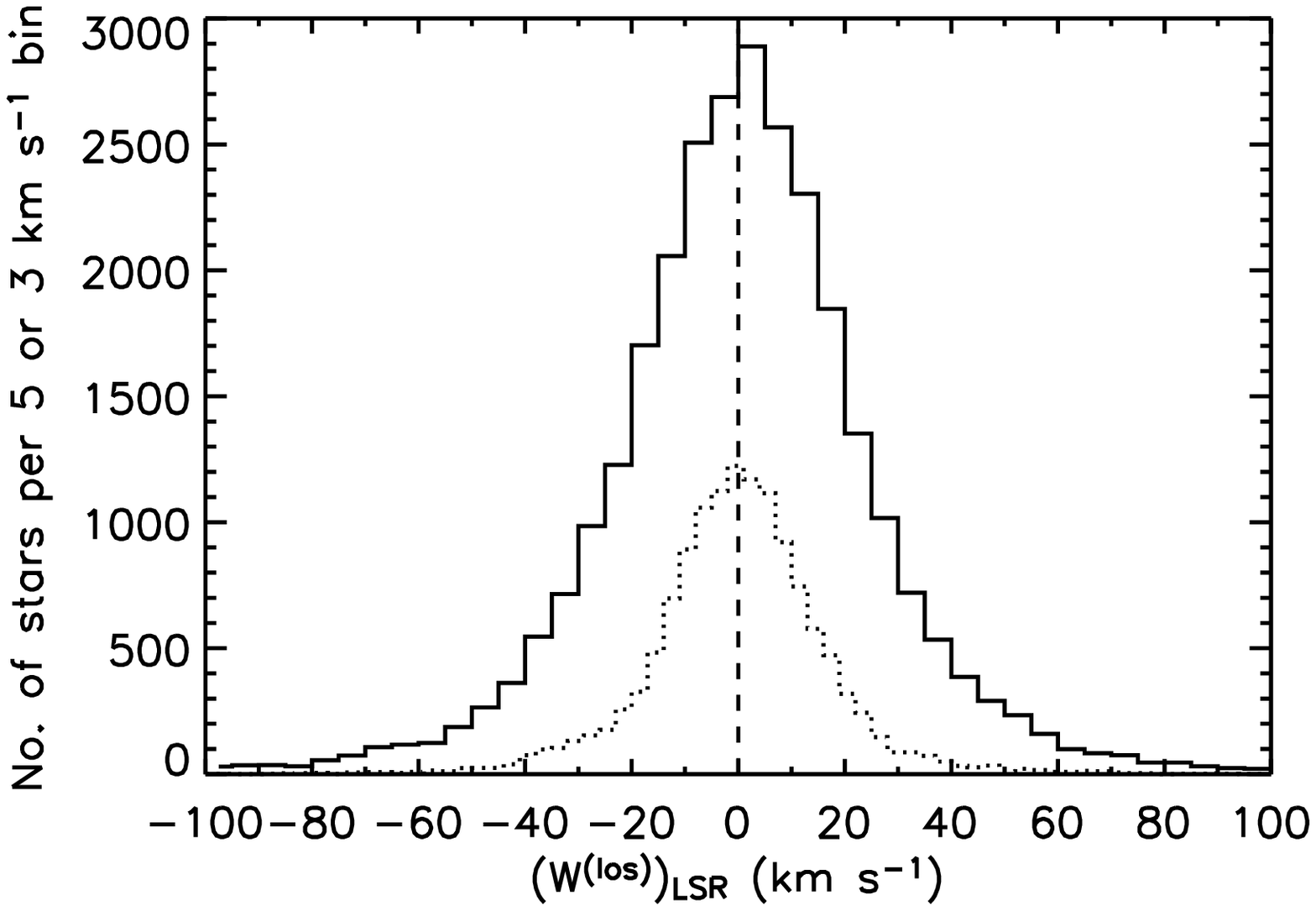,width=1.\columnwidth}
            \psfig{figure=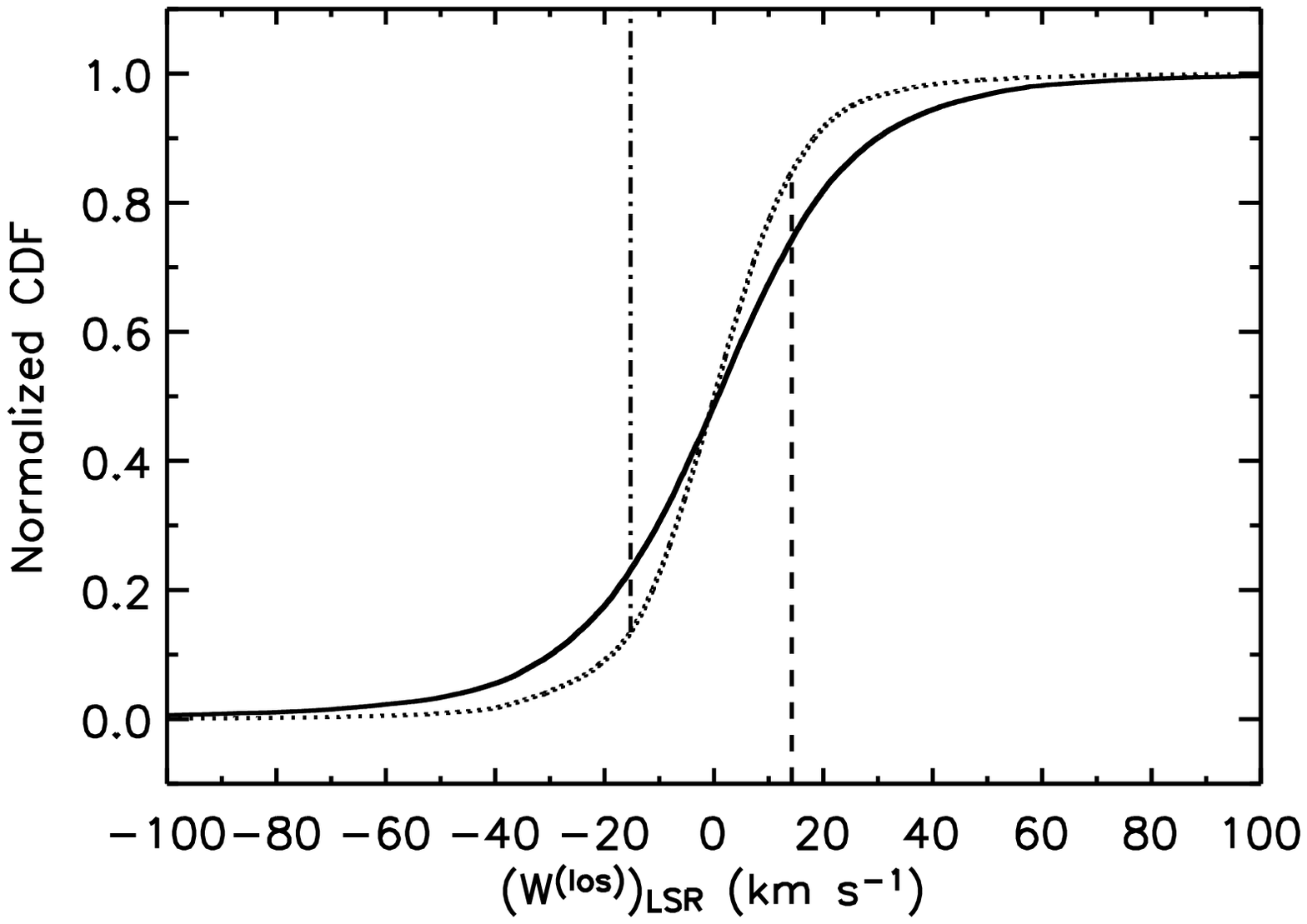,width=1.\columnwidth}
              \caption{{\it Top:} $W^{(los)}_{LSR}$ distribution of the 28 868 RAVE stars
        (solid histogram) and the 13 240 CORAVEL dwarfs (dotted histogram).  {\it Bottom:} Normalized CDF of RAVE (solid line) and CORAVEL  dwarfs (dotted line) as a function of $W^{(los)}_{LSR}$, where the maximum differences between them are indicated by vertical lines: $D_{+}$ (dot-dashed line) and $D_{-}$ (dashed line).}  
                                  \label{fig:rave_dwarfs}
                                            \end{figure}

\begin{figure}
\centering
 \psfig{figure=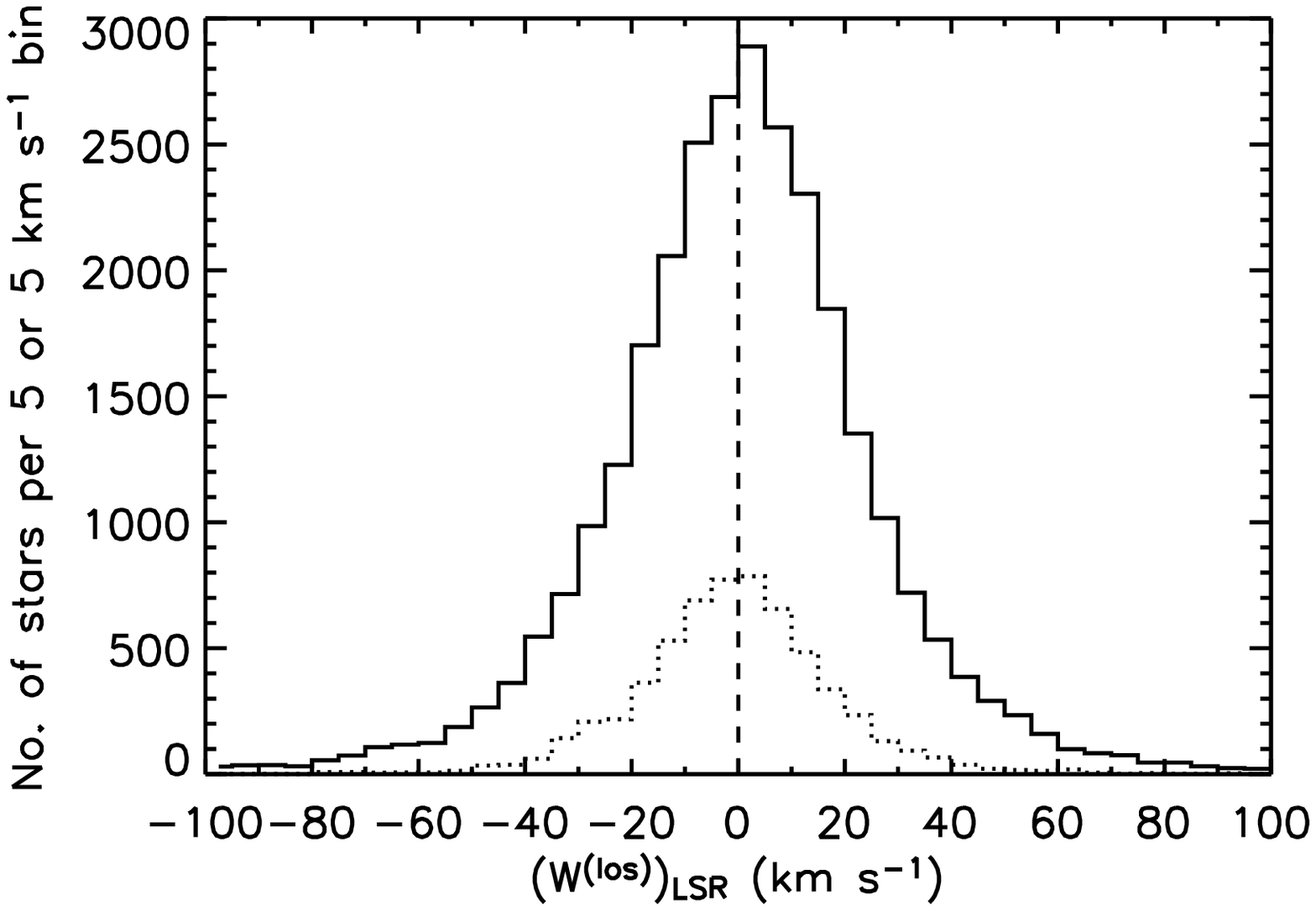,width=1.\columnwidth}
 \psfig{figure=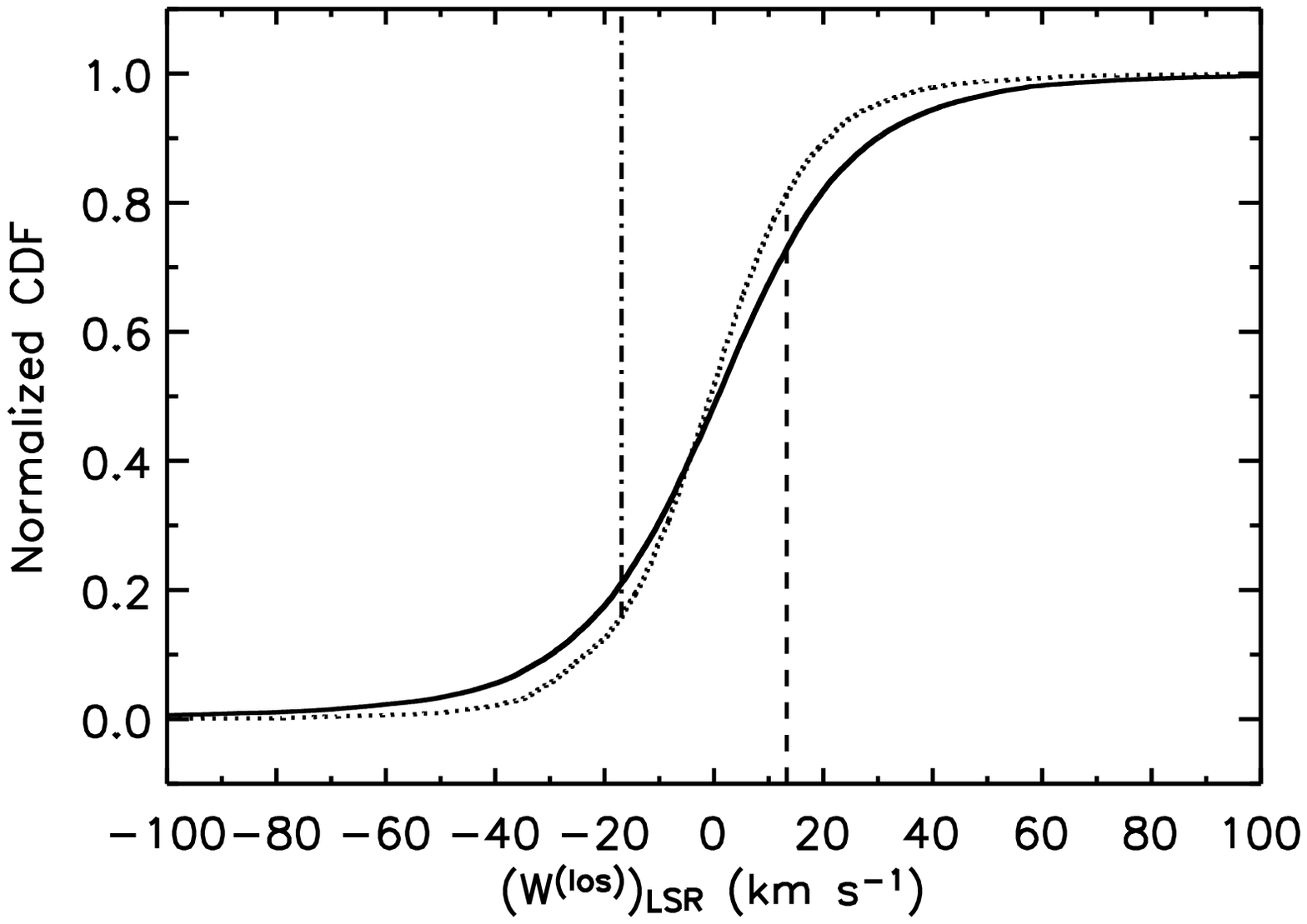,width=1.\columnwidth}
              \caption{{\it Top:} $W^{(los)}_{LSR}$ distribution of the 28 868 RAVE stars
        (solid histogram) and the 6030 CORAVEL giants (dotted histogram).  {\it Bottom:} Normalized CDF of RAVE (solid line) and CORAVEL  giants (dotted line) as a function of $W^{(los)}_{LSR}$, where the maximum differences between them are indicated by vertical lines: $D_{+}$ (dot-dashed line) and $D_{-}$ (dashed line).}  
                                  \label{fig:rave_giants}
                                            \end{figure}

\begin{figure}
\centering
        \psfig{figure=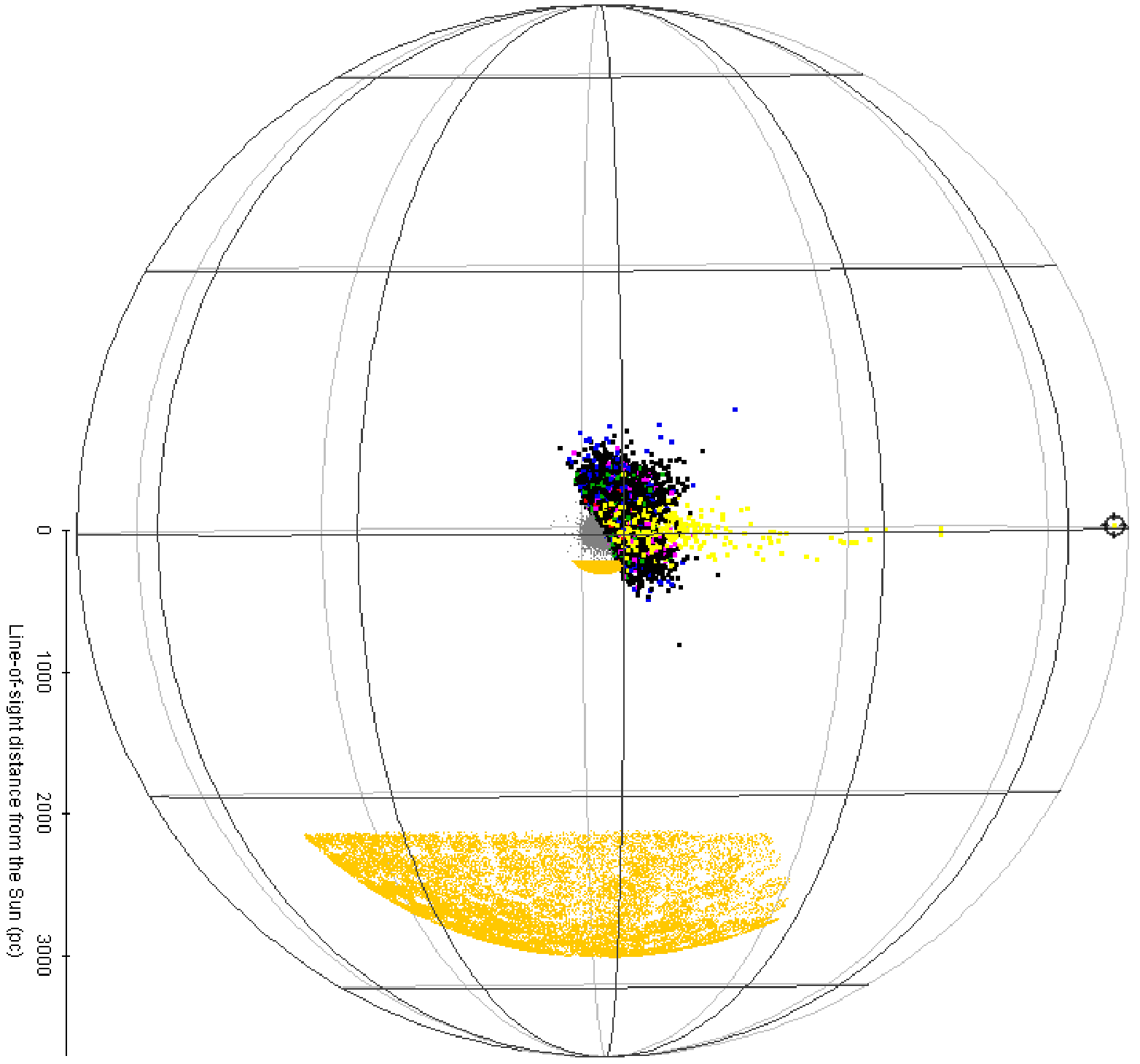,width=1.0\columnwidth}
 \caption{Three dimensional Galactic sky distribution of all the 13 092
        CORAVEL giants with $W$ space
  velocities 
(grey), the 6030 CORAVEL giants
        with $W$ space velocities, including spectroscopic binaries with
        centre-of-mass RVs, colour-coded according to their maximum
        likelihood base group
        assignment in Fig. \ref{fig:famaeyj},
  and the south Galactic
  cap RAVE sample (orange), where each RAVE star has been assigned a distance
  of 300 pc, which partitions the maximum distance range of RAVE K-M dwarfs and the minimum distance range of RAVE F-G dwarfs, and a distance of 3 kpc to represent the maximum distance range of $I$ = 12 mag RAVE K-M giants.
  The spherical polar axes have a radius of 3.7 kpc, centred on the
  Sun, chosen to include the most distant CORAVEL giant with a $W$
  velocity (3.6 kpc, marked by the black cursor) with the same viewer's angle as in Fig. \ref{fig:famaey3d}.}
                \label{fig:rave3d}
\end{figure}

Table 7 presents the Kuiper test results from comparing the entire RAVE
$W^{los}_{LSR}$ distribution with the entire CORAVEL dwarfs
(Fig. \ref{fig:rave_dwarfs}) and giants $W_{LSR}$ distributions
(Fig. \ref{fig:rave_giants}).  The test finds the largest differences at
$\sim$$\pm$1$\sigma_{W_{LSR}}$ in the CORAVEL distributions because of the
greater RAVE $\sigma_{W^{los}_{LSR}}$, due to the larger volume it samples.  A
comparison of the respective volumes sampled by the surveys is depicted in
Fig. \ref{fig:rave3d}.  It shows that only the $\sim$2000 RAVE K-M dwarfs
probe the same volume as the CORAVEL F-G dwarfs.  The $\sim$17 600 RAVE F-G
dwarfs fill in the SGP at $b < -45^{\circ}$ not probed by the CORAVEL giants.  It is the $\sim$9200 RAVE giants that really extends our tidal stream
search into the thick disc, its scaleheight being $\sim$1 kpc (Veltz et
al. 2007). 
Because of the differing distance ranges covered by the different RAVE stellar
populations, we have repeated our search procedure on each population
separately (as defined in Fig. \ref{fig:raverpm}).  This has been investigated in case e.g. the asymmetry of a thick disc or halo stream of giants is smeared out by the more numerous sample of dwarfs.  However,  a Kuiper symmetry test on each component fails to reject the null hypothesis. F-G dwarfs are found to be the most symmetric, then K-M dwarfs with the K-M giants the least symmetric.

\begin{table}
\centering
\label{tab:RAVE_CORAVEL}
  \caption{Kuiper test results comparing the RAVE $W^{los}_{LSR}$ distribution to the CORAVEL dwarfs and giants $W_{LSR}$ distributions.}
    \begin{tabular}{@{}lccccc@{}}
  \hline
CORAVEL & $N_{e}$ & $D_{-}$ & $D_{+}$ & $D$ & $Q$\\
\hline
Dwarfs & 9077 &  0.1060 &   0.0974 &    0.2034 & $<$$10^{-37}$\\
Giants & 4988 &  0.0863 &    0.0538 &     0.1401 & $<$$10^{-37}$\\
    \hline
\end{tabular}
\end{table}

\section{Discussion}
\label{s:discussion}

We have searched the CORAVEL and RAVE surveys for evidence of a net vertical
flow through the solar neighbourhood that could be associated with a tidal stream of
stars with vertically coherent kinematics.  Due to the current unavailability
of accurate distances to RAVE stars, we have used the Kuiper statistic to test
the symmetry of the vertical velocity distribution function of the samples.
In particular, the test serves as a useful first search for the presence of
prominent, kinematically coherent, vertical tidal streams in the RAVE data,
which probes further into thick disc and inner halo phase space than any other
survey of the local Galactic volume. Using the Geneva-Copenhagen stellar
binarity flag, we demonstrate that binarity has a statistically insignificant
effect on the $W$ and RV$_{LSR}\sin b$ velocity distribution
functions of the CORAVEL dwarfs, allowing us to use all the CORAVEL dwarfs and RAVE radial
velocities (RVs) in our search.

A tidal stream falling slowly through the solar neighbourhood volumes
considered (see Table \ref{tab:summary}) needs to be very
coherent for
our Kuiper symmetry test to detect it against the background $W^{(los)}_{LSR}$
distribution at $|W^{(los)}_{LSR}| \lesssim \sigma_{W^{(los)}_{LSR}}$.   
As
expected, there is a general trend that the test can detect smaller numbers of
stars in a coherent stream the further from the centre of the distribution
the stream is, out to $\sim$$\pm$3$\sigma_{W^{(los)}_{LSR}}$.  However, further into the tails of the
distribution the test becomes insensitive to the position of the stream and it
is the number of stream members that cause the $+W^{(los)}_{LSR}$ and
$-W^{(los)}_{LSR}$ CDFs to diverge sufficiently to generate a stream
detection.  This number appears to approximately saturate at
$|W^{(los)}_{LSR}| \gtrsim$ 3$\sigma_{W^{(los)}_{LSR}}$ and is set by the
detection threshold at $\sim$$\pm$3$\sigma_{W^{(los)}_{LSR}}$.  The required
number of stars is $\gtrsim$200.
Fig. \ref{fig:RAVE_kuiper_gausSgr} shows streams should be detectable by eye
in the extreme tails of a distribution where more specialized statistical
analysis could be applied to test their significance.  We have not found any
asymmetries in any of our samples that warrant such treatment. Table 8 summarizes the limits we have been able to place on the
properties of streams in the solar neighbourhood.

In contrast to the `Field of Streams' found in the outer halo
\citep{belokurov2006apaper}, we find that the volume of the solar
neighbourhood sampled by the CORAVEL and RAVE surveys is devoid of any tidal
streams with coherent vertical kinematics containing hundreds of stars.  This
suggests the Sagitarrius (Sgr) tidal stream does not pass through the solar
neighbourhood, which agrees with the latest empirical determination of its orbit
\citep{newberg2007} and dissociates it from the  \citet{helmi1999b} stream
\citep{freese2004}.  

The absence
 of the Sgr stream near the Sun is consistent with simulations of the
 disruption of Sgr in nearly spherical and prolate Galactic potentials
 \citep{helmi2004,law2005,fellhauer2006a,martinez-delgado2007} and seemingly
 inconsistent with oblate potentials \citep{law2005,martinez-delgado2007}.
 The constraints on the number of the Sgr stream stars in the solar
 neighbourhood volume searched in our samples could prove a useful tool for discriminating
 between Galactic potential models.

The CORAVEL Kuiper symmetry test results are supported by the lack of strong clumping
in orbital angular momenta space.  The limited range of the CORAVEL
giants in angular momenta space shows the \citet{majewski1992} and \citet{helmi1999b}, hereafter H99b, streams are not
present in this sample.  One of the CORAVEL
dwarfs is a member of one of the H99b streams.  There is another CORAVEL
dwarf with angular momenta and metallicity consistent with the H99b streams.  However, we consider it to be a possible outlier because its kinetic energy is too large to be consistent with the energies of the other members of the H99b streams.

The lack of a net vertical flow
through the solar neighbourhood argues against the Virgo overdensity (VOD) crossing the disc near the Sun.  This agrees with
\citet{juric2005paper} not finding it within a few kpc of the Sun.
Their preliminary analysis of 2MASS M giants did not reveal a similarly large
density enhancement to the VOD in the Southern
Galactic hemisphere.  
All this evidence is in accord with dark matter being smoothly distributed in configuration space throughout the solar neighbourhood.  

\begin{table}
\centering
\label{tab:summary}
  \caption{Summary of how many stream stars ($N_{s}$) and their density
  ($\rho_{s}$ in $N$ stars kpc$^{-3}$) are needed for the
    Kuiper symmetry test to detect them and how many stars ($N$) from the
    Sagitarrius stream (Sgr) and Virgo overdensity (VOD) are expected in our sample
    volumes (V), where the CORAVEL values correspond to the volumes within
    which they are complete and the RAVE volume is its total approximate
    volume (its volume completeness (VC) is taken into account).  Whether the Kuiper symmetry test is sensitive to the
    presence of each stream, given their non-local stellar density estimates in the
    literature is indicated: yes (y), no (n), maybe (?).}
    \begin{tabular}{@{}lrrrr@{}}
  \hline
Sample & \multicolumn{2}{c}{CORAVEL} & \multicolumn{2}{c}{RAVE}\\
                & Dwarfs   & Giants             & & \\
\hline
Section   & \ref{s:dwarf_stream} & \ref{s:giant_stream} & \multicolumn{2}{c}{\ref{s:rave_stream}}\\
V (kpc$^{3}$)   & 0.0003   & 0.0511  & \multicolumn{2}{c}{7.9052}\\
$N_{s}$ (low)       & 200      & 200 & \multicolumn{2}{c}{300} \\
$N_{s}$ (high)      & 600      & 800 & \multicolumn{2}{c}{600}\\
VC (\%)         & 100      & 100 & 5 & 15\\
$\rho_{s}$ (low)     & $0.7 \times 10^{6}$ & 4000 & 800 & 300\\
$\rho_{s}$ (high)    & $2.2 \times 10^{6}$  & 16 000 & 1500 & 500\\
$N$ Sgr (low)       & 0.1 (n) & 10 (n) & 80 (n) & 250 (?)\\ 
$N$ Sgr (high)      & 0.4 (n) & 80 (n) & 590 (y) & 1800 (y)\\
$N$ VOD             & 30 (n)   & 6000 (y) & 48 000 (y) & 144 000 (y)\\
  \hline
\end{tabular}
\end{table}

Comparison between the volume covered by SDSS photometry in the top right plot of \citet{juric2005paper}
fig. 20 and the local volume with measured phase space in Fig. \ref{fig:rave3d} highlights the disparity between the two.  
\citet{juric2005paper} extrapolate the four thin and thick
disc overdensities seen in their survey volume to the full Galactic disc ($|Z|
<$ 3 kpc, $R <$ 15 kpc) to imply that there are $\sim$20-40 
substructures of this type in the Galaxy, which could be the ``missing
satellites''.  It is not too surprising that none of these are within
the small Galactic volumes searched in this paper.

Depending on the availability and accuracy of metallicities in the second RAVE
data release (Zwitter et al. 2007, in preparation), distances to RAVE giants may be
sufficiently accurate to allow action-angle space or integrals of motion space
to be searched in the future for clustering indicative of tidal streams.
Additional photometric data in the future (e.g. from SkyMapper, 
\citealt{keller2007}) will further reduce distance errors to RAVE stars.
However, as in this study, accurate metallicity-derived distances are not
required if using RAVE stars towards the cardinal Galactic directions.  In the
rotation direction ($l \sim 270^{\circ}$), RVs are dominated by $V_{GSR}$, where the conversion between the two is only weakly dependent on distance
\citep{woolley1978,frenk1980}.  
Since $J_{Z}$ is conserved during phase-mixing, tidal streams should have similar $V_{GSR}$.  \citet{gilmore2002} found a tidal stream with coherent $V_{GSR}$ in between canonical thick disc and inner halo values.  The magnitude range of RAVE was chosen so that its giants probe the interface of the thin and thick discs and inner halo.  Hence, RAVE giants can be also used to investigate the angular momenta of the Milky Way's stellar components and its intervening stream(s?) (Seabroke et al. 2007, in preparation).

\section*{Acknowledgments}

GMS gratefully acknowledges financial support from a Particle Physics and Astronomy Research Council PhD
Studentship, the Gordon Wigan Fund, Gonville and Caius College and the Cambridge Philosophical Society Research Studentship Fund.

Funding for RAVE has been provided by the Anglo-Australian Observatory, by the Astrophysical Institute Potsdam, by the Australian Research Council, by the German Research Foundation, by the National Institute for Astrophysics at Padova, by The Johns Hopkins University, by the Netherlands Research School for Astronomy, by the Natural Sciences and Engineering Research Council of Canada, by the Slovenian Research Agency, by the Swiss National Science Foundation, by the National Science Foundation of the USA, by the Netherlands Organisation for Scientific Research, by the Particle Physics and Astronomy Research Council of the UK, by Opticon, by Strasbourg Observatory, and by the Universities of Basel, Cambridge, and Groningen.  
The RAVE web site is at www.rave-survey.org.

This publication makes use of data products from the Two Micron All Sky Survey, which is a joint project of the University of Massachusetts and the Infrared Processing and Analysis Center/California Institute of Technology, funded by the National Aeronautics and Space Administration and the National Science Foundation.

The DENIS project has been partly funded by the SCIENCE and the HCM plans of
the European Commission under grants CT920791 and CT940627.
It is supported by INSU, MEN and CNRS in France, by the State of Baden-W\"urttemberg 
in Germany, by DGICYT in Spain, by CNR in Italy, by FFwFBWF in Austria, by FAPESP in Brazil,
by OTKA grants F-4239 and F-013990 in Hungary, and by the ESO C\&EE grant A-04-046.  Jean Claude Renault from IAP was the Project manager.  Observations were  
carried out thanks to the contribution of numerous students and young 
scientists from all involved institutes, under the supervision of  P. Fouqu\'e,  
survey astronomer resident in Chile.  

\bibliographystyle{mn2e}
\bibliography{references}

\appendix
\label{s:appendix}

\section[]{Historical parallels}

\begin{table*}
   \centering
   \begin{minipage}{140mm}
 \label{tab:uvw}
   \caption{Comparison of the \citet{homann1886} solar space velocity components compared
     to more modern estimates.}.
   \begin{tabular}{@{}lllll@{}} 
   \hline 
 Author(s) & Data used & $U^{\odot}_{LSR}$ & $V^{\odot}_{LSR}$ & $W^{\odot}_{LSR}$ 
  \\
 & & (km s$^{-1}$) & (km s$^{-1}$) & (km s$^{-1}$) \\ 
 \hline      
  \citet{homann1886} & \citet{seabroke1879} RVs & 17.4 $\pm$ 11.2 & 16.9 $\pm$ 10.9 & 3.6 $\pm$ 2.3\\ 
 \citet{dehnen1998a} & {\it Hipparcos} $\pi$ \& $\mu$ & 10.00 $\pm$ 0.36 & 5.25 $\pm$ 0.62 & 7.17 $\pm$ 0.38\\
 This paper & CORAVEL dwarfs  & & & 7.0\\ 
 This paper & CORAVEL giants  & & & 7.0\\ 
 F05 & CORAVEL giants  & 10.25 $\pm$ 0.15 & & 7.98 $\pm$ 0.09\\
                    & Excluding streams & 2.78 $\pm$ 1.07 & & 8.26 $\pm$ 0.38\\
 Veltz et al. (2007) & RAVE RVs, UCAC2\footnote{UCAC2 is the Second US Naval Observatory CCD Astrograph
     Catalogue \citep{zacharias2004a}} $\mu$ & 8.5 $\pm$ 0.3 & & 11.1 $\pm$ 1.0\\
 \hline
 \end{tabular}
 \end{minipage}
 \end{table*}

\begin{figure}
\centering
        \psfig{figure=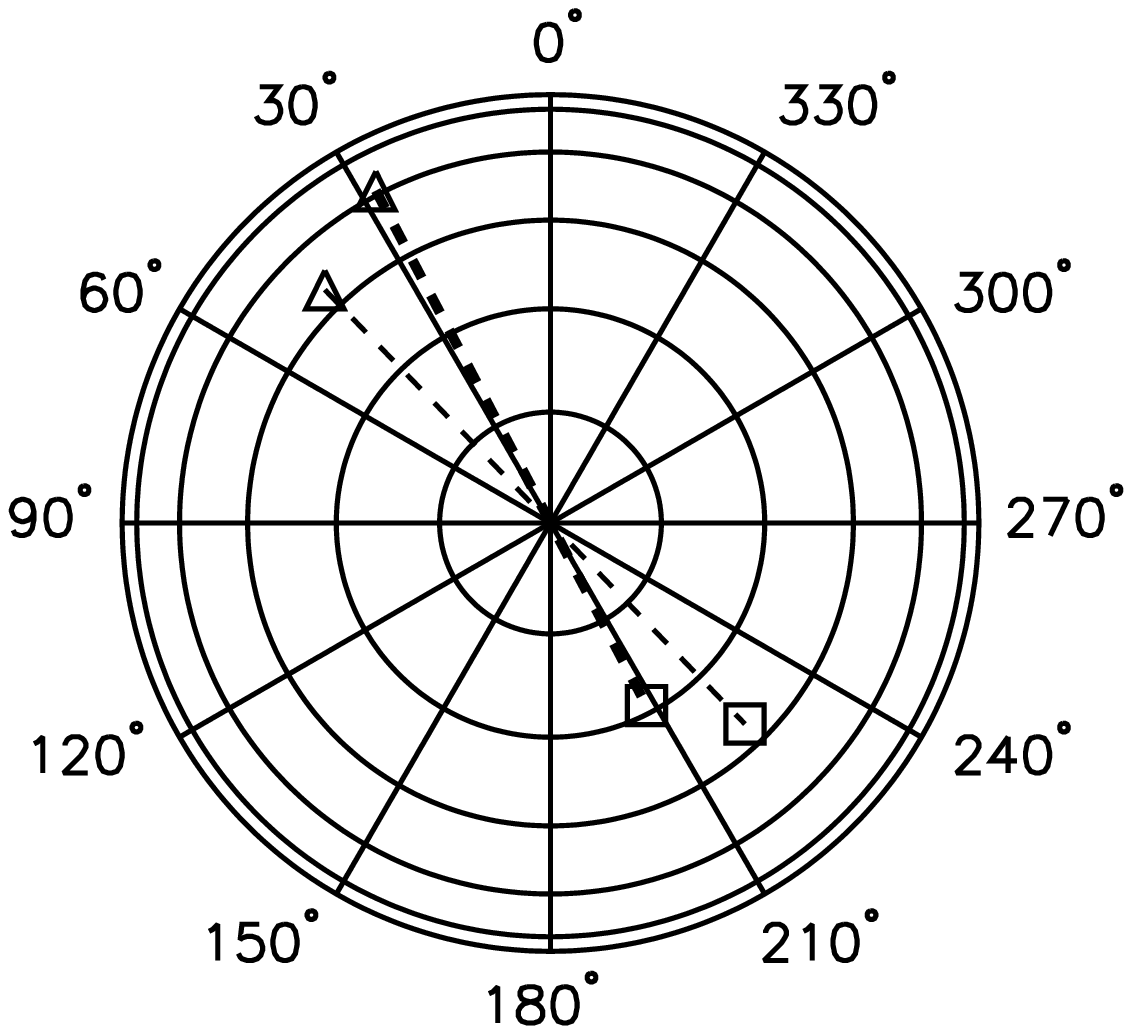,width=1.0\columnwidth}
       \caption{Lambert polar (equal area corresponds to equal solid angle on
        the sky) projections of the entire Galactic sky, where
        the SGP ($b = -90^{\circ}$) is at the centre and concentric circles correspond to
        constant $b$ at every 30$^{\circ}$ to the outermost circle at
        +90$^{\circ}$ (NGP), around which the radial lines of constant $l$ are
        labelled.  The Galactic directions of estimated values of the solar
        apex (triangles) and antapex (squares) are plotted. The thick dashed
        line connects the solar motion direction calculated from the
        \citet{dehnen1998a} solar space velocities in Table A1. The thin
        dashed line connects the solar motion direction calculated by
        \citet{homann1886}, calculated from \citet{seabroke1879} RVs.}
\label{fig:appendix}
\end{figure}

 Soon after  \citet{vogel1873}\footnote{A predecessor as director of Potsdam Observatory to MS, co-author of this paper and PI of the RAVE project.} measured the first 
  reliable RVs of Sirus and Procyon,  \citet{seabroke1879}\footnote{The
  great-great-grandfather of GMS, first author of this paper.} performed one of the first RV surveys,  measuring 68 RVs for  29 stars, followed by
  699 RVs for 40 stars \citep{seabroke1887,seabroke1887book} and 866 RVs for 49
  stars  \citep{seabroke1889}.  Although these visually determined RVs were individually inaccurate, when
 analysed statistically they showed that the Sun
  was moving relative to nearby stars at 24.5 $\pm$ 15.8 km s$^{-1}$
  \citep{homann1886}.  Within the (admittedly large) error, this measurement
  of solar speed with respect to the LSR agrees with the standard modern
  measurement of 13.38 $\pm$ 0.12 km s$^{-1}$ (calculated from the
  \citet{dehnen1998a} solar space velocities in Table A1).  Fig. \ref{fig:appendix} shows the direction of the \citep{homann1886} solar apex towards $l=44.1^{\circ}$, $b=8.5^{\circ}$ is approximately in the same direction as the \citet{dehnen1998a} values: $l = 27.72^{\circ}$, $b = 32.43^{\circ}$.
  
  We have resolved the \citet{homann1886} solar motion into its Galactic cardinal direction vectors,
  $U^{\odot}_{LSR}$, $V^{\odot}_{LSR}$ and $W^{\odot}_{LSR}$, assuming the solar apex direction is error-free.
  It almost certainly is not but \citet{homann1886} does not
  quote a directional error so the solar space velocity errors in Table A1 should be
  considered a lower limit.  The \citet{homann1886} $U^{\odot}_{LSR}$ agrees with more modern estimates within the
  errors while $V^{\odot}_{LSR}$ and $W^{\odot}_{LSR}$ are close to agreeing within the
  errors.

Over a century after the \citet{homann1886} derivation of the solar motion, the debate on this subject
 in the literature has been revived by the CORAVEL surveys.  F05 interpreted the solar neighbourhood substructure in the $U$-$V$ plane found in their sample of CORAVEL giants and in CORAVEL dwarfs by N04 (and in {\it Hipparcos} $\pi$, $\mu$ and RVs by \citealt{skuljan1999}) as dynamical streams.  Table A1 shows
 that when F05 excluded all the dynamical streams, they found a different $U^{\odot}_{LSR}$ from when they included all the CORAVEL giants, which agrees with the \citet{dehnen1998a} value.  This discrepancy raises the question of
 how to derive the solar motion in the presence of dynamical perturbations
 altering the kinematics of the solar neighbourhood.  They posed the question
 of whether a subset of stars exists in the solar neighbourhood that has no net
 radial motion, which can used as a reference against which to measure the
 solar motion.  The simulations of \citet{desimone2004} demonstrate that transient spiral waves can excite a net radial motion through the solar neighbourhood of $\sim$10 km s$^{-1}$.  
 
 Table A1 also shows that F05 found
 that the presence of dynamical streams in the solar neighbourhood has a
 negligible affect on $W^{\odot}_{LSR}$.  This suggests that the dynamical streams do
 not have a net vertical motion because they are well phase-mixed.  In this
 paper, we have shown that there is no net vertical flow through the solar
 neighbourhood, suggesting $W^{\odot}_{LSR}$ is an appropriate measure of the
 vertical solar motion, unbiased by the presence of in-falling tidal streams
 on to the Milky Way disc near the Sun.
 
\bsp

\label{lastpage}

\end{document}